\documentclass[prb,twocolumn,floatfix,superscriptaddress, longbibliography]{revtex4-1}
\usepackage[pdftex,plainpages=false,colorlinks=true,linkcolor=blue, citecolor=blue, urlcolor=blue]{hyperref}
\usepackage{amsfonts}
\usepackage{amsmath}
\usepackage{amssymb}
\usepackage{graphicx}% Include figure files
\usepackage{natbib}
\usepackage{color}
\usepackage{placeins}
\usepackage{physics}

\begin{document}
 
\title{Superconductivity in the two-dimensional Hubbard model with cellular dynamical mean-field theory: a quantum impurity model analysis}
\author{C. Walsh}
\affiliation{Department of Physics, Royal Holloway, University of London, Egham, Surrey, UK, TW20 0EX}
\author{M. Charlebois}
\affiliation{D\'epartement de Chimie, Biochimie et Physique, Institut de Recherche sur l’Hydrog\`ene, Universit\'e du Qu\'ebec \`a Trois-Rivi\`eres, Trois-Rivi\`eres, Qu\'ebec, Canada G9A 5H7}
\author{P. S\'emon}
\affiliation{D\'epartement de physique, Institut quantique \& RQMP, Universit\'e de Sherbrooke, Sherbrooke, Qu\'ebec, Canada J1K 2R1}
\author{A.-M. S. Tremblay}
\affiliation{D\'epartement de physique, Institut quantique \& RQMP, Universit\'e de Sherbrooke, Sherbrooke, Qu\'ebec, Canada J1K 2R1}
\author{G. Sordi}
\email[corresponding author: ]{giovanni.sordi@rhul.ac.uk}
\affiliation{Department of Physics, Royal Holloway, University of London, Egham, Surrey, UK, TW20 0EX}
\date{\today}

\begin{abstract}
Doping a Mott insulator gives rise to unconventional superconducting correlations. Here we address the interplay between d-wave superconductivity and Mott physics using the two-dimensional Hubbard model with cellular dynamical mean-field theory on a $2\times2$ plaquette. 
Our approach is to study superconducting correlations from the perspective of a cluster quantum impurity model embedded in a self-consistent bath. 
At the level of the cluster, we calculate the probabilities of the possible cluster electrons configurations. 
Upon condensation we find an increased probability that cluster electrons occupy a four-electron singlet configuration, enabling us to identify this type of short-range spin correlations as key to superconducting pairing. The increased probability of this four-electron singlet comes at the expense of a reduced probability of a four-electron triplet with no significant probability redistribution of fluctuations of charges. This allows us to establish that superconductivity at the level of the cluster primarily involves a reorganisation of short-range spin correlations rather than charge correlations. 
We gain information about the bath by studying the spectral weight of the hybridization function. Upon condensation, we find a transfer of spectral weight leading to the opening of a superconducting gap. 
We use these insights to interpret the signatures of superconducting correlations in the density of states of the system and in the zero-frequency spin susceptibility. 
\end{abstract}
 
\maketitle

 \section{Introduction}

The nature of the correlations leading to superconducting pairing in doped Mott insulators remains a central challenge to the understanding of high temperature cuprate superconductors~\cite{lee, Norman2011, keimerRev, Anderson:1987}. The two-dimensional (2D) Hubbard model~\cite{Hubbard1963, ArovasAnnuRev2022, QinAnnuRev2022}, which encodes both electrons hopping $t$ and Coulomb onsite repulsion $U$, is the simplest model to study superconducting correlations arising from a purely electronic mechanism~\cite{AMJulich}. The large value of the interaction strength $U$, which is necessary to open a Mott gap in the half filled model, requires the use of nonperturbative approaches. Cluster extensions~\cite{maier, kotliarRMP, tremblayR} of dynamical mean-field theory~\cite{rmp} have proved to be powerful tools for exploring strongly correlated superconductivity in the 2D Hubbard model~\cite{AMJulich, QinAnnuRev2022}. 

Over the years, many cluster DMFT studies have established that upon doping the Mott insulator realised by the 2D Hubbard model, a d-wave superconducting state occurs, with a dome-like shape in the temperature-doping phase diagram~\cite{maierSC, lkAF, maierSystem, tremblayR, hauleDOPING, kancharla, sshtSC, Gull:2013, Chen:2015, massimoAF, sakai2023}. This indicates that in a strongly correlated superconducting state the short range superexchange interaction leads electrons to pair up into coherent Cooper pairs. 
Doping a Mott insulator generates a rich phase diagram with many states that are close in energy~\cite{Dagotto:Science2005}, and ongoing effort is devoted to clarify whether or not the zero-temperature ground state of the system is superconducting, and for what parameters~\cite{Zheng:Science2017, Qin:PRX2020, Chung:PRB2020}. Despite the fact that long-range order is excluded by thermal fluctuations in 2D ~\cite{MWtheorem}, and irrespective of whether superconductivity is a true zero-temperature ground state, superconductivity obtained in cluster DMFT is a locally stable state of physical relevance. For example, the study of the emergence of superconductivity from the underlying normal state upon reducing the temperature gives information about the pairing mechanism. Furthermore, tweaking temperature and model parameters (such as third dimensionality and frustration) may cause superconductivity to become the state with lowest free energy~\cite{AMJulich}.

Within cluster DMFT, there is a variety of ways by which we can gain insights on this microscopic mechanism. 
One possible approach to the study of superconducting correlations in the 2D Hubbard model takes an energetic viewpoint, and thus analyses the relative change in potential and kinetic energy between the superconducting and the underlying normal state. The rationale of this approach is to identify whether the condensation energy arises from a gain in potential energy (potential energy driven superconductivity), as described by conventional BCS theory, or from a gain in kinetic energy (kinetic energy driven superconductivity). 
Intense cluster DMFT investigations~\cite{maierENERGY, carbone2006, millisENERGY, LorenzoSC} have shown that upon doping a Mott insulator, the kinetic energy decreases upon superconducting condensation and the doping interval where this kinetic energy driven mechanism occurs progressively extends to larger doping with increasing the interaction strength $U$. 

Another approach to study superconducting correlations in the 2D Hubbard model is through the dynamics of pairing. Within cluster DMFT, this approach comprises the study of the superconducting gap, of the frequency dependence of the anomalous part of the self-energy, of the anomalous spectral weight and its cumulative order parameter~\cite{maierPRL2008, Kyung:2009, civelli1, civelli2, senechalPRB2013, reymbautPRB2016}, and of several fluctuation diagnostic techniques~\cite{DongPNAS2022, DongNatPhys2022}. 
Overall, cluster DMFT studies on the dynamics of pairing have indicated the importance of short-range spin fluctuations for the pairing mechanism in the doped Hubbard model, at least at moderate coupling. 

Yet another approach to the study of the superconducting correlations in the 2D Hubbard model is from a quantum information perspective. The rationale of this approach is to characterise the entanglement properties of superconductivity~\cite{amicoRMP2008,zhengBOOK}. Common measures of quantum and classical correlations are entanglement entropies and quantum mutual information. Ref.~\onlinecite{CaitlinPNAS2021} has shown that the local entropy reflects the source of condensation energy and the quantum mutual information is enhanced in the superconducting state. 

\begin{figure}[ht!]
\centering{
\includegraphics[width=1.\linewidth]{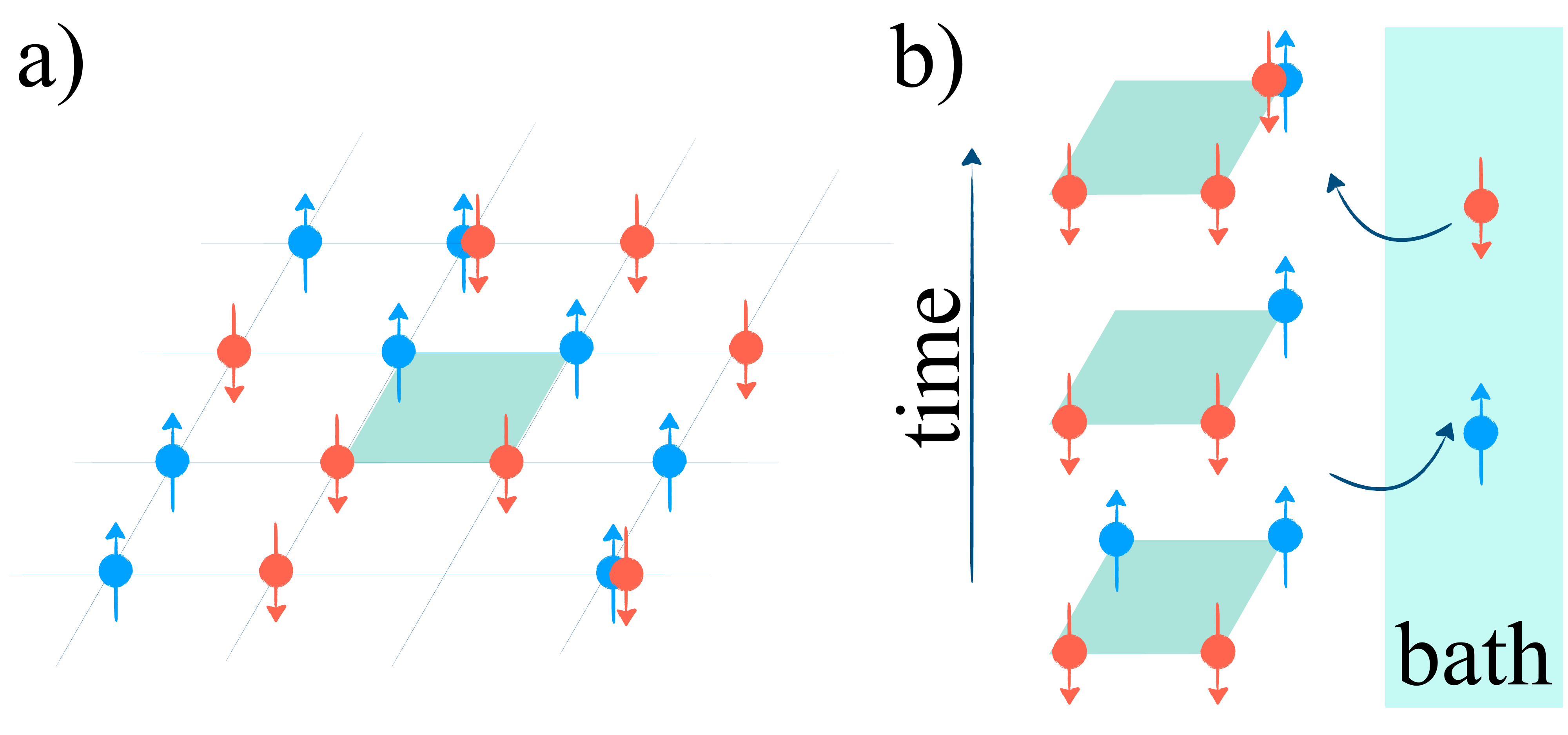}}
\caption{Real-space sketch of the cluster DMFT framework. (a) Sketch of the 2D Hubbard model on a square lattice, up and down spins are represented by spheres with arrows. (b) Cluster DMFT maps the lattice model onto a cluster impurity ($2\times2$ shaded plaquette) embedded in a self-consistent bath. As a function of time, the cluster exchanges electrons with the bath. Our work shows how superconducting correlations emerge both at the level of the cluster and of the bath. At the level of the cluster we can extract the relative time the cluster spends in a given electronic configuration; at the level of the bath we can extract the bath hybridization function which encodes the dynamics of the processes between cluster and bath.}
\label{fig:embedding}
\end{figure}

In this article, we study the superconducting correlations in the 2D Hubbard model at finite temperature from a complementary perspective, that of a cluster impurity embedded in a self-consistent bath, which underlies the cluster DMFT method. This is emphasised in Figure~\ref{fig:embedding}.
Cluster DMFT maps the lattice system (panel (a)) onto a cluster quantum impurity model fulfilling a self-consistent condition (panel (b)). Hence, in cluster DMFT one focuses on a cluster coupled to a bath of electrons which describes the rest of the lattice. The cluster fluctuates among different electronic configurations and exchanges electrons with the self-consistent bath (illustrated in the Figure), so that both spatial fluctuations (within the cluster) and temporal fluctuations are taken into account.

Therefore in cluster DMFT we can focus on the properties of {\it both} the cluster and the bath. 
At the level of the cluster, we can analyse the probability that cluster electrons are found in a given configuration, i.e. the relative time the cluster electrons spend in a given configuration~\cite{hauleCTQMC, hauleDOPING, shim:nature}. At the level of the bath, we can analyse the hybridization function, which fully encodes the dynamics of the hopping processes between cluster and bath. 

The strategy of analysing both the probabilities of the impurity configurations and the hybridization function is a standard and often used approach in single-site DMFT studies~\cite{rmp}, and especially for multi-orbital systems, where electrons fluctuations among different atomic configurations can be related to a generalised concept of valence~\cite{shim:nature, Gabi:Science2018}. 
However, this approach has not been explored much in the context of cluster DMFT studies, and there is little knowledge on the signatures of superconductivity on both the cluster electrons configurations and bath hybridization function. Our work addresses this problem, i.e. the effects of superconducting correlations on {\it both} cluster electron configurations and bath hybridization function. 

More precisely, at the level of the cluster impurity, several cluster DMFT studies have revealed that cluster electrons are locked into short-range singlets in the {\it normal state} of the doped Mott insulator~\cite{hauleDOPING, gullEPL, michelEPL, michelPRB, sht, sht2, LorenzoPRB2022} realised by the 2D Hubbard model. 
However, to our knowledge, only a few studies have investigated the impact of superconducting pairing on the cluster electrons configurations: Haule and Kotliar~\cite{hauleAVOIDED, hauleDOPING} have demonstrated amplified short-range singlet correlations for the superconducting state of the $t-J$ model around optimal doping. However, it is still not clear to what extent this mechanism extends to the 2D Hubbard model, and how the onsite interaction strength, doping, and temperature affect this mechanism. Hence, we present here a study of the cluster electrons configurations in the superconducting state of the 2D Hubbard model, for a wide range of doping levels and of interaction strength. We shall show that upon entering the superconducting state, for all values of $U$ and doping, the cluster electrons spend more time in a four-electron singlet configuration, allowing us to identify short-range spin correlations in the form of singlets as key to superconducting pairing. Increased probability of the four-electron singlet comes at the expenses of reduced probability of the four-electron triplet and no significant probability redistribution of fluctuations of charges. This allows us to establish that at the level of the cluster superconductivity primarily involves a reorganisation of short-range spin correlations rather than charge correlations.

At the level of the dynamics of the fluctuations between cluster and bath, few existing cluster DMFT studies have analysed the behavior of the hybridization function in the normal state of the 2D Hubbard model~\cite{hauleDOPING, michelPRB, sht2}. They revealed a mild momentum and doping dependence, in sharp contrast with the Green's function of the system which showed a marked dependence. 
Even less work exists on the behavior of the hybridization function in the superconducting state~\cite{hauleDOPING}, and it is primarily focused on the $t-J$ model. A detailed characterization of the hybridization function is still missing for the 2D Hubbard model in the superconducting state. 
Hence, we present here a systematic analysis of the interaction strength, doping and temperature dependence of the spectral function of the bath in the superconducting state. We shall show that upon condensation, there is a redistribution of spectral weight leading to the opening of a superconducting gap. This enables us to infer that singlet pairs propagate coherently throughout the lattice. 

Our work is organised as follows. In Section~\ref{sec:model} we briefly outline the cellular extension of DMFT (CDMFT) used in our study and how the probabilities of the different electronic configurations of the cluster can be extracted. In Section~\ref{sec:phasediagram} we overview the salient features of the established CDMFT solution of the 2D Hubbard model in the superconducting state. In Section~\ref{sec:probabilities} we analyse the probabilities of the different electronic configurations of the cluster in the superconducting state. In Section~\ref{sec:bath} we analyse the behavior of the spectral function of the self-consistently determined bath in the superconducting state. Then in Section~\ref{sec:discussion} we discuss the insights that can be gained on the behavior of the density of state of the system and of the zero-frequency spin susceptibility in the superconducting state. Finally, Section~\ref{sec:conclusions} summarises our main findings.

\section{Model and method}
\label{sec:model}

\subsection{2D Hubbard model}

The 2D Hubbard model on a square lattice is
\begin{align}
H & = - \sum_{ij \sigma} t_{ij} c_{i\sigma}^{\dagger} c_{j\sigma} + U\sum_{i} n_{i\uparrow} n_{i\downarrow} -\mu \sum_{i\sigma} n_{i\sigma} . 
\label{eq:HM}
\end{align}
Here, $t_{ij}$ is the hopping amplitude between nearest neighbor sites $i,j$ only, $U$ is the onsite Coulomb repulsion, $\mu$ is the chemical potential, $c_{i\sigma}$ and $c_{i\sigma}^{\dagger}$ respectively destroy and create an electron at site $i$ with spin $\sigma$, and $n=c_{i\sigma}^{\dagger}c_{i\sigma}$ is the number operator. $t_{ij}=t=1$ fixes our units.

\subsection{CDMFT}
\label{subsec:cdmft}

We solve Eq.~\ref{eq:HM} at finite temperature with the cellular~\cite{maier, kotliarRMP} extension of dynamical mean field-theory~\cite{rmp} (CDMFT). For the purpose of discussing the approach from a cluster plus bath perspective of our work, we outline the CDMFT procedure. In this subsection we focus on the CDMFT equations in the normal state; the generalisation to the superconducting state will be discussed in subsection~\ref{subsec:SC}. CDMFT partitions the lattice into a superlattice of clusters, singles out one (any) cluster of size $N_c$ from the lattice and embeds it in a self-consistent bath of noninteracting electrons. Hence CDMFT relies on the self-consistent solution of a cluster quantum impurity model. 

The cluster quantum impurity Hamiltonian (cluster plus bath) is 
\begin{align}
H_{\rm imp} &  = H_{\rm cl} + H_{\rm hyb} +H_{\rm hyb}^{\dagger} + H_{\rm bath} ,  
\end{align}
where $H_{\rm cl}=H_{\rm cl}(d_{i\sigma}, d_{i\sigma}^{\dagger})$ is the Hamiltonian of the cluster described by the operators $d_{i\sigma}, d_{i\sigma}^{\dagger}$, $H_{\rm bath}  = \sum_{\mu\sigma} \epsilon_\mu a_{\mu\sigma}^\dagger a_{\mu\sigma}$ is the Hamiltonian of the bath described by the bath energies $\epsilon_{\mu}$ and operators $a_{\mu\sigma}, a_{\mu\sigma}^{\dagger}$, and $H_{\rm hyb} = \sum_{i\mu\sigma} V_{\mu i } a_{\mu\sigma}^\dagger d_{i\sigma}$ is the Hamiltonian describing the hybridization between the cluster and the bath via the amplitude $V_{\mu i}$ for an electron to hop from the cluster to the bath. 

By integrating out the bath degrees of freedom, the action of the cluster quantum impurity (cluster plus bath) is: 
\begin{align}
S & = S_{\rm cl} (\boldsymbol{\hat{\psi}}^{\dagger}, \boldsymbol{\hat{\psi}} ) + \int_{0}^{\beta} d\tau \int_{0}^{\beta} d\tau'  \boldsymbol{\hat{\psi}}^{\dagger} (\tau) {\bf \Delta}(\tau,\tau') \boldsymbol{\hat{\psi}}(\tau'), 
\label{eq:S1}
\end{align}
where $S_{\rm cl}$ is the action of the cluster resulting from the tiling of the lattice, $\boldsymbol{\hat{\psi}} = (\hat{d}_{1 \uparrow} \, \cdots \, \hat{d}_{N_c \uparrow} \, \hat{d}_{1 \downarrow} \, \cdots \, \hat{d}_{N_c \downarrow})^T$ is a vector of the Grassmann variables $\hat{d}_{i\sigma}$ corresponding to the operators on the lattice, 
and ${\bf \Delta}(\tau,\tau')$ is the hybridization matrix function.  
It describes the amplitude processes of hopping via any bath orbital from the cluster site $i$ at time $\tau$ to the cluster site $j$ at time $\tau'$. Eq.~\ref{eq:S1} can be rewritten as
\begin{align}
S = & - \int_{0}^{\beta} d\tau  \int_{0}^{\beta} d\tau' \,  \boldsymbol{\hat{\psi}}^{\dagger}(\tau)  \boldsymbol{\cal G}_0^{-1} (\tau,\tau') \boldsymbol{\hat{\psi}}(\tau') \nonumber \\
& + U \int_{0}^{\beta} d\tau  \, {\bf \hat{n}}_{i\uparrow}(\tau) {\bf \hat{n}}_{i\downarrow}(\tau), 
\label{eq:S2}
\end{align}
where the Green's function of the noninteracting impurity, $\boldsymbol{\cal G}_0$, has been introduced as
\begin{align}
\boldsymbol{\cal G}_0^{-1} (i\omega_n) & = (i\omega_n + \mu) {\bf I} -{\bf t_{\rm cl}} - {\bf \Delta} (i\omega_n) .
\end{align}
Here ${\bf t_{\rm cl}}$ is the cluster hopping matrix ${\bf t_{\rm cl}} = \int d \tilde{k} \, {\bf t}(\tilde{k})$ with ${\bf t}(\tilde{k})$ the lattice hopping matrix in the supercell notation and with $\tilde{k}$ running over the reduced Brillouin zone of the superlattice. The elements of the hybridization matrix function ${\bf \Delta}(i\omega_n)$ can be written in the form
\begin{align}
\Delta_{ij} (i\omega_n) & = \sum_{\mu} \frac{V_{i\mu} V_{\mu j}^\dagger}{i\omega_n - \epsilon_{\mu}}
\end{align}
i.e. as a function of the bath degrees of freedom $\epsilon_\mu$, $V_{\mu i}$. 

For a given ${\bf \Delta}(i\omega_n)$, the solution of the cluster quantum impurity model Eq.~\ref{eq:S2} gives the cluster Green's function 
\begin{align}
{\bf G}_{\rm cl} (\tau -\tau') & = - \langle T_\tau \,  \boldsymbol{\hat{\psi}}(\tau) \boldsymbol{\hat{\psi}}^{\dagger} (\tau') \rangle_{S} . 
\label{eq:Gcl}
\end{align}
The Dyson equation defines the cluster self-energy as
\begin{align}
{\bf \Sigma_{\rm cl}} (i\omega_n) & = \boldsymbol{\cal G}_0^{-1} (i\omega_n) -  {\bf G}_{\rm cl}^{-1} (i\omega_n) .
\end{align}

To fix ${\bf \Delta} $, we need to apply the self-consistency condition. The self-consistency condition requires that the cluster Green's function ${\bf G}_{\rm cl}$ computed from the cluster quantum impurity model coincides with the projection onto the cluster of the lattice Green's function ${\bf G}_{\rm latt}$, i.e. the superlattice averaged Green's function $\bar{\bf G}$:
\begin{align}
\bar{\bf G} (i\omega_n) & = \int d \tilde{k} \, {\bf G}_{\rm latt} (\tilde{k}, i\omega_n) \\
& =  \int d \tilde{k} \, \left[ (i\omega_n +\mu){\bf I} -{\bf t}(\tilde{k}) -{\bf \Sigma}_{\rm latt} (\tilde{k}, i\omega_n) \right]^{-1} .
\end{align}
The approximation that allows one to identify ${\bf G}_{\rm cl}$ with $\bar{\bf G}$ is that ${\bf \Sigma}_{\rm latt} (\tilde{k}, i\omega_n) \approx {\bf \Sigma}_{\rm cl} (i\omega_n)$, i.e.
\begin{align}
\bar{\bf G} (i\omega_n) & \approx  \int d \tilde{k} \, \left[ (i\omega_n +\mu){\bf I} -{\bf t}(\tilde{k}) -{\bf \Sigma}_{\rm cl} (i\omega_n) \right]^{-1} .
\label{eq:Gbar}
\end{align}
The self-consistency condition can then be written as
\begin{align}
{\bf \Delta} (i\omega_n) = &  (i\omega_n +\mu){\bf I} -{\bf t_{\rm cl}} -{\bf \Sigma_{\rm cl}} (i\omega_n) - \bar{\bf G}^{-1}(i\omega_n) .
\label{eq:SCC}
\end{align}
In practice, we solve the CDMFT equations with an iterative procedure: starting from an initial guess for $\bf{\Delta}$, we solve the cluster quantum impurity model to obtain $\bf{G}_{\rm cl}[\bf{\Delta}]$, and then compute $\bar{\bf G}$ with Eq.~\ref{eq:Gbar}. From $\bar{\bf G}$ we obtain an updated hybridization matrix ${\bf \Delta}$ using Eq.~\ref{eq:SCC}, and we iterate the process until convergence is reached.

\subsection{CT-HYB impurity solver}
\label{subsec:cthyb}

We solve the cluster quantum impurity model Eq.~\ref{eq:Gcl} using the hybridization expansion continuous-time quantum Monte Carlo method (CT-HYB)~\cite{millisRMP, Werner:2006, hauleCTQMC, patrickSkipList, Hebert:2015}. Here we limit ourselves to outline the key aspects of the CT-HYB algorithm that are relevant for our discussion. 

To reduce the size of the matrices involved and to speed up the calculation, we choose a single-particle basis that transforms as the irreducible representations of the cluster Hamiltonian symmetries~\cite{hauleCTQMC}. For a $2\times 2$ plaquette with vertices $1,2, 3, 4$ oriented counter-clockwise with $1$ on the left bottom corner, the point group symmetry $C_{2v}$ with mirrors along the plaquette axes leads to the following single-particle basis (cluster momentum basis):
\begin{equation}
\begin{aligned} \label{eq:basis}
d_{A_1, \sigma} & = d_{(0,0), \sigma}  = \frac{1}{2}\left(d_{1\sigma}+d_{2\sigma}+d_{3\sigma}+d_{4 \sigma}\right) \\
d_{B_1, \sigma} & = d_{(\pi,0), \sigma} = \frac{1}{2}\left(d_{1\sigma}-d_{2\sigma}-d_{3\sigma}+d_{4 \sigma}\right) \\
d_{B_2, \sigma} & = d_{(0,\pi), \sigma} = \frac{1}{2}\left(d_{1\sigma}+d_{2\sigma}-d_{3\sigma}-d_{4 \sigma}\right) \\
d_{A_2, \sigma} & = d_{(\pi,\pi), \sigma} = \frac{1}{2}\left(d_{1\sigma}-d_{2\sigma}+d_{3\sigma}-d_{4 \sigma}\right) ,
\end{aligned}
\end{equation}
where $A_1, B_1, B_2, A_2$ are the irreducible representations of $C_{2v}$, respectively denoted with the cluster momenta ${\bf K} = \left\{ (0,0), (\pi,0), (0,\pi), (\pi,\pi) \right\}$. In the previous section, every operator was expressed in the position basis, but here they are expressed in this new ${\bf K}$ basis. In this basis, the $8 \times 8$ hybridization matrix ${\bf \Delta}$ acquires a block diagonal form:
\begin{align}
{\bf \Delta} & = \left( \begin{array}{@{}cc@{}}
{\bf \Delta}_{\uparrow} & {\bf 0} \\
{\bf 0} & {\bf \Delta}_{ \downarrow}
\end{array} \right) , 
\end{align}
with 
\begin{align}
{\bf \Delta}_{\sigma} = \left( \begin{array}{cccc}
\Delta_{(0,0)} & 0 & 0 & 0 \\
0 & \Delta_{(\pi,0)} & 0 & 0 \\
0 & 0 & \Delta_{(0,\pi)} & 0 \\
0 & 0 & 0 & \Delta_{(\pi,\pi)}
\end{array} \right) . 
\label{eq:Delta_K_NS}
\end{align}

Furthermore, time-reversal symmetry restricts the matrix blocks ${\bf \Delta}_{\uparrow}, {\bf \Delta}_{\downarrow}$ to take the same value, and $C_4$ symmetry ($\pi/2$ rotation) prescribes that $\Delta_{(0,\pi)} = \Delta_{(\pi,0)}$. As a result, there are only 3 independent components of the hybridization matrix ${\bf \Delta}$. 

The CT-HYB algorithm writes the impurity partition function $Z_{\rm imp} = \Tr e^{-\beta H_{\rm imp}}$ in the interaction representation and expands it in powers of the hybridization, obtaining
\begin{align}
Z_{\rm imp} &  = \int {\cal D} \left[ \hat{d}, \hat{d}^\dagger \right] e^{-S}  \\
& = Z_{\rm bath} \sum_{k=0}^{\infty} \int_{0}^{\beta} d\tau_{1} \cdots d\tau_{k} \int_{0}^{\beta} d\tau_{1}^{\prime} \cdots d\tau_{k}^{\prime} \nonumber \\
& \times \sum_{{\bf K}_{1} \cdots {\bf K}_{k}} \sum_{{\bf K}_{1}^{\prime} \cdots {\bf K}_{k}^{\prime}} w \{ \cal{C} \} , 
\end{align}
where the integrands
\begin{align}
w \{ \cal{C} \} & = 
\det_{1 \le m, n \le |{\cal{C}}|} 
\left[ 
\Delta_{{\bf K}_{m} {\bf K}_{n}^{\prime}} (\tau_{m} - \tau_{n}^{\prime}) 
\right]
\nonumber \\
& \times \Tr_{\rm cl} \left[ T_{\tau}  e^{-\beta H_{\rm cl}} \prod_{r=1}^{|{\cal{C}}|} d_{{\bf K}_r} (\tau_r) d_{{\bf K}_r^{\prime}}^{\dagger} (\tau_r^{\prime}) \right] 
\label{eq:weightsZ}
\end{align}
are the weights of a distribution over the configuration space ${\cal{C}}= \{ ({\bf K}_{1}, \tau_{1}), ({\bf K}_{1}^{\prime}, \tau_{1}^{\prime}) \cdots ({\bf K}_{k}, \tau_{k}), ({\bf K}_{k}^{\prime}, \tau_{k}^{\prime}) \} $. This configuration space is sampled with Markov chain Monte Carlo. In order reuse matrix products previously calculated, we use the Lazy Skip List algorithm~\cite{patrickSkipList}.

\subsection{Probabilities of the plaquette sectors}
\label{subsec:prob}

In this work we are interested in the reduced density matrix of the cluster, $\rho_{\rm cl}$. Within the CT-HYB algorithm it is possible to measure the diagonal elements of $\rho_{\rm cl}$. This procedure was developed in Ref.~\onlinecite{hauleCTQMC}. Here we limit ourselves to outline the key aspects of the CT-HYB algorithm that are relevant for our discussion. Further details can be found in Refs.~\onlinecite{hauleCTQMC, patrickSkipList}. 

For a $2\times 2$ plaquette, the cluster Hamiltonian $H_{\rm cl}$ conserves charge, spin and cluster momentum. 
We can group the 256 eigenstates $\left\{ \ket{\mu} \right\}$ of $H_{\rm cl}$ according to the quantum numbers $N= \sum_{i} n_{i\uparrow} + n_{i\downarrow}$, $S_z = \sum_{i} (n_{i\uparrow} - n_{i\downarrow})/2$ and cluster momentum ${\bf K}$, so that both $H_{\rm cl}$ and $\rho_{\rm cl}$ become block diagonal. This grouping results in 84 blocks where each block, or sector, is labeled by the set of quantum numbers $N,S_z, {\bf K}$. Let $m$ be the index of the sector (or matrix block) and $\left\{ \ket{\mu} \right\}_m$ be the set of cluster eigenstates belonging to $m$. Table~\ref{table} in Appendix~\ref{sec:eigenstates} lists the sectors $m$ grouped according to the quantum numbers $N$, $S_z$ and ${\bf K}$. 

The reduced density matrix is $\rho_{\rm cl} = \Tr_{\rm bath} [e^{-\beta H_{\rm imp}}/Z_{\rm imp}]$, and the estimator for its diagonal elements is~\cite{hauleCTQMC, patrick21}
\begin{align}
(\rho_{\rm cl})_{\mu \mu} &  = \frac{ \bra{\mu} T_{\tau}  e^{-\beta H_{\rm cl}} \prod_{r=1}^{|{\cal{C}}|} d_{{\bf K}_r} (\tau_r) d_{{\bf K}_r^{\prime}}^{\dagger} (\tau_r^{\prime}) \ket{\mu} }{\Tr_{\rm cl}  \left[  T_{\tau} e^{-\beta H_{\rm cl}} \prod_{r=1}^{|{\cal{C}}|} d_{{\bf K}_r} (\tau_r) d_{{\bf K}_r^{\prime}}^{\dagger} (\tau_r^{\prime}) \right] } . 
\end{align}
The probability associated to the cluster eigenstates belonging to a given sector $m$ is 
\begin{align}
P_{m} & = \sum_{\mu \in \left\{ \ket{\mu} \right\}_m} (\rho_{\rm cl})_{\mu \mu} .
\end{align}
The probabilities $\{ P_m \}$ will be analysed in Sec.~\ref{sec:probabilities}.

\subsection{d-wave superconducting state}
\label{subsec:SC}

The CDMFT formalism outlined in subsections~\ref{subsec:cdmft}, \ref{subsec:cthyb}, \ref{subsec:prob} applies to the normal state. 
For the d-wave superconducting state, the CDMFT method can be generalised~\cite{lkAF, maier, kancharla, hauleCTQMC, patrickERG} as follows. 

It is useful to introduce the Nambu basis $\Psi_{\bf K} = (d_{{(0,0)}\uparrow} \, d_{{(\pi,0)}\uparrow} \, d_{{(0,\pi)}\uparrow}  \, d_{{(\pi,\pi)}\uparrow} \, d_{{(0,0)}\downarrow}^\dagger \, d_{{(\pi,0)}\downarrow}^\dagger \, d_{{(0,\pi)}\downarrow}^\dagger \, d_{{(\pi,\pi)}\downarrow}^\dagger )^T$, where we use that $-(0,0) = (0,0)$, $-(\pi,0) = (\pi,0)$, etc. because of Umklapp processes. In this basis the $8 \times 8$ matrix hybridization function becomes
\begin{align}
{\bf \Delta}(\tau) & = \left(\begin{array}{@{}cc@{}}
 {\bf \Delta}_{\uparrow}(\tau) & {\bf F}(\tau) \\
%\hline
{\bf F}^\dagger(\tau) & -{\bf \Delta}_{\downarrow}(-\tau)
\end{array}\right) , 
\label{eq:DeltaSC}
\end{align}
where ${\bf \Delta}_{\sigma}$ has the same structure of Eq.~\ref{eq:Delta_K_NS}, and the anomalous component ${\bf F}$ is block diagonal.

Note that the $2 \times 2$ cluster Hamiltonian $H_{\rm cl}$ has $C_{4v}$ point group symmetry. 
The d-wave superconducting order parameter breaks the $C_4$ symmetry (i.e. it changes sign under a $\pi/2$ rotation), but preserves the $C_{2v}$ symmetry with mirrors along the plaquette axes, of the original $C_{4v}$ group.
Therefore the superconducting order parameter has the same $C_{2v}$ symmetry as the one particle cluster basis (Eq.~\ref{eq:basis}). 
This choice implies that the d-wave superconducting order parameter transforms in space as the $A_1$ representation of the $C_{2v}$ symmetry group with mirrors along the plaquette axes~\footnote{This has to be contrasted with Ref.~\onlinecite{Hebert:2015}, where, for the anisotropic Hubbard model within $2 \times 2$ CDMFT, the $C_{2v}$ symmetry with mirrors along the plaquette diagonals was used as a representation for the one particle basis. This choice dictates that the superconducting order parameter transforms in space as the $A_2$ representation of the $C_{2 v}$ symmetry group with mirrors along the plaquette diagonals.}. 
Hence only the entries in ${\bf F}$ transforming as $A_1$ can be finite, implying that only diagonal components of ${\bf F}$ are nonzero. 
Furthermore, the d-wave superconducting order parameter changes sign under a $\pi/2$ rotation, imposing the constraints that $F_{(0,0)\uparrow, (0,0)\downarrow}$ and $F_{(\pi,\pi)\uparrow, (\pi,\pi)\downarrow}$ are zero and $F_{(0,\pi) \uparrow, (0,\pi) \downarrow} = - F_{(\pi,0) \uparrow, (\pi,0) \downarrow}$. Thus, written explicitly, the anomalous component of the matrix hybridization function reads
\begin{align}
{\bf F}= \left( \begin{array}{cccc}
0 & 0 & 0 & 0 \\
0 & F_{(\pi,0) \uparrow, (\pi,0) \downarrow} & 0 & 0 \\
0 & 0 & -F_{(\pi,0) \uparrow, (\pi,0) \downarrow} & 0 \\
0 & 0 & 0 & 0
\end{array} \right) . 
\label{eq:F_K_SC}
\end{align}

To allow for a superconducting solution, we start the CDMFT loop scheme with a guess for ${\bf \Delta}$ containing a finite off-diagonal component $F_{(\pi,0) \uparrow, (\pi,0) \downarrow}$. In all subsequent iterations ${\bf \Delta}$ evolves unconstrained 
and $F_{(\pi,0) \uparrow, (\pi,0) \downarrow}$ will either survive or vanish. When self-consistency is reached and $F_{(\pi,0) \uparrow, (\pi,0) \downarrow}$ survives, the solution is superconducting.

The d-wave symmetry is broken in the bath (${\bf \Delta}$) but not in the cluster. In other words, the superconducting state breaks the $C_4$ symmetry but this symmetry is still present in the cluster Hamiltonian $H_{\rm cl}$. This broken symmetry in ${\bf \Delta}$ (corresponding to the non vanishing $F_{(\pi,0) \uparrow, (\pi,0) \downarrow}$ component) propagates to ${\bf G}_{\rm cl}$ and ${\bf \Sigma}_{\rm cl}$ of the cluster through Eq.~\eqref{eq:SCC}, even though $H_{\rm cl}$ still has the full $C_{4v}$ symmetry. 

From the point of view of the impurity solver, the calculation of the Monte Carlo weight of each configuration is influenced by the bath, whose degrees of freedom are in the determinant in Eq.~\ref{eq:weightsZ}. That bath has components that are off-diagonal in Nambu space. The trace on the cluster in Eq.~\ref{eq:weightsZ} however always conserves the number of particles since the symmetry is never explicitly broken in the cluster. It is through the bath that the cluster Green's function acquires off-diagonal components. 

Finally, as demonstrated in Ref.~\onlinecite{patrickERG}, the ergodicity of the CT-HYB algorithm in the d-wave superconducting state can only be obtained by allowing four operator updates in the sampling of the Markov chain.

\section{Superconducting state phase diagram with plaquette CDMFT}
\label{sec:phasediagram}

\begin{figure*}[th!]
\centering{
\includegraphics[width=1.\linewidth]{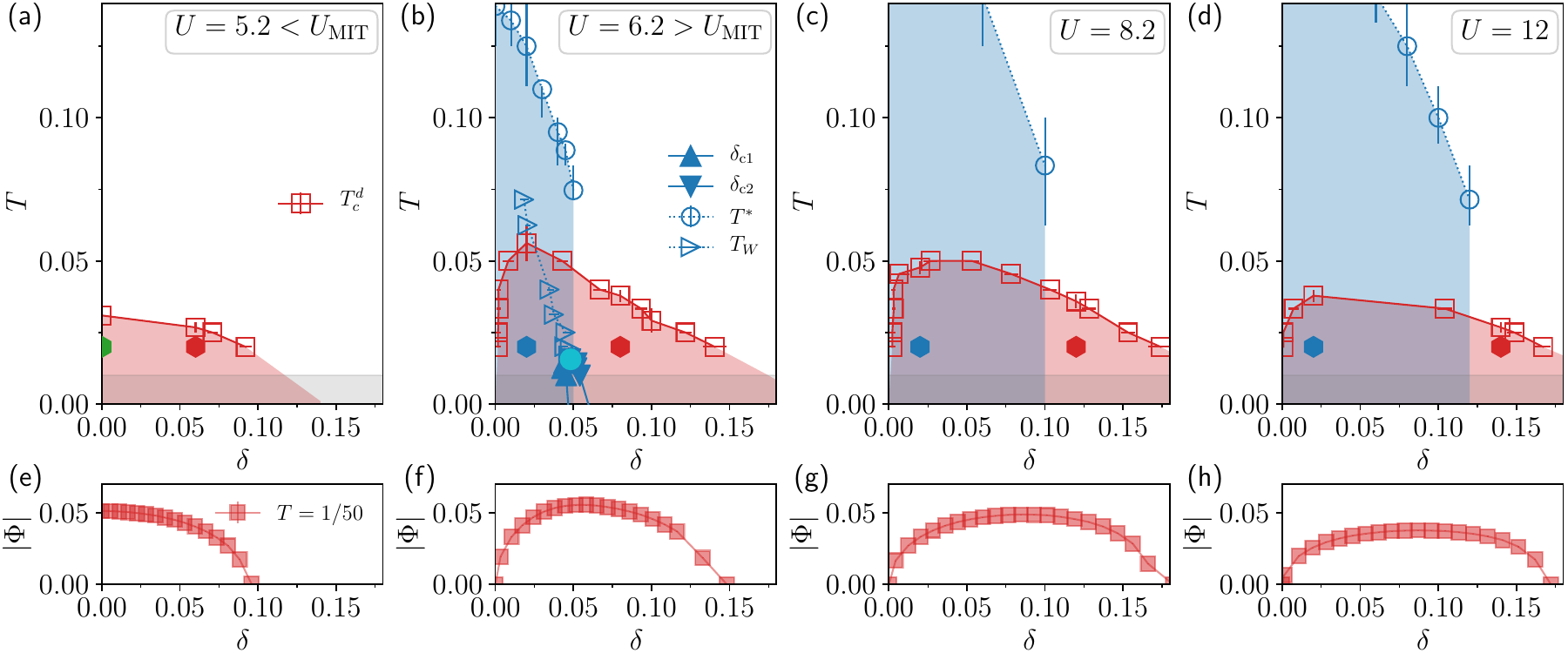}
}
\caption{(a), (b), (c), (d): temperature versus doping phase diagram of the plaquette CDMFT solution of 2D Hubbard model, for several values of the interaction strength $U$. Only normal and superconducting phases are shown. Data are taken from Ref.~\onlinecite{CaitlinPNAS2021}. Filled hexagons indicate the dataset analysed in Figures~\ref{fig2} to \ref{fig9}. Open red squares denote the CDMFT superconducting transition temperature $T_c^d$, below which the superconducting order parameter $\Phi$ is nonzero. Open blue circles indicate the crossover temperature $T^*$ corresponding to the opening of the pseudogap, as determined by the drop in the spin susceptibility vs $T$ at fixed doping. Red and blue shaded areas are a guide to the eye for the superconducting and pseudogap phases. In (b), the superconducting dome hides a pseudogap to metal transition in the underlying normal state. This transition is first order and bounded by spinodal lines (filled blue triangles), terminates in a critical endpoint (filled cyan circle), from which a supercritical crossover emerges (Widom line~\cite{water1, supercritical, ssht}), as determined here by the loci of the maximum of the charge compressibility vs $\delta$. Shaded grey area corresponds to the low temperature region that is not accessible to our calculations because of the sign problem. (e), (f), (g), (h): superconducting order parameter $|\Phi|$ versus $\delta$, for several values of the interaction strength $U$ and at temperature $T=1/50$. Data are taken from Ref.~\onlinecite{CaitlinPNAS2021}.
}
\label{fig1}
\end{figure*}
Prior work revealed the finite temperature aspects of the superconducting phase diagram of the 2D Hubbard model on a square lattice with $2\times 2$ plaquette CDMFT~\cite{hauleDOPING, sshtSC, LorenzoSC}. Here, we briefly survey two key features of the superconducting state that are relevant for our discussion: the behavior of the superconducting transition temperature $T^{d}_{c}$ (where $d$ emphasises it is the critical temperature at the cluster dynamical mean-field level), and the link between $T^{d}_{c}$ and the onset temperature of the pseudogap $T^{*}$. 
Although in 2D long-range order is excluded by thermal fluctuations~\cite{MWtheorem}, $T_c^d$ physically denotes when superconducting pairs develop within the $2\times 2$ plaquette~\cite{sshtSC}. Also, here we focus only on the superconducting and normal states only, and competition with other states is not considered. 

Figure~\ref{fig1}a-d shows the temperature hole-doping phase diagram for four different values of the interaction strength $U$, ranging from $U=5.2$ to $U=12$. Within $2\times 2$ plaquette CDMFT, the value of $U$ needed to transform a metal to a Mott insulator at half-filling ($\delta=0$), is $U_{\textrm{MIT}} \approx 5.95$~\cite{CaitlinSb}. All data points in Figure~\ref{fig1} are extracted from our previous work of Ref.~\onlinecite{CaitlinPNAS2021}, which employs the same methodology as used in this work. 

The superconducting state is indicated in red. It is bounded by $T_{c}^{d}$, and it is the region below which the superconducting order parameter $\Phi = \langle d^{\dagger}_{(0,\pi) \uparrow}d^{\dagger}_{(0,\pi)\downarrow}\rangle$ is nonzero. Figure~\ref{fig1}e-h shows $\Phi(\delta)$ for $T=1/50$ as a sample of the calculations performed across the $U-T-\delta$ space. 

By systematically varying the interaction strength $U$ and doping $\delta$, interesting trends emerge, from which insights on microscopic mechanisms of superconductivity can be derived~\cite{LorenzoSC}. 
Below $U_{\textrm{MIT}}$, $T_{c}^{d}(\delta)$ decreases with increasing doping; above $U_{\textrm{MIT}}$, $T_{c}^{d}(\delta)$ has the shape of a dome, with the highest $T_{c}^{d}$ just above $U_{\textrm{MIT}}$; the superconducting dome is asymmetric in doping with a steep slope upon doping the parent Mott state.  

\begin{figure*}[ht!]
\centering{
\includegraphics[width=1.\linewidth]{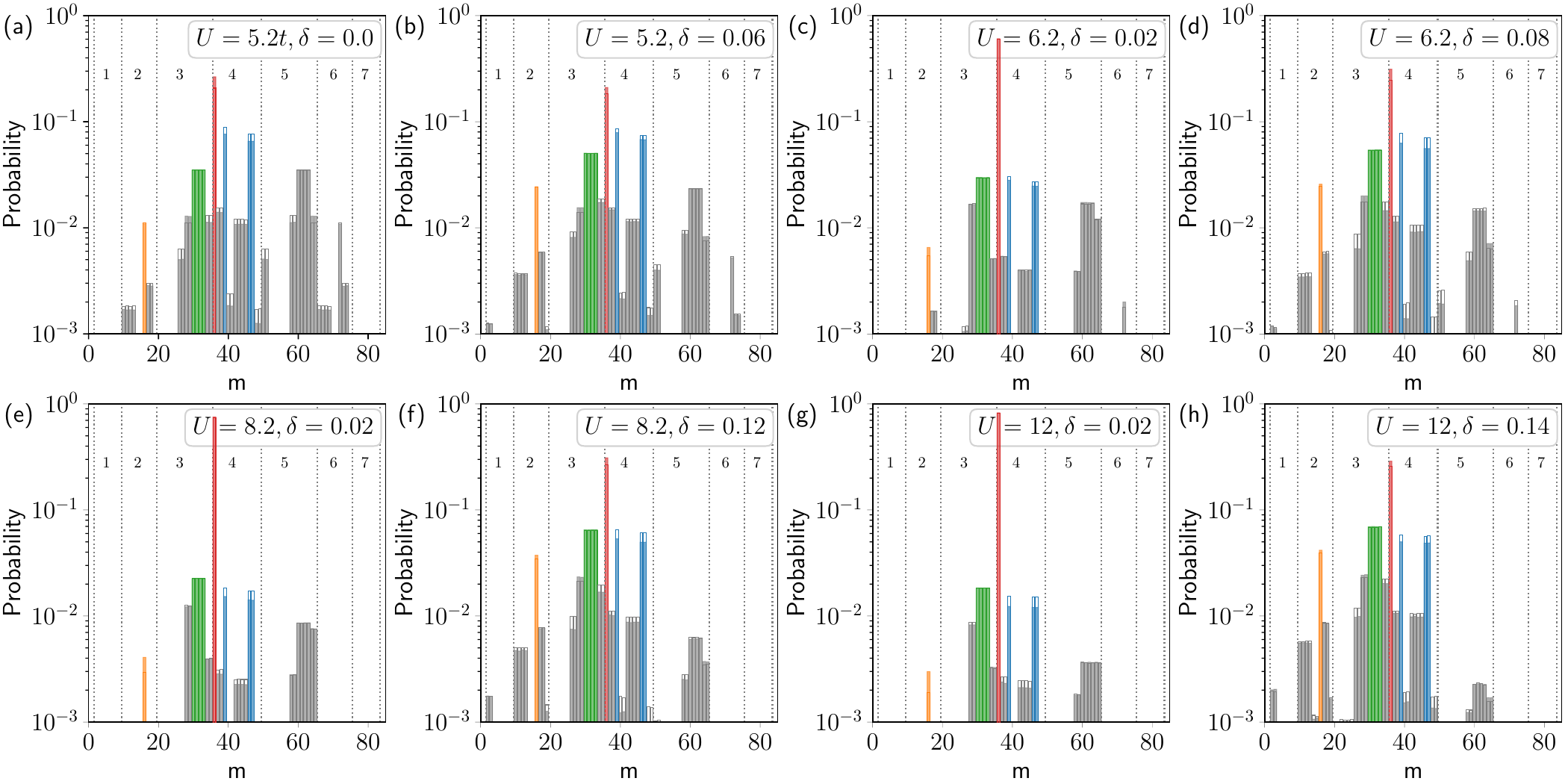}
}
\caption{Histograms of the probability of the plaquette sectors $\{ m \}$. Data are shown at $T=1/50$ for several values of interaction strength $U$ and doping $\delta$, corresponding to the filled hexagons in the $T-\delta$ phase diagrams of Figure~\ref{fig1}. The index on the $x$ axis refers to the label of the plaquette sectors in the last column of Table~\ref{table}. Vertical dotted lines bound the sectors with same quantum number $N$. Empty bars correspond to the normal state CDMFT solution, whereas filled bars correspond to the superconducting state solution. The sectors with highest probabilities for a given $N$ and $S_{z}$ are shown in color, and correspond to the sets plotted in Fig.~\ref{fig3}. The 'singlets' with two and four electrons with $S_z=0$ in the cluster momentum ${\bf K}=(0,0)$ (indexes 16 and 36 in Table~\ref{table}), are shown in yellow and red, respectively. The 'triplet' with four electrons with $S_z=0,\pm 1$ in the cluster momentum ${\bf K}=(\pi,\pi)$ (indices 39, 46 and 47 in Table~\ref{table}), is shown in blue. The 'doublet' with three electrons with $S_z=\pm 1/2$ in the cluster momentum ${\bf K}=(\pi,0)$ and its degenerate counterpart ${\bf K}=(0,\pi)$ (indexes 30-33 in Table~\ref{table}), is shown in green. 
}
\label{fig2}
\end{figure*}

Additional insights can be gained from contrasting the superconducting phase and pseudogap phase (shown in blue)~\cite{sshtSC}. The pseudogap is bounded by a crossover $T^{*}(\delta)$, which can be calculated by the drop in the zero-frequency spin susceptibility as a function of $T$~\cite{sshtRHO} (also see Section~\ref{chi}). It is a strongly correlated phase that only appears for $U> U_{\textrm{MIT}}$. This suggests a link to the superconducting phase, which has a dome-like shape for $U> U_{\textrm{MIT}}$ only. However at large doping superconductivity can emerge from a metal in the absence of a pseudogap implying they are two distinct phenomena~\cite{sshtSC}. This is because the doping at which the pseudogap ends is contained within the centre of the superconducting dome. 
This is where a hidden strongly correlated pseudogap - correlated metal transition occurs. The nature of this transition is first order, and it is a purely electronic transition without symmetry breaking~\cite{sht, sht2, ssht}. Upon increasing temperature, this transition ends at a critical endpoint which gives way to crossover lines~\cite{ssht, sshtRHO}. These crossovers mark anomalies~\cite{ssht} in observables as a function doping, including electronic specific heat \cite{Giovanni:PRBcv}, charge compressibility \cite{ssht}, nonlocal density fluctuations \cite{CaitlinOpalescence}, entanglement entropy \cite{Caitlin:PRXQ2020}, velocity of sound~\cite{CaitlinSoundVelocity}. $T^{*}(\delta)$ is a high temperature precursor of such crossovers~\cite{sshtRHO}.
Furthermore by studying the difference in kinetic and potential energy between the normal and superconducting states, Ref.~\onlinecite{LorenzoSC} shows that the doping at which the hidden transition occurs correlates with the largest condensation energy. 

Finally, Ref.~\onlinecite{Lorenzo3band} shows that the main features obtained in the 2D Hubbard model reviewed here (the superconducting dome, the pseudogap to correlated metal hidden transition and its associated supercritical crossovers, and the source of pairing energy), are also found in the three-band Emery model, suggesting they are emergent phenomena of doped Mott insulators, robust against microscopic details.

\section{Fluctuations between cluster eigenstates in the superconducting and normal phases}
\label{sec:probabilities}

This work aims at obtaining new insights on the superconducting correlations in the 2D Hubbard model by taking the perspective of the cluster impurity embedded in a self consistent bath. 
This section focuses on the properties of the embedded cluster (here, a $2\times 2$ plaquette). Next section will focus on the properties of the bath. Here we calculate the probability that the electrons in the cluster are in any of the cluster eigenstates $\left\{ \ket{\mu} \right\}_m$ characterised by the quantum numbers $N,S_z, {\bf K}$ of the sector $m$. This probability can be viewed~\cite{hauleCTQMC,hauleDOPING, shim:nature} to represent the relative time the plaquette electrons occupy the cluster eigenstates $\left\{ \ket{\mu} \right\}_m$.

\subsection{Probability distribution of plaquette sectors}

\begin{figure*}[ht!]
\centering{
\includegraphics[width=1.\linewidth]{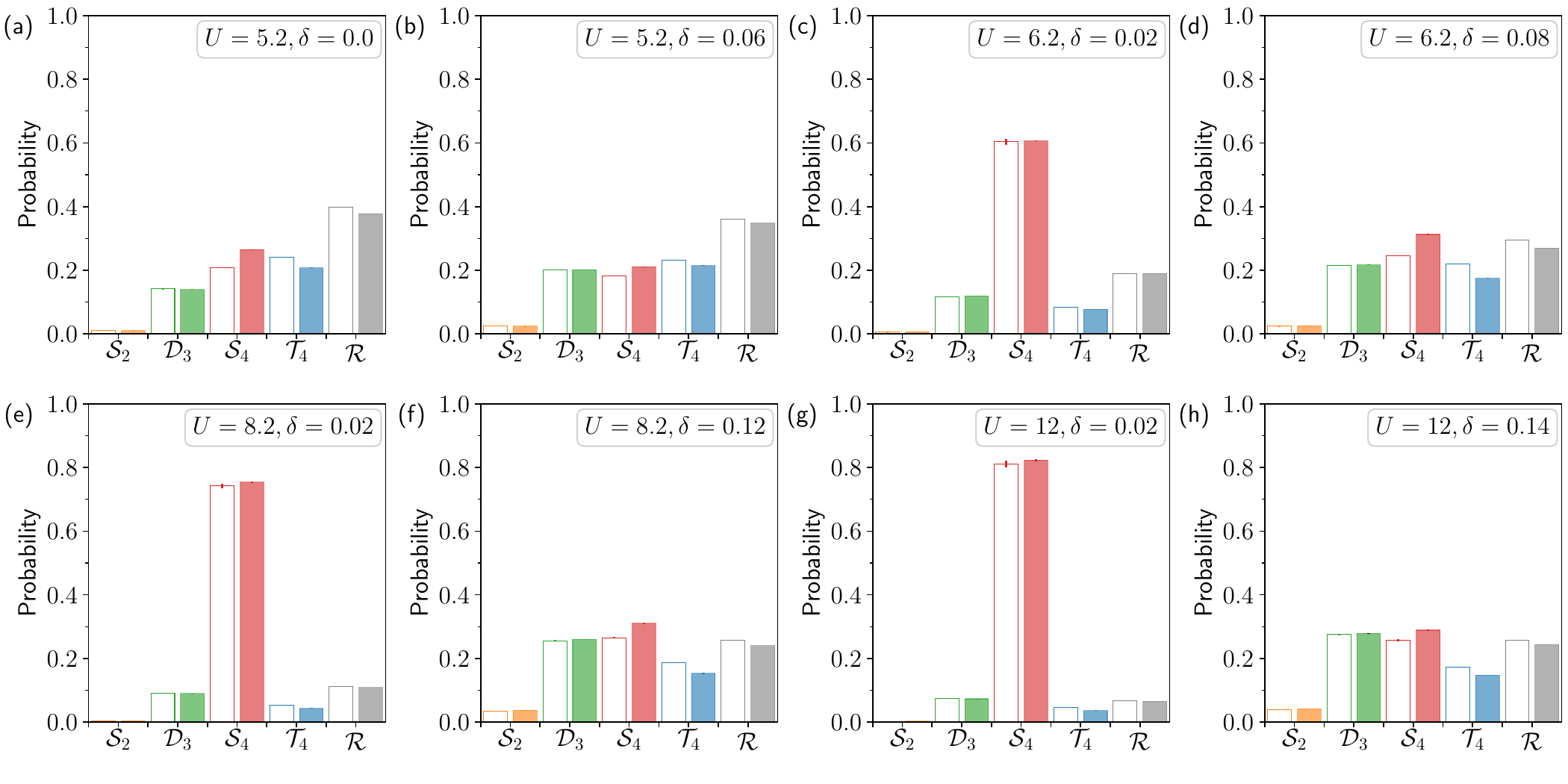}
}
\caption{Histogram showing superconducting and normal state probabilities for selected key plaquette sets of sectors (see Eq.~\ref{keyeigenstates}). They are the four-electron singlet $\mathcal{S}_4$ (red), the four-electron triplet $\mathcal{T}_4$ (blue), the three-electron doublet $\mathcal{D}_3$ (green), the two-electron singlet $\mathcal{S}_2$ (yellow), the remnant eigenstates $\mathcal{R}$ (grey). 
Data are shown for the same values of $U$ and $\delta$ as Fig.~\ref{fig2} corresponding to the filled hexagons of Fig.~\ref{fig1}, at $T=1/50$. Filled and unfilled bars indicate the superconducting and normal phases respectively. Error bars indicate the root-mean-square error over the last 20 CDMFT iterations.
\label{fig3}}
\end{figure*}
The behavior of the probabilities of the plaquette sectors for the normal state solution has been analysed in Ref.~\onlinecite{sht2}. In this work we report the behavior of the probabilities in the superconducting phase, and we contrast with that in the normal state. 

Figure~\ref{fig2} shows the histogram of the probability of the plaquette sectors $\left\{ m \right\}$, for the $T-U-\delta$ values indicated by filled hexagons in Figure~\ref{fig1}. Data are at $T=1/50$, chosen as it is well below $(T_c^d)_{\rm max}$ for each $U$. The $x$-axis shows the index $m$ of each sector (see Table~\ref{table}). Each bar of the histogram has two solutions that are superimposed: the superconducting solution for each sector $m$ is shown with a filled bar, whereas the normal state solution is shown by an unfilled bar.

From this analysis a few trends emerge. Firstly, of the 84 available sectors $\left\{ m \right\}$ of the plaquette, very few have a large probability~\cite{hauleDOPING, gullEPL, sht, sht2} (note the logarithmic scale of the $y$-axis). Secondly, for each $U> U_{\textrm{MIT}}$, there are fewer highly probable states at low doping than at high doping, and even fewer upon increasing $U$. We note that for $U< U_{\textrm{MIT}}$ at $\delta=0$, the system is particle-hole symmetric, which is reflected in the probabilities of the sectors. The overall difference in the probability for a given $\left\{ m \right\}$ between the normal and the superconducting phase is small (less than 0.1). The small difference in the probabilities between the superconducting and normal state is consistent with the small superconducting condensation energy, which in Ref.~\onlinecite{LorenzoSC} has been estimated as smaller than $0.01t$ for the same range of $U$ and $T$ considered here.

To gain further insights, we highlight sectors $\left\{ m \right\}$ with the highest probabilities for given $N$ and $S_{z}$ in color in Figure~\ref{fig2}.
Upon inspection, we find that they can be grouped in the following sets:
\begin{equation}
\begin{aligned} \label{keyeigenstates}
    &\mathcal{S}_{2}= \{ N=2,~S_{z}=0,~\textbf{K}=(0,0) \} \\
    &\mathcal{D}_{3}= \{N=3,~S_{z}=\pm1/2,~\textbf{K}=(0,\pi), (\pi,0) \} \\
    &\mathcal{S}_{4}= \{N=4,~S_{z}=0,~\textbf{K}=(0,0) \} \\
    &\mathcal{T}_{4}= \{N=4,~S_{z}=0,~\pm 1,~\textbf{K}=(\pi,\pi) \} .
\end{aligned}
\end{equation}
They represent the following sets: $\mathcal{S}_{2}$ and $\mathcal{S}_{4}$ denote the set of all the eigenstates with two and four electrons with $S_z=0$ and in the cluster momentum ${\bf K}=(0,0)$. As a shorthand notation, we call these sets 'two-electron singlet' and 'four-electron singlet' respectively, because they contain only the sector where $S_z=0$. 
The set $\mathcal{T}_{4}$ denotes the set of all the eigenstates with four electrons with $S_z=0, \pm 1$ in the cluster momentum ${\bf K}=(\pi,\pi)$. We call this set 'four-electron triplet' as it contains the 3 sectors with $S_z=0, \pm 1$. 
The set $\mathcal{D}_{3}$ contains 4 sectors and denotes the set of all the eigenstates with three-electrons with $S_z=\pm 1/2$ in the cluster momenta ${\bf K}=(\pi,0), (0,\pi)$.  We call this set 'three-electron doublet' as it contains the 2 sectors with $S_z=\pm 1/2$. Our naming convention follows Refs.~\onlinecite{hauleDOPING, sht, sht2}. 

The probabilities of these four key sets are shown in Figure~\ref{fig3} for both the normal and superconducting states. The set of the remnant sectors are grouped and denoted $\mathcal{R}$.

\begin{figure*}[ht!]
\centering{
\includegraphics[width=1.\linewidth]{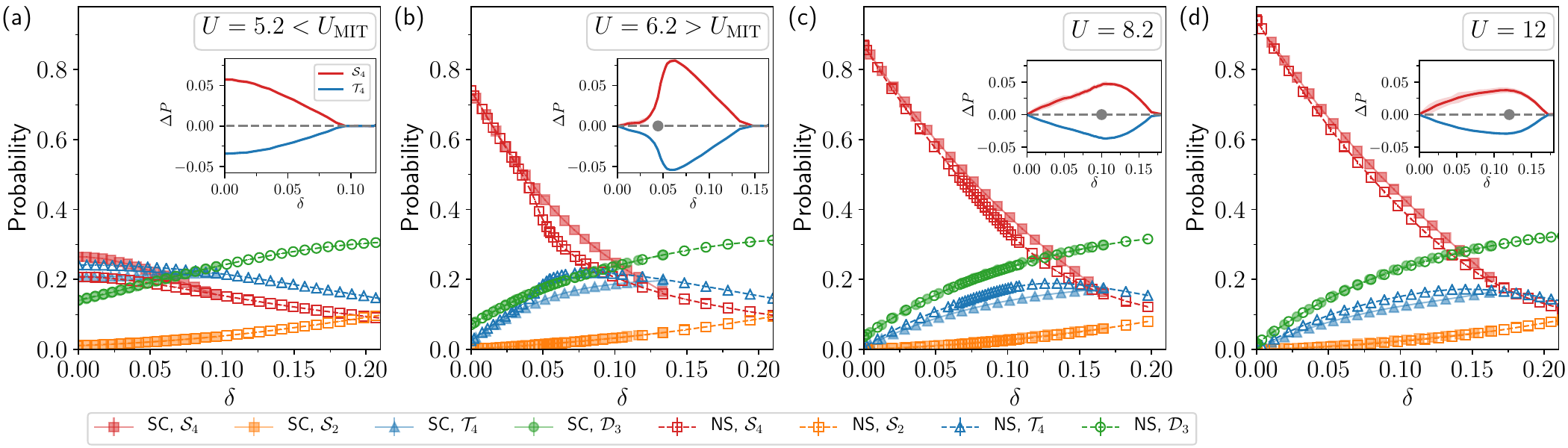}
}
\caption{Probability of selected superconducting and normal state plaquette sets of sectors as a function of doping, for several values of the interaction strength $U$, at $T=1/50$. Selected plaquette sets are those shown in Figure~\ref{fig3}, using the same color code and listed in Eq.~\ref{keyeigenstates}: $\mathcal{S}_4$, $\mathcal{S}_2$, $\mathcal{T}_4$, and $\mathcal{D}_3$. Filled symbols correspond to the superconducting state solution. Empty symbols correspond to the normal state solution. Insets contain the difference between the superconducting and normal state probabilities of the singlet $\mathcal{S}_4$ (red) and triplet $\mathcal{T}_4$ (blue) respectively. The shaded area around each curve shows the root-mean-square error. For each $U$ above $U_{\rm MIT}$, a grey filled circle indicates the estimate of the doping level of the pseudogap to metal critical endpoint in the normal state (see Fig.~\ref{fig1}).
}
\label{fig4}
\end{figure*}

For $U> U_{\textrm{MIT}}$, previous studies~\cite{hauleDOPING, gullEPL, sht, sht2} have demonstrated that the dominant sector in the normal state pseudogap phase is the four-electron singlet $\mathcal{S}_{4}$. Physically, this is because superexchange locks the electrons of the plaquette into one prevailing singlet configuration. When superconductivity emerges from a pseudogap (panels c, e, g), the probabilities of the plaquette sectors do not undergo a drastic change, and in particular the four-electron singlet  $\mathcal{S}_{4}$ remains the dominant configuration with a slightly increased probability. 

On the other hand, when superconductivity emerges from a metal without a pseudogap (panels d, f, h for $U> U_{\textrm{MIT}}$, as well as a, b for $U< U_{\textrm{MIT}}$) there is a more marked redistribution of plaquette probabilities, although still overall small, and in particular there is an increase in the probability of the four-electron singlet $\mathcal{S}_{4}$ at the expense of the probability of the four-electron triplet $\mathcal{T}_{4}$. In this regime the $\mathcal{S}_{4}$ singlet is largest but not dominant, i.e. its probability is comparable with that of other sectors, suggesting the electrons of the plaquette spend a similar amount of time in the other sectors. Regardless of if superconductivity emerges from a pseudogap or a metal, the probability of the doublet $\mathcal{D}_{3}$ remains essentially unchanged upon superconducting condensation.

\subsection{Doping evolution of plaquette sectors probabilities}

Figure~\ref{fig4} shows the probabilities of the four key cluster sets of sectors identified in Eq.~\ref{keyeigenstates} as a function of doping, in the superconducting and normal states (filled and unfilled symbols, respectively), for different values of $U$ at $T=1/50$. 
The doping evolution of the plaquette sectors probabilities in the normal state has been discussed in Refs.~\onlinecite{sht,sht2}, here we report the behavior in the superconducting state and compare with the normal state.

Hole doping enables both charge and spin fluctuations. Compatible with those previous reports, for $U>U_{\textrm{MIT}}$, we find: 1) the probability of the four-electron singlet $\mathcal{S}_4$ decreases rapidly upon doping the parent Mott insulating state, where superexchange is largest; 2) the probability of the three-electron doublet $\mathcal{D}_3$ increases, in line with charge fluctuations introduced by hole-doping,  which break the singlet bonds; 3) the probability of the four-electron triplet $\mathcal{T}_4$ first increases due to the decay of superexchange with doping, and then decreases as the total number of electrons in the system is reduced; 4) the two-electron singlet $\mathcal{S}_2$ undergoes a slow but steady increase as the system evolves away from the dominant four-electron singlet sector with doping. On the other hand, for $U<U_{\textrm{MIT}}$, the reduction of superexchange physics causes the overall depletion of the four-electron singlet $\mathcal{S}_4$, and so at low doping the probability of the four-electron triplet $\mathcal{T}_4$, and to a lesser degree the probability of the three-electron doublet $\mathcal{D}_3$, is no longer suppressed. 

Let us now turn to the doping evolution of the probabilities of the cluster sectors in the superconducting state (filled symbols). Overall, upon condensation the probability of the four-electron singlet $\mathcal{S}_4$ increases at the expense of the probability of the four-electron triplet $\mathcal{T}_4$, whereas the probabilities of the three-electron doublet $\mathcal{D}_3$ and the two-electron singlet $\mathcal{S}_2$ do not change appreciably.

Physically, this means that, for all values of $U$, upon condensation the system lowers its energy by a redistribution of mainly short-range spin, but not charge, excitations - electrons in Cooper pairs are locked into short-range spin singlets due to superexchange. 
These four-electron $\mathcal{S}_4$ singlets are already the dominant configuration in the underlying normal state pseudogap. As we shall discuss in Section~\ref{sec:bath}, upon condensation these singlets propagate coherently in the lattice. Our results complement the findings for the $t-J$ model around optimal doping of Refs.~\onlinecite{hauleAVOIDED, hauleDOPING}.

To better understand the doping evolution of the plaquette fluctuations between singlet $\mathcal{S}_4$ and triplet $\mathcal{T}_4$, we show the difference $\Delta P$ between the probabilities of each of these two sets between the normal and the superconducting phases (red and blue respectively) in the insets of each panel. For $U>U_{\textrm{MIT}}$, this difference is nonmonotonic with doping.
Examining panel (b) for $U=6.2$, upon condensation the probability of the $\mathcal{S}_4$ singlet shows minimal change at low doping, with a rapid increase as the doping is increased past the critical endpoint of the normal-state pseudogap-metal first order transition (grey dot). The difference $\Delta P$ eventually decreases again approaching the end of the superconducting dome to recover the normal phase probabilities. At higher $U$, the trend of the difference for the singlet is similar, but instead shows a gradual increase to a broad maximum. This is possibly because of the temperature dependence of the probabilities in the normal state - indeed, the critical endpoint of the pseudogap-metal transition in the normal state shifts to lower temperature and higher doping with increasing $U$. Note that for $U=6.2$, $T=1/50$ is in close proximity to this transition.

Overall, our analysis on the plaquette sectors provides two main insights on the superconducting correlations. First, it establishes that at the level of the cluster, superconductivity mainly entails a reorganisation of short-range spin correlations rather than charge correlations. Second, it identifies short-range spin correlations in the form of four-electrons singlets as key to superconducting pairing. 

Note that, even if isolated plaquettes show tendencies to singlet formation~\cite{scalapino1996, Danilov:2022}, it is only when these singlets are immersed in the self-consistent bath, which takes into account the effect of the infinite lattice, that superconductivity can arise. The behavior of the bath is thus analysed in the next session.

\begin{figure*}%[ht!]
\centering{
\includegraphics[width=1.\linewidth]{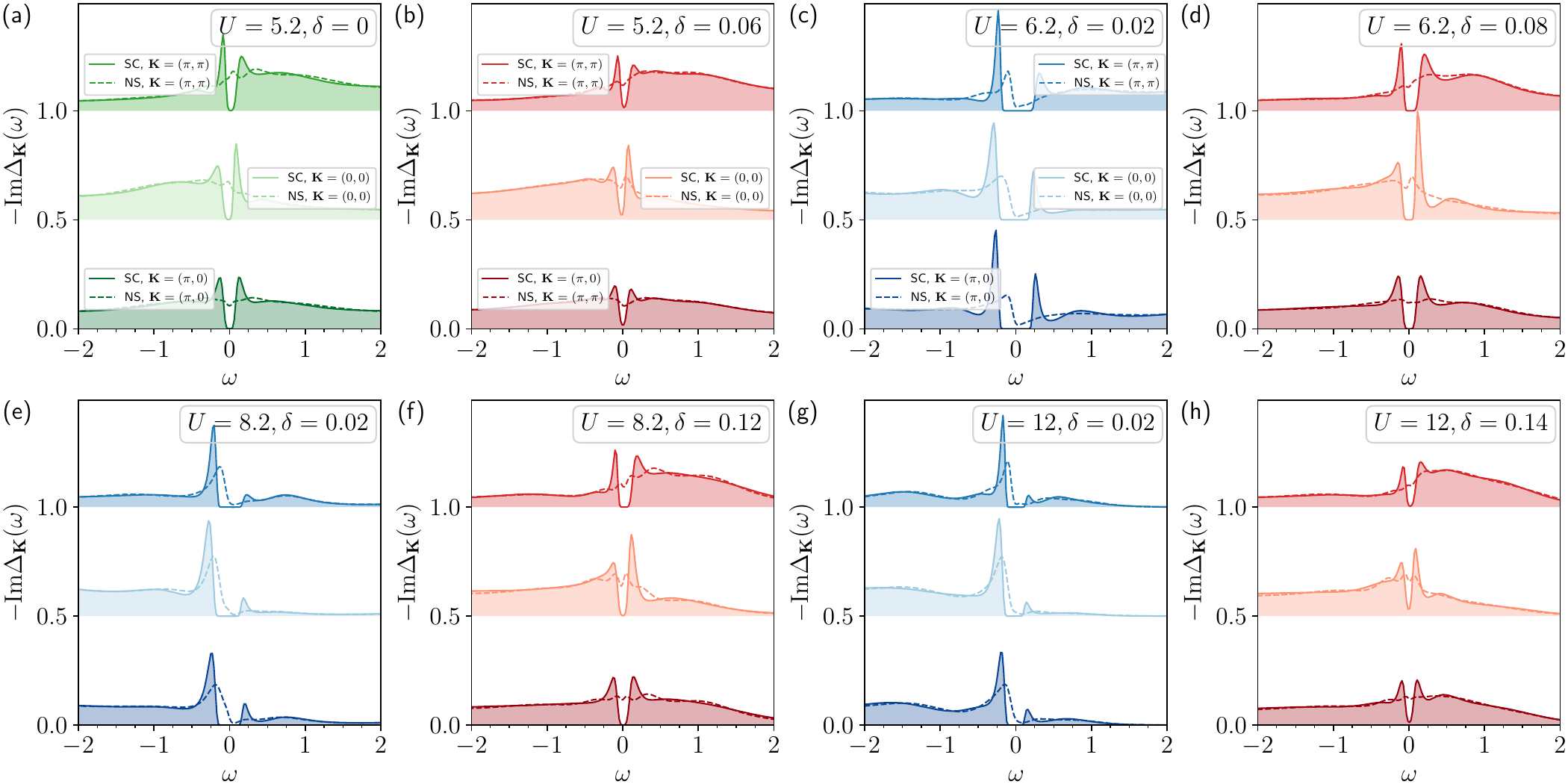}
}
\caption{Imaginary part of the hybridization function $-\textrm{Im}\Delta_{\textbf{K}}(\omega)$ for several values of $U$ and $\delta$, corresponding to the filled hexagons in Fig.~\ref{fig1}. The dashed line shows the normal state solution, whereas the filled line with shaded region denotes the normal component of the superconducting state solution. Each panel shows the spectral function corresponding to cluster momenta $\textbf{K}=(\pi, 0), (0, 0), (\pi,\pi)$ (shifted by $0.5$ each for a better visualisation). Analytical continuation has been done using the method of Ref.~\onlinecite{DominicMEM}. Each spectral function is normalised to $1$.}
\label{fig5}
\end{figure*}
\begin{figure*}%[ht!]
\centering{
\includegraphics[width=1.\linewidth]{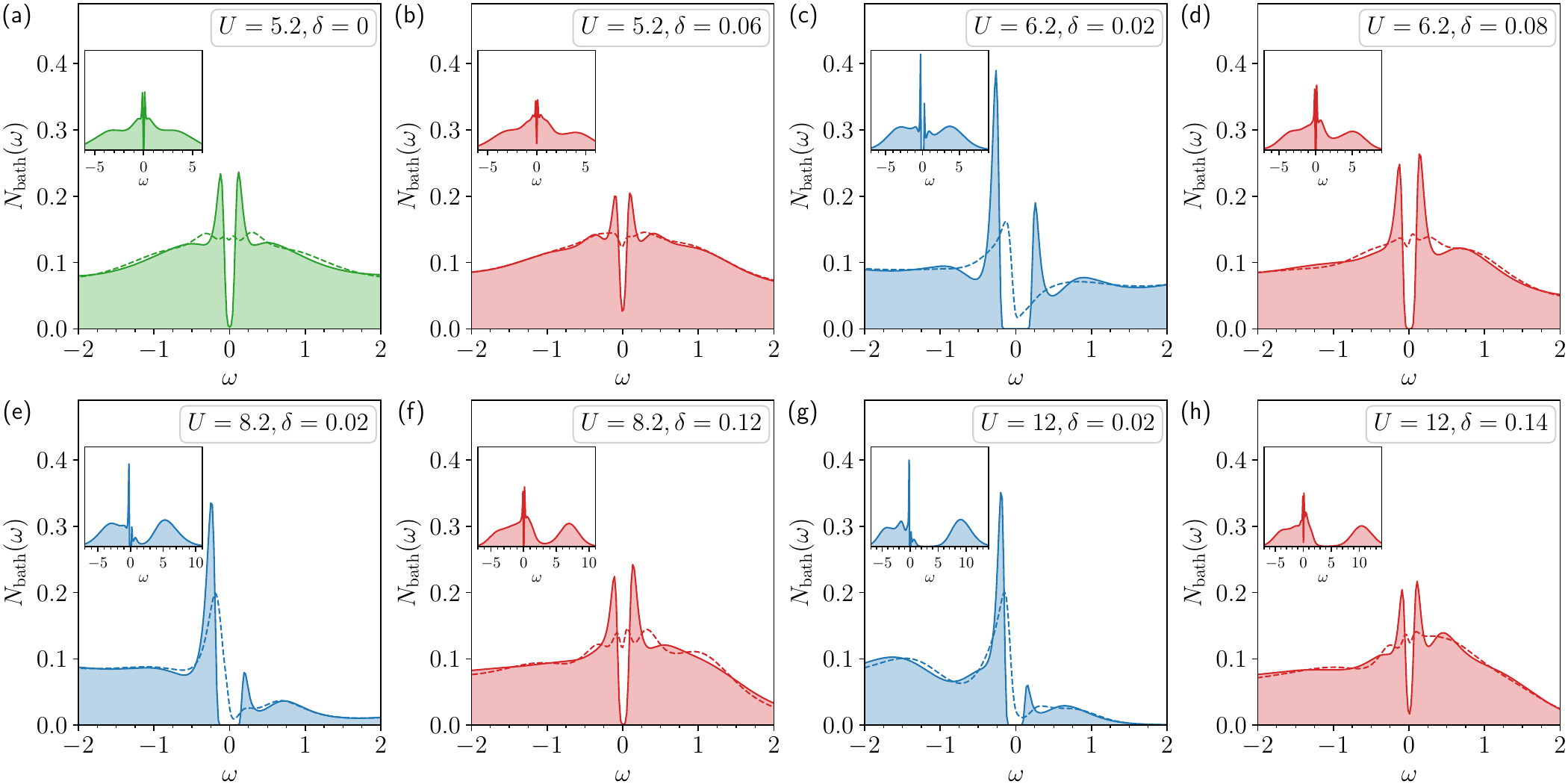}
}
\caption{Local density of states of the bath $N_{\textrm{bath}}(\omega)=-\textrm{Im}\Delta_{\textbf{R}=(0,0)}(\omega)$ for the same values of $U$ and $\delta$ as in Fig.~\ref{fig5}, which correspond to the filled hexagons in Fig.~\ref{fig1}. The dashed line shows the normal state solution, whereas the filled line with shaded region denotes the superconducting state solution. Insets in each panel show $N_{\rm bath}(\omega)$ in the superconducting state on the full spectrum of the frequency. As in Fig.~\ref{fig5}, the normalisation of each density of states is equal to $1$.}
\label{fig6}
\end{figure*}

\section{Self-consistent bath hybridization function}
\label{sec:bath}

The preceding section demonstrated that upon condensation there is a redistribution of the probability of the plaquette sectors. This redistribution is small and involves mainly short range spin (singlet $\mathcal{S}_4$, triplet $\mathcal{T}_4$) but not charge (doublet $\mathcal{D}_3$) excitations. In CDFMT however, the plaquette is not isolated but is embedded in a self-consistent bath of noninteracting electrons. The plaquette exchanges electrons with the bath, and it is through this exchange that the plaquette is able to make transitions between the 256 available plaquette eigenstates. 
Furthermore, the d-wave symmetry is broken in the bath.   
The goal of this section is therefore to analyse the behavior of the bath upon condensation, which is described by the Nambu diagonal hybridization matrix function ${\bf \Delta}$. Note that although we separate the discussion of the behavior of the bath from that of the plaquette for practical purposes, they are not independent quantities: the plaquette is immersed in the bath which is self-consistently determined. Therefore the behavior of the plaquette influences that of the bath and vice versa. 

Figure~\ref{fig5} shows $-\textrm{Im}\Delta_{\textbf{K}}(\omega)$ both in the normal and superconducting states (dashed and filled lines respectively). In the superconducting case, $\Delta_{\textbf{K}}(\omega)$ is the Nambu diagonal (i.e. normal) hybridization function (see Eq.~\ref{eq:DeltaSC}). For both normal and superconducting cases, $-\textrm{Im}\Delta_{\textbf{K}}(\omega)$ gives the (normal) spectral function of the bath resolved in the cluster momentum ${\bf K}$. This quantity is shown for the values of interaction, doping, and temperature corresponding to the filled hexagons of Fig.~\ref{fig1}. 
We perform the analytical continuation from imaginary to real frequencies using the method of Ref.~\onlinecite{DominicMEM} and plot the independent diagonal components $\textbf{K}= (0,0), (\pi, 0), (\pi, \pi)$ of the matrix ${\bf \Delta}$, in the energy window $\omega \in (-2,2)$. 
The behavior of the bath has been described in the normal state in Ref.~\onlinecite{hauleDOPING, sht2}, and in Ref.~\onlinecite{michelPRB} for a 2-site cluster. The focus of the present work is to analyse the behavior in the superconducting state and contrast it with that in the normal state. Hence, for a better visualisation we show the superconducting solution shaded. 

We shall first briefly review the properties of the ${\bf K}$-resolved spectral function of the bath in the normal state, as prior work~\onlinecite{hauleDOPING, sht2} mostly focused on the spectral function of the bath on the Matsubara frequency. The bath shows finite spectral weight close to the Fermi energy, displaying metallic (panels a, b, d, f, h) or pseudogap (panels c, e, g) behavior. The latter takes a markedly asymmetric shape, with a more pronounced peak below the Fermi energy. In all cases, the bath is weakly $\textbf{K}$-dependent close to the Fermi energy. 

Upon entering into the superconducting phase, the Nambu diagonal spectral function of the bath shows a dramatic redistribution of spectral weight at low frequency. 
A  redistribution of the spectral weight in the bath is expected because in CDMFT, d-wave symmetry is broken in the bath but not in the cluster. 
For all interaction strengths and dopings considered, the bath opens a superconducting gap. Superconductivity emerging from a pseudogap (panels c, e, g) leaves a distinct signature in the spectral function of the bath, in the form of an inherited asymmetry of the superconducting gap. The position of the superconducting coherence peaks is different from the position of the peaks of the pseudogap. Upon doping, the size of the gap narrows (panels b, d, f, h) and becomes more symmetric, particularly for $\textbf{K}= (\pi,0)$. Similarly to the normal state, the bath in the superconducting state shows a weak $\textbf{K}$-dependence at low frequency. 

The results of Figure~\ref{fig5} can be summarised by computing the local density of states of the bath $N_{\textrm{bath}}(\omega)=-\textrm{Im}\Delta_{\textbf{R}=(0,0)}(\omega)$. Its low frequency part is shown in Figure~\ref{fig6} with the same color code as Figure~\ref{fig5}. The inset of each panel shows the full frequency range of $N_{\textrm{bath}}(\omega)$, where the lower and upper Hubbard bands can be seen. 

To conclude this section on the behavior of the (Nambu diagonal) spectral function and local density of states of the bath, our main finding is therefore a dramatic spectral weight redistribution at low frequency upon superconducting condensation, which leads to the opening of a superconducting gap. 
The bath hybridization function $\Delta_{\textbf{K}}(\omega)$ describes the hopping processes of the electrons between the plaquette and the bath, therefore a superconducting gap in $\Delta_{\textbf{K}}(\omega)$ suggests no dissipation of the dynamics of these one-particle hopping processes. 
The increased coherence in the superconducting state can also be deduced from the suppressed electronic entropy~\cite{CaitlinPNAS2021}. 
This study also paves the way for future investigations of the Nambu off-diagonal (i.e. anomalous) component of the spectral function of the bath, which gives information about the pairing dynamics. Since this anomalous component is not positive definite, analytical continuation is more challenging~\cite{AlexisPRB2015MEM, reymbautPRB2016, yue2023maximum}.

\section{Consequences on the density of states of the system and spin susceptibility}
\label{sec:discussion}

Sections \ref{sec:probabilities} and \ref{sec:bath} have shown how superconductivity is realised in CDMFT at the level of the cluster quantum impurity problem. We found that upon entering the superconducting state, there is a redistribution of the probabilities of the plaquette sectors and of the spectral weight of the bath hybridization function. In this section, we show how this analysis can provide new insights of the behavior of the density of states of the system and of the zero-frequency spin susceptibility.

\subsection{Density of states of the system}

\begin{figure*}[ht!]
\centering{
\includegraphics[width=1.\linewidth]{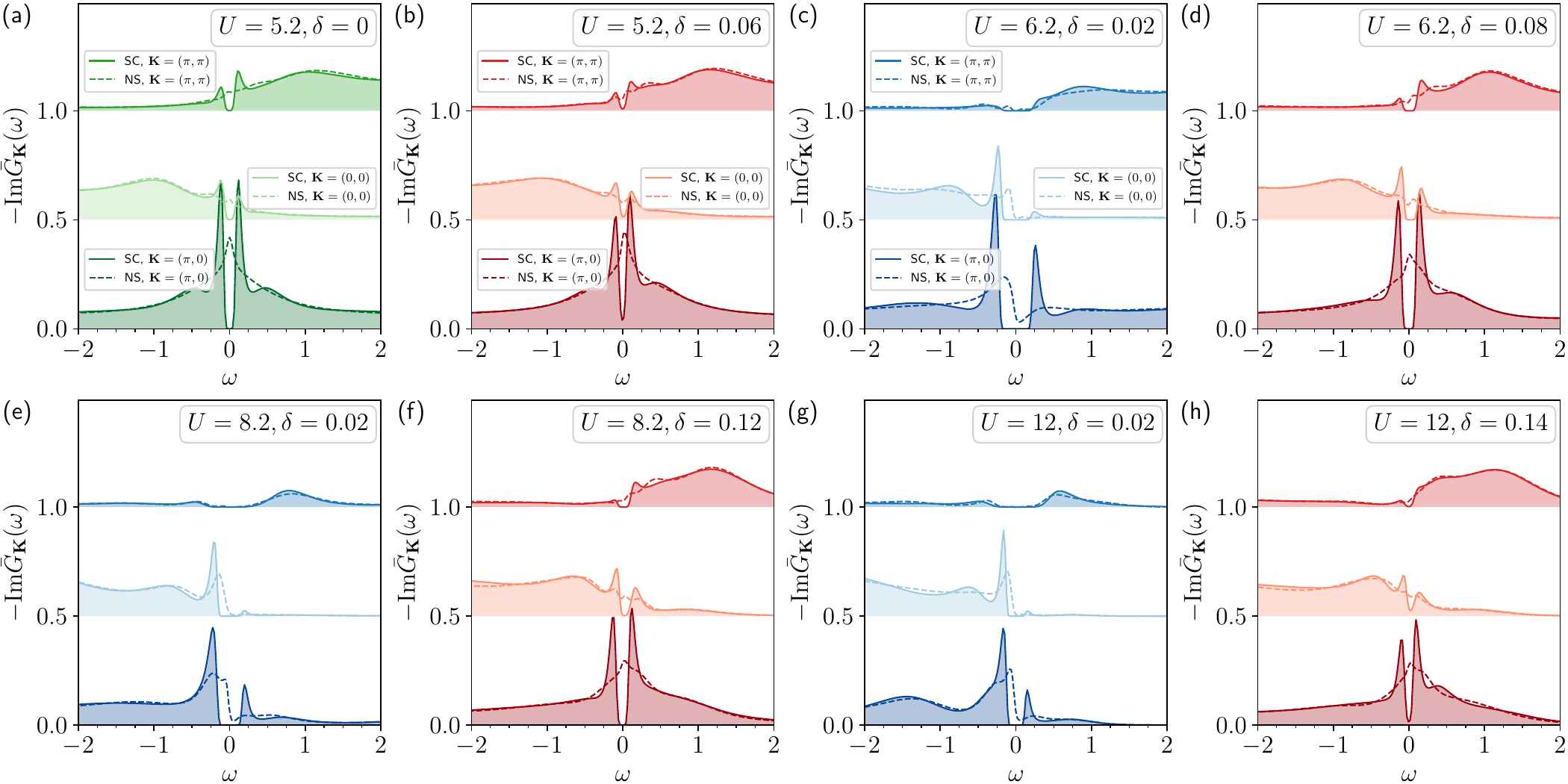}
}
\caption{Imaginary part of the Green's function $-\textrm{Im}\bar{G}_{\bf K}(\omega)$ for different values of $U$ and $\delta$ corresponding to the filled hexagons in Fig.~\ref{fig1}. The dashed line shows the normal state solution, whereas the filled line with shaded region denotes the normal component of the superconducting state solution. Each panel shows the spectral function corresponding to cluster momenta $\textbf{K}=(\pi, 0), (0, 0), (\pi,\pi)$ (shifted by $0.5$ each for a better visualisation). Each spectral function is normalised to $1$.}
\label{fig7}
\end{figure*}
\begin{figure*}[ht!]
\centering{
\includegraphics[width=1.\linewidth]{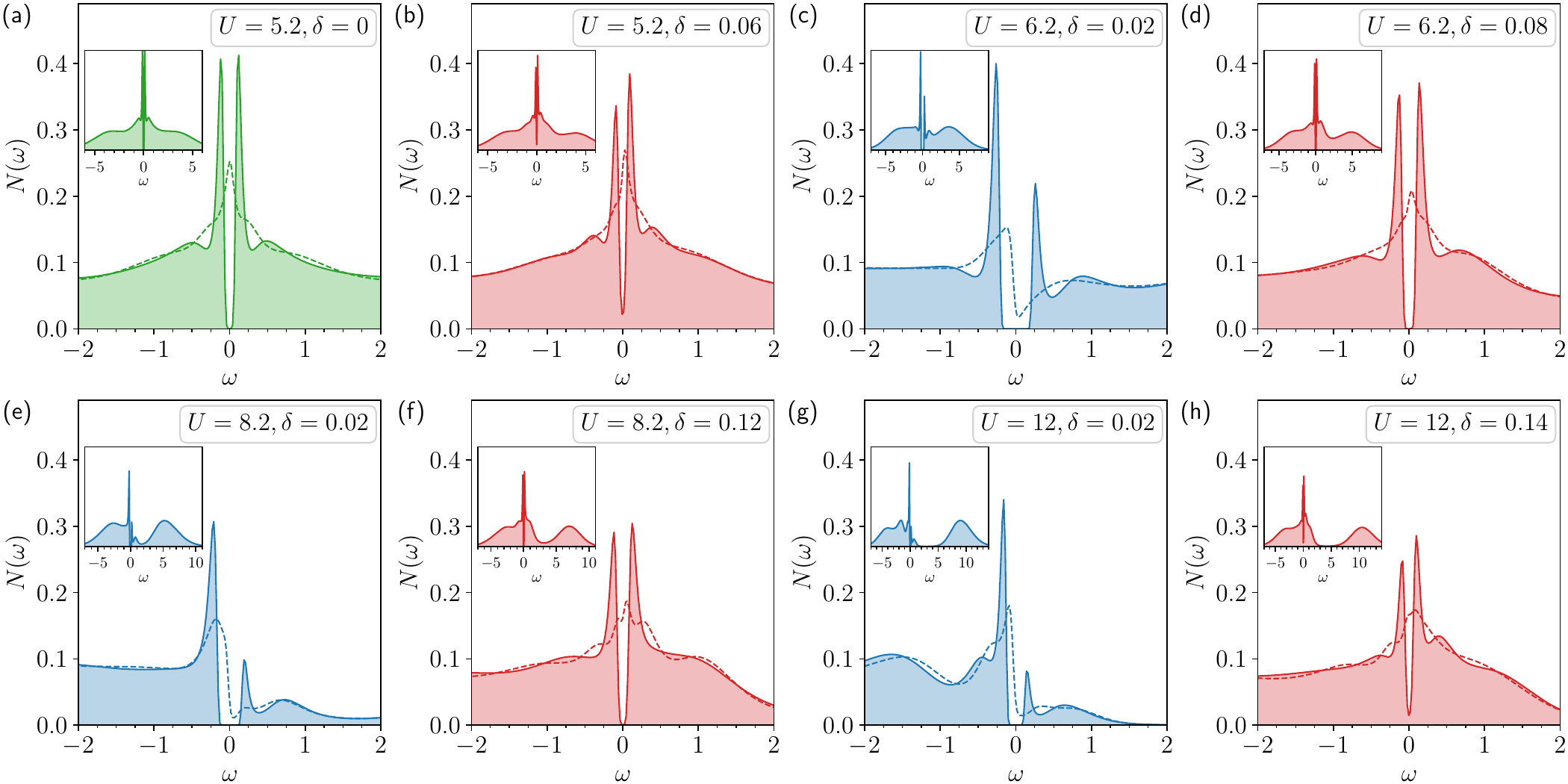}
}
\caption{Low frequency part of the local density of states $N(\omega)$ of the system for the values of $U$, $\delta$ and $T$ indicated by filled hexagons in Fig.~\ref{fig1}. Solid line with shaded region indicates the superconducting state, dashed line the normal state. Insets show the density of states of the system $N(\omega)$ in the superconducting state on the full spectrum of the frequency. Each density of states is normalised to $1$. 
}
\label{fig8}
\end{figure*}
Figure~\ref{fig7} shows the imaginary part of the (normal) Green's function $-\textrm{Im}\bar{G}_{\textbf{K}}(\omega)$, which gives the (normal) spectral function of the system. Figure~\ref{fig8} shows the resulting local density of states of the system $N(\omega)=-\textrm{Im}\bar{G}_{\textbf{R}=(0,0)}(\omega)$. Data are shown for both the normal and superconducting state (dashed lines and filled lines with shaded regions, respectively). Previous studies have analysed the behavior of these quantities~\cite{hauleDOPING, sshtSC, Gull:2013}, but the new insight brought by our study lies in the comparison of the Green's function and the hybridization function, in the superconducting state. In single-site DMFT on the Bethe lattice, the self-consistency condition requires $\Delta= t^{2}\bar{G}$, so the bath is directly proportional to the Green's function. Therefore in CDMFT, differences between ${\bf \Delta}$ and $\bar{\bf G}$ should be ascribed to the short range correlations that are incorporated in the cluster.

First, we will briefly recap the behavior of the spectral function of the system in the normal state, analysed in Refs.~\onlinecite{sht2, ssht}. In sharp contrast with Fig.~\ref{fig5} for the bath, the Green's function is strongly $\textbf{K}$-dependent. At low doping, the spectral function of the bath in Fig.~\ref{fig5} shows an asymmetric pseudogap for all $\textbf{K}$-components. In Fig.~\ref{fig7} there is instead an asymmetric pseudogap in the $\textbf{K}=(\pi,0)$ component only, with the $\textbf{K}=(\pi,\pi)$ component showing insulating-like behavior. Therefore the strong $\textbf{K}$-differentiation is linked to short range correlations within the cluster. At high doping, the flat behavior of the bath hybridization function at low frequency for all $\textbf{K}$ in Fig.~\ref{fig5} is replaced in Fig.~\ref{fig7} by a quasiparticle peak in the $\textbf{K}=(\pi,0)$ component. 

We now turn to the analysis of the superconducting phase. Again, for all values considered here, the Nambu diagonal spectral function of the system shows marked cluster momentum differentiation, in contrast with the behavior of the bath, emphasising again the importance of short-range correlations included in the cluster. 
The bath hybridization function in Fig.~\ref{fig5} shows superconducting coherence peaks for all $\textbf{K}$ components. In Fig.~\ref{fig7} the coherence peaks manifest predominantly in the $\textbf{K}=(\pi,0)$ component and to a lesser degree in the $\textbf{K}=(0,0)$ component. At low doping for $U>U_{\textrm{MIT}}$ (panels c, e, g), the spectral function in the superconducting state reflects the inherited particle-hole asymmetry of the underlying normal-state pseudogap. 

We can summarise the results of Fig.~\ref{fig7} in the local density of states, Fig.~\ref{fig8}. Upon condensation, the density of states develops a superconducting gap across the Fermi energy. The redistribution of spectral weight between normal and superconducting state occurs over a range of frequency larger than the gap - a typical signature of strongly correlated superconductivity. The asymmetry in the superconducting state is inherited from the asymmetry in the pseudogap. However as has already been observed in Refs.~\onlinecite{Gull:2013, Verret:PRB2019}, the magnitude of the superconducting gap differs from that of the pseudogap, implying they are two distinct phenomena. The width of the superconducting gap decreases with increasing doping. Note that the system is a $d_{x^{2}-y^{2}}$ superconductor, however a cluster larger than a $2\times2$ plaquette is needed to resolve the nodes along the diagonals of the Brillouin zone.
If we compare Fig.~\ref{fig8} to Fig.~\ref{fig6}, the overall shape is similar, however the magnitude of the coherence peaks is enhanced in $\bar{\bf G}$ compared to ${\bf \Delta}$.  
This reflects the fact that superconducting fluctuations are also present in the cluster through the Nambu off-diagonal hybridization function and self-energy. 

Overall, the comparison between the spectral functions of ${\bf \Delta}$ and $\bar{\bf G}$ enabled us to identify key signatures of short-range correlations in the spectral functions of the system: enhanced coherence pics and strong cluster momentum dependence.

\subsection{Zero-frequency spin susceptibility \label{chi}}

\begin{figure*}[ht!]
\centering{
\includegraphics[width=1.\linewidth]{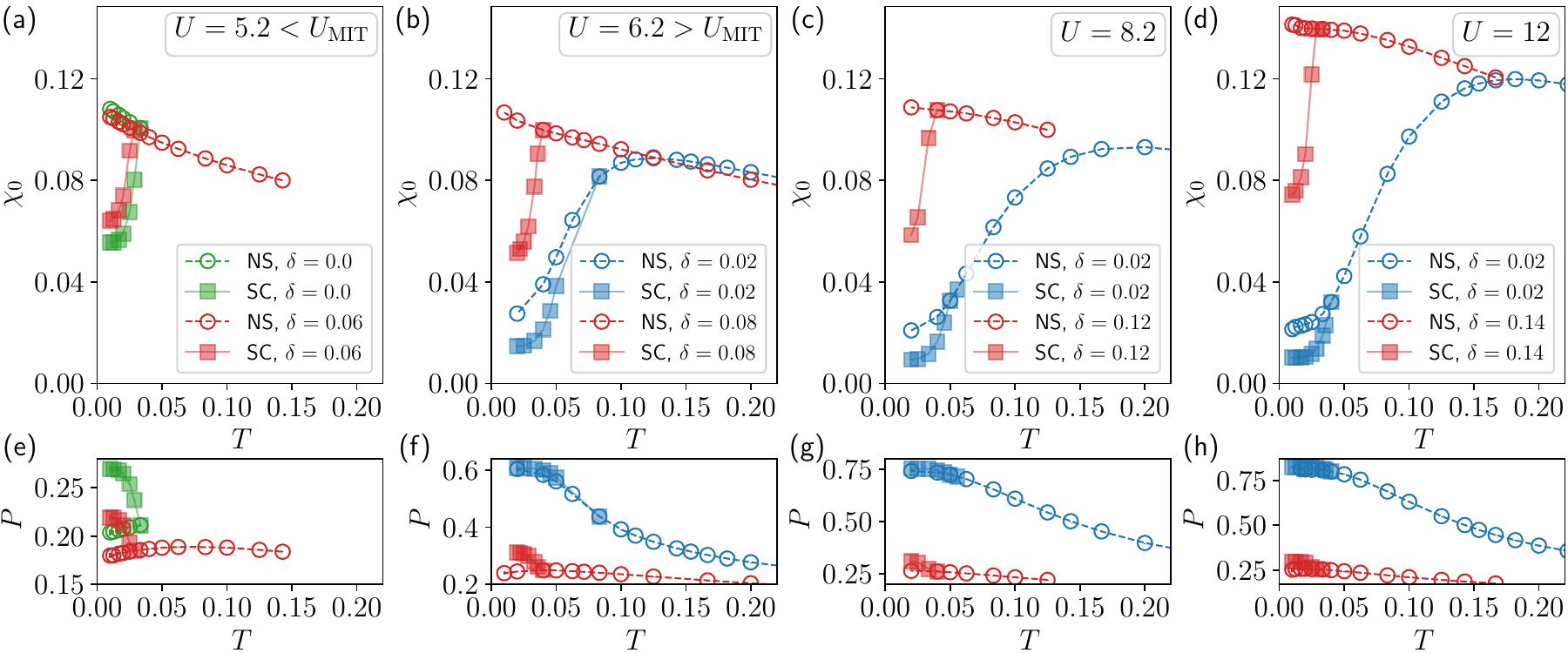}
}
\caption{(a), (b), (c), (d): zero-frequency spin susceptibility $\chi_0$ versus temperature $T$. Data are shown for several values of $U$ and $\delta$, corresponding to the filled hexagons in the $T-\delta$ phase diagram of Figure~\ref{fig1}. Filled squares correspond to the superconducting state solution. Empty circles correspond to the normal state solution. 
(e), (f), (g), (h): temperature dependence of the superconducting and normal state probabilities of the four-electron singlet $\mathcal{S}_4$. Data are shown for the same values of top panels. 
}
\label{fig9}
\end{figure*}

Next, we turn to the signatures of the superconducting correlations in the zero frequency spin susceptibility. This quantity is defined by $\chi_0 (T) = \int_0^\beta \langle S_z(\tau) S_z(0) \rangle d\tau$, where $S_z$ is the projection of the total spin of the plaquette along the $z$ direction.  Figure~\ref{fig9} a, b, c, d shows $\chi_0(T)$ as a function of temperature $T$, both in the superconducting and normal states (filled and open symbols, respectively). Data are shown for several values of $U$ and $\delta$, corresponding to the color-coded hexagons in Figure~\ref{fig1}.

The temperature and doping behavior of $\chi_0$ in the normal state has been discussed in Refs.~\onlinecite{sht2, ssht, sshtRHO}. In the normal state, $\chi_0(T)$ is Pauli-like in the correlated metal found below $U_{\rm MIT}$ (green and red circles in panel (a)) and above $U_{\rm MIT}$ at large $\delta$ (red circles in panels (b),(c), (d)). In contrast, for $U>U_{\rm MIT}$ $\chi_0(T)$ and small doping (blue circles in panels (b),(c), (d)) $\chi_{0}(T)$ shows a low temperature drop.   
By looking at the behavior of the probabilities of the cluster sectors as a function of temperature, it can be seen that the drop in the spin susceptibility coincides with an increase in the probability of the four-electron singlet (see lower panels e, f, g, h)  as noted in Ref.~\onlinecite{ssht}. The maximum in $\chi_0(T)$ signals the crossover temperature $T^*$ in Fig.~\ref{fig1}, which indicates the opening of the strongly correlated pseudogap.

Let us now turn to the superconducting state. Our main finding is that $\chi_0(T)$ dramatically drops below the superconducting temperature $T_c^d$, for all doping levels and all values of $U$ considered here (filled squares in Figure~\ref{fig9} panels a, b, c, d). 
The drop can be associated to the further increase of the probability of the four-electron singlet that occurs upon entering the superconducting state below $T_{c}^{d}$ (see panels e, f, g, h). Physically, Cooper pairs in the superconducting state are locked into singlets, and hence spin fluctuations to other cluster configurations are reduced. 
When superconductivity emerges from a metallic state (red curve) there is a more pronounced drop in the spin susceptibility compared to superconductivity emerging from a pseudogap (blue curve). This is because singlet formation already occurred at $T^{*}$ in the underlying normal state. 

Note that in the superconducting state at low temperature, $\chi_{0}(T)$ saturates at a nonzero value, mirroring the saturation of the singlet probability with temperature (see lower panels). This saturation is due to the persisting probability of states such as the four-electron triplet $\mathcal{T}_{4}$ and the three-electron doublet $\mathcal{D}_{3}$ at low temperature, and hence residual spin flipping due to fluctuations between configurations require that $\chi_{0}\neq 0$ at finite doping. The spin susceptibility $\chi_0(T\rightarrow 0)$ drops to zero in the normal state Mott insulator at $\delta=0$ and large $U$ (see Ref.~\onlinecite{sht2, ssht}). This is because in the Mott state at low $T$ and large $U$ the four-electron singlet probability approaches 1. 
In the superconducting state, the magnitude of $\chi_{0}(T)$ when superconductivity condenses from a metal exceeds that condensing from a pseudogap. This is due to the doping evolution of the probabilities of the plaquette sectors, where upon doping the high probability of the dominant $\mathcal{S}_{4}$ singlet is redistributed across other plaquette sectors, meaning an increase in fluctuations between different configurations, and thus an increase in $\chi_{0}(\delta)$.

The results of $\chi_{0}$ should be considered only as a proxy for understanding the trend of the Knight shift in NMR~\cite{Alloul:1989, Takigawa:1989}. In a pure singlet superconductor, $\chi_{0}$ drops to zero at zero temperature. This is because electrons in the Cooper pairs locked into singlets cannot be polarised by an applied magnetic field and thus $\chi_0(T)$ drops below the superconducting critical temperature. However, the $\chi_{0}$ we calculate here is a cluster quantity, and as seen $\chi_0(T)$ cannot drop to zero at finite doping. 

Overall, our comparison between the zero-frequency susceptibility and the cluster sectors probabilities allowed us to link key features of $\chi_0$ to the redistribution of short-range correlations with temperature and doping.

\section{Conclusions}
\label{sec:conclusions}

We address the interplay between superconducting correlations and Mott physics in the two-dimensional Hubbard model solved with CDMFT on a $2\times2$ plaquette. Our approach takes the perspective of a cluster quantum impurity model embedded in a self-consistent bath. Thus we focus on the properties of {\it both} the cluster and the bath. To unveil microscopic trends in the superconducting correlations, our analysis (a) compares the superconducting state with the underlying normal state and (b) covers a wide range of interaction strength, doping, and temperature.
 
First, at the level of the plaquette, we compute the probabilities that cluster electrons are found in the cluster sectors. We observe that few states have high probability. Upon entering the superconducting state, the cluster electrons spend more time in the four-electron singlet set of sectors $\mathcal{S}_4$, suggesting the electrons in the Cooper pairs are bound into short-range spin singlets owing to superexchange mechanism. This finding enables us to identify short-range spin correlations in the form of singlets as central to superconducting pairing. 
Furthermore, our results show an increase in the probability of the four-electron singlet mostly at the expense of a decrease of the four-electron triplet probability, with a negligible probability redistribution of the charge fluctuations. 
The implication of this finding is that superconductivity at the level of the cluster mainly involves a reorganisation of short-range spin correlations but not of charge correlations.

At the level of the self-consistent bath, upon entering the superconducting state we find a redistribution of the spectral weight of the cluster-momentum-resolved (normal) spectral function and of the resulting local density of states of the bath. The most notable feature is the appearance of a superconducting gap and the weak ${\bf K}$ dependence of the diagonal bath hybridization function. 

Our analysis from the perspective of a cluster quantum impurity model in a self-consistent bath can help us to unveil the links between superconducting correlations and some features of the spectral function and of the density of states of the system, as well as the zero-frequency spin susceptibility of the plaquette. 
In the superconducting state, short-range correlations give rise to a marked $\textbf{K}$-dependence of the spectral function of the system, and pronounced coherence peaks in the density of states. Upon superconducting condensation the spin susceptibility drops, mirroring the increase in the probability of the four-electron singlet state. 

Overall, our work underscores the importance of short-range spin correlations in the formation of Cooper pairs in a doped Mott insulator. This suggests the possibility of controlling superconducting properties by tuning the probability of the four-electron singlet, for example by introducing frustration at the level of the hopping or the lattice geometry. 

From a broader perspective, our work illustrates the value of the approach of analysing both cluster impurity and bath~\cite{micheleJPCM2007}. 
Thus our work may open up a new direction to analyse other strongly correlated models with cluster DFMT methods, from the perspective of a quantum impurity model embedded in self-consistent bath. 

This work may also contribute to the goal of understanding quantum phases of matter using measures of entanglement \cite{amicoRMP2008, Caitlin:PRL2019, CaitlinSb, Caitlin:PRXQ2020, CaitlinPNAS2021}. Further refinement of the method may enable access to the off-diagonal elements of the reduced density matrix~\cite{Udagawa_Motome:2015, Humeniuk2019}, which would be key to enabling the calculation of more sophisticated measures of quantum entanglement~\cite{amicoRMP2008, zhengBOOK}.

\begin{acknowledgments}
This work has been supported by the Canada First Research Excellence Fund. Simulations were performed on computers provided by the Canadian Foundation for Innovation, the Minist\`ere de l'\'Education des Loisirs et du Sport (Qu\'ebec), Calcul Qu\'ebec, and Compute Canada.
\end{acknowledgments}

%\clearpage
\appendix*
\section{List of plaquette sectors}
\label{sec:eigenstates}

Table~\ref{table} shows the list of the plaquette sectors. 

%
%\FloatBarrier
\begin{table*}[ht!]
\caption{\label{table} List of the plaquette sectors.}
\begin{ruledtabular}
\begin{tabular}{llllll}
Number of particles $N$ & Total cluster spin $S_z$ & Cluster momentum ${\bf K}$ & Sector dimension &  Sector index $m$ \\
\hline
0 & 0 &  $(0,0)$ & 1  & 1 \\
\hline
1 & $\pm 1/2$ &  $(0,0)$ & 1  & 2, 3 \\
1 & $\pm 1/2$ &  $(0,\pi)$ & 1  & 4, 5 \\
1 & $\pm 1/2$ &  $(\pi,0)$ & 1  & 6, 7 \\
1 & $\pm 1/2$ &  $(\pi,\pi)$ & 1  & 8, 9 \\
\hline
2 & $\pm 1$ &  $(0,\pi)$ & 2  & 10, 11 \\
2 & $\pm 1$ &  $(\pi,0)$ & 2  & 12, 13 \\
2 & $\pm 1$ &  $(\pi,\pi)$ & 2  & 14, 15 \\
2 & 0 & $(0,0)$ & 4  & 16 \\
2 & 0 & $(0,\pi)$ & 4  & 17 \\
2 & 0 & $(\pi,0)$ & 4  & 18 \\
2 & 0 & $(\pi,\pi)$ & 4  & 19 \\
\hline
3 & $\pm 3/2$ &  $(0,0)$ & 1  & 20, 21 \\
3 & $\pm 3/2$ &  $(0,\pi)$ & 1  & 22, 23 \\
3 & $\pm 3/2$ &  $(\pi,0)$ & 1  & 24, 25 \\
3 & $\pm 3/2$ &  $(\pi,\pi)$ & 1  & 26, 27 \\
3 & $\pm 1/2$ &  $(0,0)$ & 6  & 28, 29 \\
3 & $\pm 1/2$ &  $(0,\pi)$ & 6  & 30, 31 \\
3 & $\pm 1/2$ &  $(\pi,0)$ & 6  & 32, 33 \\
3 & $\pm 1/2$ &  $(\pi,\pi)$ & 6  & 34, 35 \\
\hline
4 & 0 &  $(0,0)$ & 12  & 36 \\
4 & 0 &  $(0,\pi)$ & 8  & 37 \\
4 & 0 &  $(\pi,0)$ & 8  & 38 \\
4 & 0 &  $(\pi,\pi)$ & 8  & 39 \\
4 & $\pm 1$ &  $(0,0)$ & 4  & 40, 41 \\
4 & $\pm 1$ &  $(0,\pi)$ & 4  & 42, 43 \\
4 & $\pm 1$ &  $(\pi,0)$ & 4  & 44, 45 \\
4 & $\pm 1$ &  $(\pi,\pi)$ & 4  & 46, 47 \\
4 & $\pm 2 $ &  $(0,0)$ & 1  & 48, 49 \\
\hline
5 & $\pm 3/2$ &  $(0,0)$ & 1  & 50, 51 \\
5 & $\pm 3/2$ &  $(0,\pi)$ & 1  & 52, 53 \\
5 & $\pm 3/2$ &  $(\pi,0)$ & 1  & 54, 55 \\
5 & $\pm 3/2$ &  $(\pi,\pi)$ & 1  & 56, 57 \\
5 & $\pm 1/2$ &  $(0,0)$ & 6  & 58, 59 \\
5 & $\pm 1/2$ &  $(0,\pi)$ & 6  & 60, 61 \\
5 & $\pm 1/2$ &  $(\pi,0)$ & 6  & 62, 63 \\
5 & $\pm 1/2$ &  $(\pi,\pi)$ & 6  & 64, 65 \\
\hline
6 & $\pm 1$ &  $(0,\pi)$ & 2  & 66, 67 \\
6 & $\pm 1$ &  $(\pi,0)$ & 2  & 68, 69 \\
6 & $\pm 1$ &  $(\pi,\pi)$ & 2  & 70, 71 \\
6 & 0 & $(0,0)$ & 4  & 72 \\
6 & 0 & $(0,\pi)$ & 4  & 73 \\
6 & 0 & $(\pi,0)$ & 4  & 74 \\
6 & 0 & $(\pi,\pi)$ & 4  & 75 \\
\hline
7 & $\pm 1/2$ &  $(0,0)$ & 1  & 76, 77 \\
7 & $\pm 1/2$ &  $(0,\pi)$ & 1  & 78, 79 \\
7 & $\pm 1/2$ &  $(\pi,0)$ & 1  & 80, 81 \\
7 & $\pm 1/2$ &  $(\pi,\pi)$ & 1  & 82, 83 \\
\hline
8 & 0 &  $(0,0)$ & 1  & 84 \\
\end{tabular}
\end{ruledtabular}
\end{table*}
%\FloatBarrier
%

%\clearpage



\begin{thebibliography}{81}%
\makeatletter
\providecommand \@ifxundefined [1]{%
 \@ifx{#1\undefined}
}%
\providecommand \@ifnum [1]{%
 \ifnum #1\expandafter \@firstoftwo
 \else \expandafter \@secondoftwo
 \fi
}%
\providecommand \@ifx [1]{%
 \ifx #1\expandafter \@firstoftwo
 \else \expandafter \@secondoftwo
 \fi
}%
\providecommand \natexlab [1]{#1}%
\providecommand \enquote  [1]{``#1''}%
\providecommand \bibnamefont  [1]{#1}%
\providecommand \bibfnamefont [1]{#1}%
\providecommand \citenamefont [1]{#1}%
\providecommand \href@noop [0]{\@secondoftwo}%
\providecommand \href [0]{\begingroup \@sanitize@url \@href}%
\providecommand \@href[1]{\@@startlink{#1}\@@href}%
\providecommand \@@href[1]{\endgroup#1\@@endlink}%
\providecommand \@sanitize@url [0]{\catcode `\\12\catcode `\$12\catcode
  `\&12\catcode `\#12\catcode `\^12\catcode `\_12\catcode `\%12\relax}%
\providecommand \@@startlink[1]{}%
\providecommand \@@endlink[0]{}%
\providecommand \url  [0]{\begingroup\@sanitize@url \@url }%
\providecommand \@url [1]{\endgroup\@href {#1}{\urlprefix }}%
\providecommand \urlprefix  [0]{URL }%
\providecommand \Eprint [0]{\href }%
\providecommand \doibase [0]{http://dx.doi.org/}%
\providecommand \selectlanguage [0]{\@gobble}%
\providecommand \bibinfo  [0]{\@secondoftwo}%
\providecommand \bibfield  [0]{\@secondoftwo}%
\providecommand \translation [1]{[#1]}%
\providecommand \BibitemOpen [0]{}%
\providecommand \bibitemStop [0]{}%
\providecommand \bibitemNoStop [0]{.\EOS\space}%
\providecommand \EOS [0]{\spacefactor3000\relax}%
\providecommand \BibitemShut  [1]{\csname bibitem#1\endcsname}%
\let\auto@bib@innerbib\@empty
%</preamble>
\bibitem [{\citenamefont {Lee}\ \emph {et~al.}(2006)\citenamefont {Lee},
  \citenamefont {Nagaosa},\ and\ \citenamefont {Wen}}]{lee}%
  \BibitemOpen
  \bibfield  {author} {\bibinfo {author} {\bibfnamefont {Patrick~A.}\
  \bibnamefont {Lee}}, \bibinfo {author} {\bibfnamefont {Naoto}\ \bibnamefont
  {Nagaosa}}, \ and\ \bibinfo {author} {\bibfnamefont {Xiao-Gang}\ \bibnamefont
  {Wen}},\ }\bibfield  {title} {\enquote {\bibinfo {title} {Doping a mott
  insulator: Physics of high-temperature superconductivity},}\ }\href {\doibase
  10.1103/RevModPhys.78.17} {\bibfield  {journal} {\bibinfo  {journal} {Rev.
  Mod. Phys.}\ }\textbf {\bibinfo {volume} {78}},\ \bibinfo {eid} {17}
  (\bibinfo {year} {2006})}\BibitemShut {NoStop}%
\bibitem [{\citenamefont {Norman}(2011)}]{Norman2011}%
  \BibitemOpen
  \bibfield  {author} {\bibinfo {author} {\bibfnamefont {Michael~R.}\
  \bibnamefont {Norman}},\ }\bibfield  {title} {\enquote {\bibinfo {title} {The
  challenge of unconventional superconductivity},}\ }\href {\doibase
  10.1126/science.1200181} {\bibfield  {journal} {\bibinfo  {journal}
  {Science}\ }\textbf {\bibinfo {volume} {332}},\ \bibinfo {pages} {196--200}
  (\bibinfo {year} {2011})}\BibitemShut {NoStop}%
\bibitem [{\citenamefont {Keimer}\ \emph {et~al.}(2015)\citenamefont {Keimer},
  \citenamefont {Kivelson}, \citenamefont {Norman}, \citenamefont {Uchida},\
  and\ \citenamefont {Zaanen}}]{keimerRev}%
  \BibitemOpen
  \bibfield  {author} {\bibinfo {author} {\bibfnamefont {B.}~\bibnamefont
  {Keimer}}, \bibinfo {author} {\bibfnamefont {S.~A.}\ \bibnamefont
  {Kivelson}}, \bibinfo {author} {\bibfnamefont {M.~R.}\ \bibnamefont
  {Norman}}, \bibinfo {author} {\bibfnamefont {S.}~\bibnamefont {Uchida}}, \
  and\ \bibinfo {author} {\bibfnamefont {J.}~\bibnamefont {Zaanen}},\
  }\bibfield  {title} {\enquote {\bibinfo {title} {From quantum matter to
  high-temperature superconductivity in copper oxides},}\ }\href {\doibase
  10.1038/nature14165} {\bibfield  {journal} {\bibinfo  {journal} {Nature}\
  }\textbf {\bibinfo {volume} {518}},\ \bibinfo {pages} {179--186} (\bibinfo
  {year} {2015})}\BibitemShut {NoStop}%
\bibitem [{\citenamefont {Anderson}(1987)}]{Anderson:1987}%
  \BibitemOpen
  \bibfield  {author} {\bibinfo {author} {\bibfnamefont {P.~W.}\ \bibnamefont
  {Anderson}},\ }\bibfield  {title} {\enquote {\bibinfo {title} {{The
  resonating valence bond state in La$_2$CuO$_4$ and superconductivity}},}\
  }\href {\doibase 10.1126/science.235.4793.1196} {\bibfield  {journal}
  {\bibinfo  {journal} {Science}\ }\textbf {\bibinfo {volume} {235}},\ \bibinfo
  {pages} {1196--1198} (\bibinfo {year} {1987})}\BibitemShut {NoStop}%
\bibitem [{\citenamefont {Hubbard}(1963)}]{Hubbard1963}%
  \BibitemOpen
  \bibfield  {author} {\bibinfo {author} {\bibfnamefont {J.}~\bibnamefont
  {Hubbard}},\ }\bibfield  {title} {\enquote {\bibinfo {title} {Electron
  correlations in narrow energy bands},}\ }\href {\doibase
  10.1098/rspa.1963.0204} {\bibfield  {journal} {\bibinfo  {journal}
  {Proceedings of the Royal Society of London Series A}\ }\textbf {\bibinfo
  {volume} {276}},\ \bibinfo {pages} {238--257} (\bibinfo {year}
  {1963})}\BibitemShut {NoStop}%
\bibitem [{\citenamefont {Arovas}\ \emph {et~al.}(2022)\citenamefont {Arovas},
  \citenamefont {Berg}, \citenamefont {Kivelson},\ and\ \citenamefont
  {Raghu}}]{ArovasAnnuRev2022}%
  \BibitemOpen
  \bibfield  {author} {\bibinfo {author} {\bibfnamefont {Daniel~P.}\
  \bibnamefont {Arovas}}, \bibinfo {author} {\bibfnamefont {Erez}\ \bibnamefont
  {Berg}}, \bibinfo {author} {\bibfnamefont {Steven~A.}\ \bibnamefont
  {Kivelson}}, \ and\ \bibinfo {author} {\bibfnamefont {Srinivas}\ \bibnamefont
  {Raghu}},\ }\bibfield  {title} {\enquote {\bibinfo {title} {The hubbard
  model},}\ }\href {\doibase 10.1146/annurev-conmatphys-031620-102024}
  {\bibfield  {journal} {\bibinfo  {journal} {Annual Review of Condensed Matter
  Physics}\ }\textbf {\bibinfo {volume} {13}},\ \bibinfo {pages} {239--274}
  (\bibinfo {year} {2022})}\BibitemShut {NoStop}%
\bibitem [{\citenamefont {Qin}\ \emph {et~al.}(2022)\citenamefont {Qin},
  \citenamefont {Sch\"{a}fer}, \citenamefont {Andergassen}, \citenamefont
  {Corboz},\ and\ \citenamefont {Gull}}]{QinAnnuRev2022}%
  \BibitemOpen
  \bibfield  {author} {\bibinfo {author} {\bibfnamefont {Mingpu}\ \bibnamefont
  {Qin}}, \bibinfo {author} {\bibfnamefont {Thomas}\ \bibnamefont
  {Sch\"{a}fer}}, \bibinfo {author} {\bibfnamefont {Sabine}\ \bibnamefont
  {Andergassen}}, \bibinfo {author} {\bibfnamefont {Philippe}\ \bibnamefont
  {Corboz}}, \ and\ \bibinfo {author} {\bibfnamefont {Emanuel}\ \bibnamefont
  {Gull}},\ }\bibfield  {title} {\enquote {\bibinfo {title} {The hubbard model:
  A computational perspective},}\ }\href {\doibase
  10.1146/annurev-conmatphys-090921-033948} {\bibfield  {journal} {\bibinfo
  {journal} {Annual Review of Condensed Matter Physics}\ }\textbf {\bibinfo
  {volume} {13}},\ \bibinfo {pages} {275--302} (\bibinfo {year}
  {2022})}\BibitemShut {NoStop}%
\bibitem [{\citenamefont {Tremblay}(2013)}]{AMJulich}%
  \BibitemOpen
  \bibfield  {author} {\bibinfo {author} {\bibfnamefont {A.-M.~S.}\
  \bibnamefont {Tremblay}},\ }\bibfield  {title} {\enquote {\bibinfo {title}
  {Strongly correlated superconductivity},}\ }in\ \href
  {http://juser.fz-juelich.de/record/137827/files/FZJ-2013-04137.pdf?version=1}
  {\emph {\bibinfo {booktitle} {Emergent Phenomena in Correlated Matter
  Modeling and Simulation}}},\ Vol.~\bibinfo {volume} {3},\ \bibinfo {editor}
  {edited by\ \bibinfo {editor} {\bibfnamefont {E.}~\bibnamefont {Pavarini}},
  \bibinfo {editor} {\bibfnamefont {E.}~\bibnamefont {Koch}}, \ and\ \bibinfo
  {editor} {\bibfnamefont {U.}~\bibnamefont {Schollw\"ock}}}\ (\bibinfo
  {publisher} {Verlag des Forschungszentrum},\ \bibinfo {address} {J\"ulich},\
  \bibinfo {year} {2013})\ Chap.~\bibinfo {chapter} {10}\BibitemShut {NoStop}%
\bibitem [{\citenamefont {Maier}\ \emph
  {et~al.}(2005{\natexlab{a}})\citenamefont {Maier}, \citenamefont {Jarrell},
  \citenamefont {Pruschke},\ and\ \citenamefont {Hettler}}]{maier}%
  \BibitemOpen
  \bibfield  {author} {\bibinfo {author} {\bibfnamefont {Thomas}\ \bibnamefont
  {Maier}}, \bibinfo {author} {\bibfnamefont {Mark}\ \bibnamefont {Jarrell}},
  \bibinfo {author} {\bibfnamefont {Thomas}\ \bibnamefont {Pruschke}}, \ and\
  \bibinfo {author} {\bibfnamefont {Matthias~H.}\ \bibnamefont {Hettler}},\
  }\bibfield  {title} {\enquote {\bibinfo {title} {Quantum cluster theories},}\
  }\href {\doibase 10.1103/RevModPhys.77.1027} {\bibfield  {journal} {\bibinfo
  {journal} {Rev. Mod. Phys.}\ }\textbf {\bibinfo {volume} {77}},\ \bibinfo
  {pages} {1027--1080} (\bibinfo {year} {2005}{\natexlab{a}})}\BibitemShut
  {NoStop}%
\bibitem [{\citenamefont {Kotliar}\ \emph {et~al.}(2006)\citenamefont
  {Kotliar}, \citenamefont {Savrasov}, \citenamefont {Haule}, \citenamefont
  {Oudovenko}, \citenamefont {Parcollet},\ and\ \citenamefont
  {Marianetti}}]{kotliarRMP}%
  \BibitemOpen
  \bibfield  {author} {\bibinfo {author} {\bibfnamefont {G.}~\bibnamefont
  {Kotliar}}, \bibinfo {author} {\bibfnamefont {S.~Y.}\ \bibnamefont
  {Savrasov}}, \bibinfo {author} {\bibfnamefont {K.}~\bibnamefont {Haule}},
  \bibinfo {author} {\bibfnamefont {V.~S.}\ \bibnamefont {Oudovenko}}, \bibinfo
  {author} {\bibfnamefont {O.}~\bibnamefont {Parcollet}}, \ and\ \bibinfo
  {author} {\bibfnamefont {C.~A.}\ \bibnamefont {Marianetti}},\ }\bibfield
  {title} {\enquote {\bibinfo {title} {{Electronic structure calculations with
  dynamical mean-field theory}},}\ }\href {\doibase 10.1103/RevModPhys.78.865}
  {\bibfield  {journal} {\bibinfo  {journal} {Rev. Mod. Phys.}\ }\textbf
  {\bibinfo {volume} {78}},\ \bibinfo {eid} {865} (\bibinfo {year}
  {2006})}\BibitemShut {NoStop}%
\bibitem [{\citenamefont {Tremblay}\ \emph {et~al.}(2006)\citenamefont
  {Tremblay}, \citenamefont {Kyung},\ and\ \citenamefont
  {S\'{e}n\'{e}chal}}]{tremblayR}%
  \BibitemOpen
  \bibfield  {author} {\bibinfo {author} {\bibfnamefont {A.-M.~S.}\
  \bibnamefont {Tremblay}}, \bibinfo {author} {\bibfnamefont {B.}~\bibnamefont
  {Kyung}}, \ and\ \bibinfo {author} {\bibfnamefont {D.}~\bibnamefont
  {S\'{e}n\'{e}chal}},\ }\bibfield  {title} {\enquote {\bibinfo {title}
  {{Pseudogap and high-temperature superconductivity from weak to strong
  coupling. Towards a quantitative theory}},}\ }\href {\doibase
  10.1063/1.2199446} {\bibfield  {journal} {\bibinfo  {journal} {Low Temp.
  Phys.}\ }\textbf {\bibinfo {volume} {32}},\ \bibinfo {pages} {424} (\bibinfo
  {year} {2006})}\BibitemShut {NoStop}%
\bibitem [{\citenamefont {Georges}\ \emph {et~al.}(1996)\citenamefont
  {Georges}, \citenamefont {Kotliar}, \citenamefont {Krauth},\ and\
  \citenamefont {Rozenberg}}]{rmp}%
  \BibitemOpen
  \bibfield  {author} {\bibinfo {author} {\bibfnamefont {Antoine}\ \bibnamefont
  {Georges}}, \bibinfo {author} {\bibfnamefont {Gabriel}\ \bibnamefont
  {Kotliar}}, \bibinfo {author} {\bibfnamefont {Werner}\ \bibnamefont
  {Krauth}}, \ and\ \bibinfo {author} {\bibfnamefont {Marcelo~J.}\ \bibnamefont
  {Rozenberg}},\ }\bibfield  {title} {\enquote {\bibinfo {title} {{Dynamical
  mean-field theory of strongly correlated fermion systems and the limit of
  infinite dimensions}},}\ }\href {\doibase 10.1103/RevModPhys.68.13}
  {\bibfield  {journal} {\bibinfo  {journal} {Rev. Mod. Phys.}\ }\textbf
  {\bibinfo {volume} {68}},\ \bibinfo {pages} {13} (\bibinfo {year}
  {1996})}\BibitemShut {NoStop}%
\bibitem [{\citenamefont {Maier}\ \emph {et~al.}(2000)\citenamefont {Maier},
  \citenamefont {Jarrell}, \citenamefont {Pruschke},\ and\ \citenamefont
  {Keller}}]{maierSC}%
  \BibitemOpen
  \bibfield  {author} {\bibinfo {author} {\bibfnamefont {Th.}\ \bibnamefont
  {Maier}}, \bibinfo {author} {\bibfnamefont {M.}~\bibnamefont {Jarrell}},
  \bibinfo {author} {\bibfnamefont {Th.}\ \bibnamefont {Pruschke}}, \ and\
  \bibinfo {author} {\bibfnamefont {J.}~\bibnamefont {Keller}},\ }\bibfield
  {title} {\enquote {\bibinfo {title} {$d$-wave superconductivity in the
  hubbard model},}\ }\href {\doibase 10.1103/PhysRevLett.85.1524} {\bibfield
  {journal} {\bibinfo  {journal} {Phys. Rev. Lett.}\ }\textbf {\bibinfo
  {volume} {85}},\ \bibinfo {pages} {1524--1527} (\bibinfo {year}
  {2000})}\BibitemShut {NoStop}%
\bibitem [{\citenamefont {Lichtenstein}\ and\ \citenamefont
  {Katsnelson}(2000)}]{lkAF}%
  \BibitemOpen
  \bibfield  {author} {\bibinfo {author} {\bibfnamefont {A.~I.}\ \bibnamefont
  {Lichtenstein}}\ and\ \bibinfo {author} {\bibfnamefont {M.~I.}\ \bibnamefont
  {Katsnelson}},\ }\bibfield  {title} {\enquote {\bibinfo {title}
  {Antiferromagnetism and d-wave superconductivity in cuprates: A cluster
  dynamical mean-field theory},}\ }\href {\doibase 10.1103/PhysRevB.62.R9283}
  {\bibfield  {journal} {\bibinfo  {journal} {Phys. Rev. B}\ }\textbf {\bibinfo
  {volume} {62}},\ \bibinfo {pages} {R9283--R9286} (\bibinfo {year}
  {2000})}\BibitemShut {NoStop}%
\bibitem [{\citenamefont {Maier}\ \emph
  {et~al.}(2005{\natexlab{b}})\citenamefont {Maier}, \citenamefont {Jarrell},
  \citenamefont {Schulthess}, \citenamefont {Kent},\ and\ \citenamefont
  {White}}]{maierSystem}%
  \BibitemOpen
  \bibfield  {author} {\bibinfo {author} {\bibfnamefont {T.~A.}\ \bibnamefont
  {Maier}}, \bibinfo {author} {\bibfnamefont {M.}~\bibnamefont {Jarrell}},
  \bibinfo {author} {\bibfnamefont {T.~C.}\ \bibnamefont {Schulthess}},
  \bibinfo {author} {\bibfnamefont {P.~R.~C.}\ \bibnamefont {Kent}}, \ and\
  \bibinfo {author} {\bibfnamefont {J.~B.}\ \bibnamefont {White}},\ }\bibfield
  {title} {\enquote {\bibinfo {title} {Systematic study of $d$-wave
  superconductivity in the 2d repulsive hubbard model},}\ }\href {\doibase
  10.1103/PhysRevLett.95.237001} {\bibfield  {journal} {\bibinfo  {journal}
  {Phys. Rev. Lett.}\ }\textbf {\bibinfo {volume} {95}},\ \bibinfo {pages}
  {237001} (\bibinfo {year} {2005}{\natexlab{b}})}\BibitemShut {NoStop}%
\bibitem [{\citenamefont {Haule}\ and\ \citenamefont
  {Kotliar}(2007{\natexlab{a}})}]{hauleDOPING}%
  \BibitemOpen
  \bibfield  {author} {\bibinfo {author} {\bibfnamefont {Kristjan}\
  \bibnamefont {Haule}}\ and\ \bibinfo {author} {\bibfnamefont {Gabriel}\
  \bibnamefont {Kotliar}},\ }\bibfield  {title} {\enquote {\bibinfo {title}
  {Strongly correlated superconductivity: A plaquette dynamical mean-field
  theory study},}\ }\href {\doibase 10.1103/PhysRevB.76.104509} {\bibfield
  {journal} {\bibinfo  {journal} {Phys. Rev. B}\ }\textbf {\bibinfo {volume}
  {76}},\ \bibinfo {eid} {104509} (\bibinfo {year}
  {2007}{\natexlab{a}})}\BibitemShut {NoStop}%
\bibitem [{\citenamefont {Kancharla}\ \emph {et~al.}(2008)\citenamefont
  {Kancharla}, \citenamefont {Kyung}, \citenamefont {S\'en\'echal},
  \citenamefont {Civelli}, \citenamefont {Capone}, \citenamefont {Kotliar},\
  and\ \citenamefont {Tremblay}}]{kancharla}%
  \BibitemOpen
  \bibfield  {author} {\bibinfo {author} {\bibfnamefont {S.~S.}\ \bibnamefont
  {Kancharla}}, \bibinfo {author} {\bibfnamefont {B.}~\bibnamefont {Kyung}},
  \bibinfo {author} {\bibfnamefont {D.}~\bibnamefont {S\'en\'echal}}, \bibinfo
  {author} {\bibfnamefont {M.}~\bibnamefont {Civelli}}, \bibinfo {author}
  {\bibfnamefont {M.}~\bibnamefont {Capone}}, \bibinfo {author} {\bibfnamefont
  {G.}~\bibnamefont {Kotliar}}, \ and\ \bibinfo {author} {\bibfnamefont
  {A.-M.~S.}\ \bibnamefont {Tremblay}},\ }\bibfield  {title} {\enquote
  {\bibinfo {title} {{Anomalous superconductivity and its competition with
  antiferromagnetism in doped Mott insulators}},}\ }\href {\doibase
  10.1103/PhysRevB.77.184516} {\bibfield  {journal} {\bibinfo  {journal} {Phys.
  Rev. B}\ }\textbf {\bibinfo {volume} {77}},\ \bibinfo {pages} {184516}
  (\bibinfo {year} {2008})}\BibitemShut {NoStop}%
\bibitem [{\citenamefont {Sordi}\ \emph
  {et~al.}(2012{\natexlab{a}})\citenamefont {Sordi}, \citenamefont {S\'emon},
  \citenamefont {Haule},\ and\ \citenamefont {Tremblay}}]{sshtSC}%
  \BibitemOpen
  \bibfield  {author} {\bibinfo {author} {\bibfnamefont {G.}~\bibnamefont
  {Sordi}}, \bibinfo {author} {\bibfnamefont {P.}~\bibnamefont {S\'emon}},
  \bibinfo {author} {\bibfnamefont {K.}~\bibnamefont {Haule}}, \ and\ \bibinfo
  {author} {\bibfnamefont {A.-M.~S.}\ \bibnamefont {Tremblay}},\ }\bibfield
  {title} {\enquote {\bibinfo {title} {{Strong Coupling Superconductivity,
  Pseudogap, and Mott Transition}},}\ }\href {\doibase
  10.1103/PhysRevLett.108.216401} {\bibfield  {journal} {\bibinfo  {journal}
  {Phys. Rev. Lett.}\ }\textbf {\bibinfo {volume} {108}},\ \bibinfo {pages}
  {216401} (\bibinfo {year} {2012}{\natexlab{a}})}\BibitemShut {NoStop}%
\bibitem [{\citenamefont {Gull}\ \emph {et~al.}(2013)\citenamefont {Gull},
  \citenamefont {Parcollet},\ and\ \citenamefont {Millis}}]{Gull:2013}%
  \BibitemOpen
  \bibfield  {author} {\bibinfo {author} {\bibfnamefont {Emanuel}\ \bibnamefont
  {Gull}}, \bibinfo {author} {\bibfnamefont {Olivier}\ \bibnamefont
  {Parcollet}}, \ and\ \bibinfo {author} {\bibfnamefont {Andrew~J.}\
  \bibnamefont {Millis}},\ }\bibfield  {title} {\enquote {\bibinfo {title}
  {Superconductivity and the pseudogap in the two-dimensional hubbard model},}\
  }\href {\doibase 10.1103/PhysRevLett.110.216405} {\bibfield  {journal}
  {\bibinfo  {journal} {Phys. Rev. Lett.}\ }\textbf {\bibinfo {volume} {110}},\
  \bibinfo {pages} {216405} (\bibinfo {year} {2013})}\BibitemShut {NoStop}%
\bibitem [{\citenamefont {Chen}\ \emph {et~al.}(2015)\citenamefont {Chen},
  \citenamefont {LeBlanc},\ and\ \citenamefont {Gull}}]{Chen:2015}%
  \BibitemOpen
  \bibfield  {author} {\bibinfo {author} {\bibfnamefont {Xi}~\bibnamefont
  {Chen}}, \bibinfo {author} {\bibfnamefont {J.~P.~F.}\ \bibnamefont
  {LeBlanc}}, \ and\ \bibinfo {author} {\bibfnamefont {Emanuel}\ \bibnamefont
  {Gull}},\ }\bibfield  {title} {\enquote {\bibinfo {title} {Superconducting
  fluctuations in the normal state of the two-dimensional hubbard model},}\
  }\href {\doibase 10.1103/PhysRevLett.115.116402} {\bibfield  {journal}
  {\bibinfo  {journal} {Phys. Rev. Lett.}\ }\textbf {\bibinfo {volume} {115}},\
  \bibinfo {pages} {116402} (\bibinfo {year} {2015})}\BibitemShut {NoStop}%
\bibitem [{\citenamefont {Capone}\ and\ \citenamefont
  {Kotliar}(2006)}]{massimoAF}%
  \BibitemOpen
  \bibfield  {author} {\bibinfo {author} {\bibfnamefont {M.}~\bibnamefont
  {Capone}}\ and\ \bibinfo {author} {\bibfnamefont {G.}~\bibnamefont
  {Kotliar}},\ }\bibfield  {title} {\enquote {\bibinfo {title} {Competition
  between $d$ -wave superconductivity and antiferromagnetism in the
  two-dimensional hubbard model},}\ }\href {\doibase
  10.1103/PhysRevB.74.054513} {\bibfield  {journal} {\bibinfo  {journal} {Phys.
  Rev. B}\ }\textbf {\bibinfo {volume} {74}},\ \bibinfo {pages} {054513}
  (\bibinfo {year} {2006})}\BibitemShut {NoStop}%
\bibitem [{\citenamefont {Sakai}(2023)}]{sakai2023}%
  \BibitemOpen
  \bibfield  {author} {\bibinfo {author} {\bibfnamefont {Shiro}\ \bibnamefont
  {Sakai}},\ }\bibfield  {title} {\enquote {\bibinfo {title} {Nonperturbative
  calculations for spectroscopic properties of cuprate high-temperature
  superconductors},}\ }\href {\doibase 10.7566/JPSJ.92.092001} {\bibfield
  {journal} {\bibinfo  {journal} {Journal of the Physical Society of Japan}\
  }\textbf {\bibinfo {volume} {92}},\ \bibinfo {pages} {092001} (\bibinfo
  {year} {2023})}\BibitemShut {NoStop}%
\bibitem [{\citenamefont {Dagotto}(2005)}]{Dagotto:Science2005}%
  \BibitemOpen
  \bibfield  {author} {\bibinfo {author} {\bibfnamefont {Elbio}\ \bibnamefont
  {Dagotto}},\ }\bibfield  {title} {\enquote {\bibinfo {title} {Complexity in
  strongly correlated electronic systems},}\ }\href {\doibase
  10.1126/science.1107559} {\bibfield  {journal} {\bibinfo  {journal}
  {Science}\ }\textbf {\bibinfo {volume} {309}},\ \bibinfo {pages} {257--262}
  (\bibinfo {year} {2005})}\BibitemShut {NoStop}%
\bibitem [{\citenamefont {Zheng}\ \emph {et~al.}(2017)\citenamefont {Zheng},
  \citenamefont {Chung}, \citenamefont {Corboz}, \citenamefont {Ehlers},
  \citenamefont {Qin}, \citenamefont {Noack}, \citenamefont {Shi},
  \citenamefont {White}, \citenamefont {Zhang},\ and\ \citenamefont
  {Chan}}]{Zheng:Science2017}%
  \BibitemOpen
  \bibfield  {author} {\bibinfo {author} {\bibfnamefont {Bo-Xiao}\ \bibnamefont
  {Zheng}}, \bibinfo {author} {\bibfnamefont {Chia-Min}\ \bibnamefont {Chung}},
  \bibinfo {author} {\bibfnamefont {Philippe}\ \bibnamefont {Corboz}}, \bibinfo
  {author} {\bibfnamefont {Georg}\ \bibnamefont {Ehlers}}, \bibinfo {author}
  {\bibfnamefont {Ming-Pu}\ \bibnamefont {Qin}}, \bibinfo {author}
  {\bibfnamefont {Reinhard~M.}\ \bibnamefont {Noack}}, \bibinfo {author}
  {\bibfnamefont {Hao}\ \bibnamefont {Shi}}, \bibinfo {author} {\bibfnamefont
  {Steven~R.}\ \bibnamefont {White}}, \bibinfo {author} {\bibfnamefont
  {Shiwei}\ \bibnamefont {Zhang}}, \ and\ \bibinfo {author} {\bibfnamefont
  {Garnet Kin-Lic}\ \bibnamefont {Chan}},\ }\bibfield  {title} {\enquote
  {\bibinfo {title} {Stripe order in the underdoped region of the
  two-dimensional hubbard model},}\ }\href {\doibase 10.1126/science.aam7127}
  {\bibfield  {journal} {\bibinfo  {journal} {Science}\ }\textbf {\bibinfo
  {volume} {358}},\ \bibinfo {pages} {1155--1160} (\bibinfo {year}
  {2017})}\BibitemShut {NoStop}%
\bibitem [{\citenamefont {Qin}\ \emph {et~al.}(2020)\citenamefont {Qin},
  \citenamefont {Chung}, \citenamefont {Shi}, \citenamefont {Vitali},
  \citenamefont {Hubig}, \citenamefont {Schollw\"ock}, \citenamefont {White},\
  and\ \citenamefont {Zhang}}]{Qin:PRX2020}%
  \BibitemOpen
  \bibfield  {author} {\bibinfo {author} {\bibfnamefont {Mingpu}\ \bibnamefont
  {Qin}}, \bibinfo {author} {\bibfnamefont {Chia-Min}\ \bibnamefont {Chung}},
  \bibinfo {author} {\bibfnamefont {Hao}\ \bibnamefont {Shi}}, \bibinfo
  {author} {\bibfnamefont {Ettore}\ \bibnamefont {Vitali}}, \bibinfo {author}
  {\bibfnamefont {Claudius}\ \bibnamefont {Hubig}}, \bibinfo {author}
  {\bibfnamefont {Ulrich}\ \bibnamefont {Schollw\"ock}}, \bibinfo {author}
  {\bibfnamefont {Steven~R.}\ \bibnamefont {White}}, \ and\ \bibinfo {author}
  {\bibfnamefont {Shiwei}\ \bibnamefont {Zhang}} (\bibinfo {collaboration}
  {Simons Collaboration on the Many-Electron Problem}),\ }\bibfield  {title}
  {\enquote {\bibinfo {title} {Absence of superconductivity in the pure
  two-dimensional hubbard model},}\ }\href {\doibase
  10.1103/PhysRevX.10.031016} {\bibfield  {journal} {\bibinfo  {journal} {Phys.
  Rev. X}\ }\textbf {\bibinfo {volume} {10}},\ \bibinfo {pages} {031016}
  (\bibinfo {year} {2020})}\BibitemShut {NoStop}%
\bibitem [{\citenamefont {Chung}\ \emph {et~al.}(2020)\citenamefont {Chung},
  \citenamefont {Qin}, \citenamefont {Zhang}, \citenamefont {Schollw\"ock},\
  and\ \citenamefont {White}}]{Chung:PRB2020}%
  \BibitemOpen
  \bibfield  {author} {\bibinfo {author} {\bibfnamefont {Chia-Min}\
  \bibnamefont {Chung}}, \bibinfo {author} {\bibfnamefont {Mingpu}\
  \bibnamefont {Qin}}, \bibinfo {author} {\bibfnamefont {Shiwei}\ \bibnamefont
  {Zhang}}, \bibinfo {author} {\bibfnamefont {Ulrich}\ \bibnamefont
  {Schollw\"ock}}, \ and\ \bibinfo {author} {\bibfnamefont {Steven~R.}\
  \bibnamefont {White}} (\bibinfo {collaboration} {The Simons Collaboration on
  the Many-Electron Problem}),\ }\bibfield  {title} {\enquote {\bibinfo {title}
  {Plaquette versus ordinary $d$-wave pairing in the
  ${t}^{\ensuremath{'}}$-hubbard model on a width-4 cylinder},}\ }\href
  {\doibase 10.1103/PhysRevB.102.041106} {\bibfield  {journal} {\bibinfo
  {journal} {Phys. Rev. B}\ }\textbf {\bibinfo {volume} {102}},\ \bibinfo
  {pages} {041106} (\bibinfo {year} {2020})}\BibitemShut {NoStop}%
\bibitem [{\citenamefont {Mermin}\ and\ \citenamefont
  {Wagner}(1966)}]{MWtheorem}%
  \BibitemOpen
  \bibfield  {author} {\bibinfo {author} {\bibfnamefont {N.~D.}\ \bibnamefont
  {Mermin}}\ and\ \bibinfo {author} {\bibfnamefont {H.}~\bibnamefont
  {Wagner}},\ }\bibfield  {title} {\enquote {\bibinfo {title} {{Absence of
  Ferromagnetism or Antiferromagnetism in One- or Two-Dimensional Isotropic
  Heisenberg Models}},}\ }\href {\doibase 10.1103/PhysRevLett.17.1133}
  {\bibfield  {journal} {\bibinfo  {journal} {Phys. Rev. Lett.}\ }\textbf
  {\bibinfo {volume} {17}},\ \bibinfo {pages} {1133--1136} (\bibinfo {year}
  {1966})}\BibitemShut {NoStop}%
\bibitem [{\citenamefont {Maier}\ \emph {et~al.}(2004)\citenamefont {Maier},
  \citenamefont {Jarrell}, \citenamefont {Macridin},\ and\ \citenamefont
  {Slezak}}]{maierENERGY}%
  \BibitemOpen
  \bibfield  {author} {\bibinfo {author} {\bibfnamefont {Th.~A.}\ \bibnamefont
  {Maier}}, \bibinfo {author} {\bibfnamefont {M.}~\bibnamefont {Jarrell}},
  \bibinfo {author} {\bibfnamefont {A.}~\bibnamefont {Macridin}}, \ and\
  \bibinfo {author} {\bibfnamefont {C.}~\bibnamefont {Slezak}},\ }\bibfield
  {title} {\enquote {\bibinfo {title} {Kinetic energy driven pairing in cuprate
  superconductors},}\ }\href {\doibase 10.1103/PhysRevLett.92.027005}
  {\bibfield  {journal} {\bibinfo  {journal} {Phys. Rev. Lett.}\ }\textbf
  {\bibinfo {volume} {92}},\ \bibinfo {pages} {027005} (\bibinfo {year}
  {2004})}\BibitemShut {NoStop}%
\bibitem [{\citenamefont {Carbone}\ \emph {et~al.}(2006)\citenamefont
  {Carbone}, \citenamefont {Kuzmenko}, \citenamefont {Molegraaf}, \citenamefont
  {van Heumen}, \citenamefont {Lukovac}, \citenamefont {Marsiglio},
  \citenamefont {van~der Marel}, \citenamefont {Haule}, \citenamefont
  {Kotliar}, \citenamefont {Berger}, \citenamefont {Courjault}, \citenamefont
  {Kes},\ and\ \citenamefont {Li}}]{carbone2006}%
  \BibitemOpen
  \bibfield  {author} {\bibinfo {author} {\bibfnamefont {F.}~\bibnamefont
  {Carbone}}, \bibinfo {author} {\bibfnamefont {A.~B.}\ \bibnamefont
  {Kuzmenko}}, \bibinfo {author} {\bibfnamefont {H.~J.~A.}\ \bibnamefont
  {Molegraaf}}, \bibinfo {author} {\bibfnamefont {E.}~\bibnamefont {van
  Heumen}}, \bibinfo {author} {\bibfnamefont {V.}~\bibnamefont {Lukovac}},
  \bibinfo {author} {\bibfnamefont {F.}~\bibnamefont {Marsiglio}}, \bibinfo
  {author} {\bibfnamefont {D.}~\bibnamefont {van~der Marel}}, \bibinfo {author}
  {\bibfnamefont {K.}~\bibnamefont {Haule}}, \bibinfo {author} {\bibfnamefont
  {G.}~\bibnamefont {Kotliar}}, \bibinfo {author} {\bibfnamefont
  {H.}~\bibnamefont {Berger}}, \bibinfo {author} {\bibfnamefont
  {S.}~\bibnamefont {Courjault}}, \bibinfo {author} {\bibfnamefont {P.~H.}\
  \bibnamefont {Kes}}, \ and\ \bibinfo {author} {\bibfnamefont
  {M.}~\bibnamefont {Li}},\ }\bibfield  {title} {\enquote {\bibinfo {title}
  {Doping dependence of the redistribution of optical spectral weight in
  ${\mathrm{bi}}_{2}{\mathrm{sr}}_{2}{\mathrm{cacu}}_{2}{\mathrm{o}}_{8+\ensuremath{\delta}}$},}\
  }\href {\doibase 10.1103/PhysRevB.74.064510} {\bibfield  {journal} {\bibinfo
  {journal} {Phys. Rev. B}\ }\textbf {\bibinfo {volume} {74}},\ \bibinfo
  {pages} {064510} (\bibinfo {year} {2006})}\BibitemShut {NoStop}%
\bibitem [{\citenamefont {Gull}\ and\ \citenamefont
  {Millis}(2012)}]{millisENERGY}%
  \BibitemOpen
  \bibfield  {author} {\bibinfo {author} {\bibfnamefont {E.}~\bibnamefont
  {Gull}}\ and\ \bibinfo {author} {\bibfnamefont {A.~J.}\ \bibnamefont
  {Millis}},\ }\bibfield  {title} {\enquote {\bibinfo {title} {Energetics of
  superconductivity in the two-dimensional hubbard model},}\ }\href {\doibase
  10.1103/PhysRevB.86.241106} {\bibfield  {journal} {\bibinfo  {journal} {Phys.
  Rev. B}\ }\textbf {\bibinfo {volume} {86}},\ \bibinfo {pages} {241106}
  (\bibinfo {year} {2012})}\BibitemShut {NoStop}%
\bibitem [{\citenamefont {Fratino}\ \emph
  {et~al.}(2016{\natexlab{a}})\citenamefont {Fratino}, \citenamefont {S\'emon},
  \citenamefont {Sordi},\ and\ \citenamefont {Tremblay}}]{LorenzoSC}%
  \BibitemOpen
  \bibfield  {author} {\bibinfo {author} {\bibfnamefont {L.}~\bibnamefont
  {Fratino}}, \bibinfo {author} {\bibfnamefont {P.}~\bibnamefont {S\'emon}},
  \bibinfo {author} {\bibfnamefont {G.}~\bibnamefont {Sordi}}, \ and\ \bibinfo
  {author} {\bibfnamefont {A.-M.~S.}\ \bibnamefont {Tremblay}},\ }\bibfield
  {title} {\enquote {\bibinfo {title} {{An organizing principle for
  two-dimensional strongly correlated superconductivity}},}\ }\href {\doibase
  10.1038/srep22715} {\bibfield  {journal} {\bibinfo  {journal} {Sci. Rep.}\
  }\textbf {\bibinfo {volume} {6}},\ \bibinfo {pages} {22715} (\bibinfo {year}
  {2016}{\natexlab{a}})}\BibitemShut {NoStop}%
\bibitem [{\citenamefont {Maier}\ \emph {et~al.}(2008)\citenamefont {Maier},
  \citenamefont {Poilblanc},\ and\ \citenamefont {Scalapino}}]{maierPRL2008}%
  \BibitemOpen
  \bibfield  {author} {\bibinfo {author} {\bibfnamefont {T.~A.}\ \bibnamefont
  {Maier}}, \bibinfo {author} {\bibfnamefont {D.}~\bibnamefont {Poilblanc}}, \
  and\ \bibinfo {author} {\bibfnamefont {D.~J.}\ \bibnamefont {Scalapino}},\
  }\bibfield  {title} {\enquote {\bibinfo {title} {Dynamics of the pairing
  interaction in the hubbard and $t\mathrm{\text{\ensuremath{-}}}j$ models of
  high-temperature superconductors},}\ }\href {\doibase
  10.1103/PhysRevLett.100.237001} {\bibfield  {journal} {\bibinfo  {journal}
  {Phys. Rev. Lett.}\ }\textbf {\bibinfo {volume} {100}},\ \bibinfo {pages}
  {237001} (\bibinfo {year} {2008})}\BibitemShut {NoStop}%
\bibitem [{\citenamefont {Kyung}\ \emph {et~al.}(2009)\citenamefont {Kyung},
  \citenamefont {S\'{e}n\'{e}chal},\ and\ \citenamefont
  {Tremblay}}]{Kyung:2009}%
  \BibitemOpen
  \bibfield  {author} {\bibinfo {author} {\bibfnamefont {B.}~\bibnamefont
  {Kyung}}, \bibinfo {author} {\bibfnamefont {D.}~\bibnamefont
  {S\'{e}n\'{e}chal}}, \ and\ \bibinfo {author} {\bibfnamefont {A.-M.~S.}\
  \bibnamefont {Tremblay}},\ }\bibfield  {title} {\enquote {\bibinfo {title}
  {Pairing dynamics in strongly correlated superconductivity},}\ }\href
  {\doibase 10.1103/PhysRevB.80.205109} {\bibfield  {journal} {\bibinfo
  {journal} {Physical Review B (Condensed Matter and Materials Physics)}\
  }\textbf {\bibinfo {volume} {80}},\ \bibinfo {eid} {205109} (\bibinfo {year}
  {2009})}\BibitemShut {NoStop}%
\bibitem [{\citenamefont {Civelli}(2009{\natexlab{a}})}]{civelli1}%
  \BibitemOpen
  \bibfield  {author} {\bibinfo {author} {\bibfnamefont {M.}~\bibnamefont
  {Civelli}},\ }\bibfield  {title} {\enquote {\bibinfo {title} {Evolution of
  the dynamical pairing across the phase diagram of a strongly correlated
  high-temperature superconductor},}\ }\href {\doibase
  10.1103/PhysRevLett.103.136402} {\bibfield  {journal} {\bibinfo  {journal}
  {Phys. Rev. Lett.}\ }\textbf {\bibinfo {volume} {103}},\ \bibinfo {pages}
  {136402} (\bibinfo {year} {2009}{\natexlab{a}})}\BibitemShut {NoStop}%
\bibitem [{\citenamefont {Civelli}(2009{\natexlab{b}})}]{civelli2}%
  \BibitemOpen
  \bibfield  {author} {\bibinfo {author} {\bibfnamefont {M.}~\bibnamefont
  {Civelli}},\ }\bibfield  {title} {\enquote {\bibinfo {title} {Doping-driven
  evolution of the superconducting state from a doped mott insulator: Cluster
  dynamical mean-field theory},}\ }\href {\doibase 10.1103/PhysRevB.79.195113}
  {\bibfield  {journal} {\bibinfo  {journal} {Phys. Rev. B}\ }\textbf {\bibinfo
  {volume} {79}},\ \bibinfo {pages} {195113} (\bibinfo {year}
  {2009}{\natexlab{b}})}\BibitemShut {NoStop}%
\bibitem [{\citenamefont {S\'en\'echal}\ \emph {et~al.}(2013)\citenamefont
  {S\'en\'echal}, \citenamefont {Day}, \citenamefont {Bouliane},\ and\
  \citenamefont {Tremblay}}]{senechalPRB2013}%
  \BibitemOpen
  \bibfield  {author} {\bibinfo {author} {\bibfnamefont {D.}~\bibnamefont
  {S\'en\'echal}}, \bibinfo {author} {\bibfnamefont {A.~G.~R.}\ \bibnamefont
  {Day}}, \bibinfo {author} {\bibfnamefont {V.}~\bibnamefont {Bouliane}}, \
  and\ \bibinfo {author} {\bibfnamefont {A.-M.~S.}\ \bibnamefont {Tremblay}},\
  }\bibfield  {title} {\enquote {\bibinfo {title} {Resilience of $d$-wave
  superconductivity to nearest-neighbor repulsion},}\ }\href {\doibase
  10.1103/PhysRevB.87.075123} {\bibfield  {journal} {\bibinfo  {journal} {Phys.
  Rev. B}\ }\textbf {\bibinfo {volume} {87}},\ \bibinfo {pages} {075123}
  (\bibinfo {year} {2013})}\BibitemShut {NoStop}%
\bibitem [{\citenamefont {Reymbaut}\ \emph {et~al.}(2016)\citenamefont
  {Reymbaut}, \citenamefont {Charlebois}, \citenamefont {Asiani}, \citenamefont
  {Fratino}, \citenamefont {S\'emon}, \citenamefont {Sordi},\ and\
  \citenamefont {Tremblay}}]{reymbautPRB2016}%
  \BibitemOpen
  \bibfield  {author} {\bibinfo {author} {\bibfnamefont {A.}~\bibnamefont
  {Reymbaut}}, \bibinfo {author} {\bibfnamefont {M.}~\bibnamefont
  {Charlebois}}, \bibinfo {author} {\bibfnamefont {M.~Fellous}\ \bibnamefont
  {Asiani}}, \bibinfo {author} {\bibfnamefont {L.}~\bibnamefont {Fratino}},
  \bibinfo {author} {\bibfnamefont {P.}~\bibnamefont {S\'emon}}, \bibinfo
  {author} {\bibfnamefont {G.}~\bibnamefont {Sordi}}, \ and\ \bibinfo {author}
  {\bibfnamefont {A.-M.~S.}\ \bibnamefont {Tremblay}},\ }\bibfield  {title}
  {\enquote {\bibinfo {title} {Antagonistic effects of nearest-neighbor
  repulsion on the superconducting pairing dynamics in the doped mott insulator
  regime},}\ }\href {\doibase 10.1103/PhysRevB.94.155146} {\bibfield  {journal}
  {\bibinfo  {journal} {Phys. Rev. B}\ }\textbf {\bibinfo {volume} {94}},\
  \bibinfo {pages} {155146} (\bibinfo {year} {2016})}\BibitemShut {NoStop}%
\bibitem [{\citenamefont {{Dong}}\ \emph
  {et~al.}(2022{\natexlab{a}})\citenamefont {{Dong}}, \citenamefont {{Del Re}},
  \citenamefont {{Toschi}},\ and\ \citenamefont {{Gull}}}]{DongPNAS2022}%
  \BibitemOpen
  \bibfield  {author} {\bibinfo {author} {\bibfnamefont {Xinyang}\ \bibnamefont
  {{Dong}}}, \bibinfo {author} {\bibfnamefont {Lorenzo}\ \bibnamefont {{Del
  Re}}}, \bibinfo {author} {\bibfnamefont {Alessandro}\ \bibnamefont
  {{Toschi}}}, \ and\ \bibinfo {author} {\bibfnamefont {Emanuel}\ \bibnamefont
  {{Gull}}},\ }\bibfield  {title} {\enquote {\bibinfo {title} {Mechanism of
  superconductivity in the hubbard model at intermediate interaction
  strength},}\ }\href {\doibase 10.1073/pnas.2205048119} {\bibfield  {journal}
  {\bibinfo  {journal} {Proceedings of the National Academy of Science}\
  }\textbf {\bibinfo {volume} {119}},\ \bibinfo {eid} {e2205048119} (\bibinfo
  {year} {2022}{\natexlab{a}})}\BibitemShut {NoStop}%
\bibitem [{\citenamefont {{Dong}}\ \emph
  {et~al.}(2022{\natexlab{b}})\citenamefont {{Dong}}, \citenamefont {{Gull}},\
  and\ \citenamefont {{Millis}}}]{DongNatPhys2022}%
  \BibitemOpen
  \bibfield  {author} {\bibinfo {author} {\bibfnamefont {Xinyang}\ \bibnamefont
  {{Dong}}}, \bibinfo {author} {\bibfnamefont {Emanuel}\ \bibnamefont
  {{Gull}}}, \ and\ \bibinfo {author} {\bibfnamefont {Andrew~J.}\ \bibnamefont
  {{Millis}}},\ }\bibfield  {title} {\enquote {\bibinfo {title} {{Quantifying
  the role of antiferromagnetic fluctuations in the superconductivity of the
  doped Hubbard model}},}\ }\href {\doibase 10.1038/s41567-022-01710-z}
  {\bibfield  {journal} {\bibinfo  {journal} {Nature Physics}\ }\textbf
  {\bibinfo {volume} {18}},\ \bibinfo {pages} {1293--1296} (\bibinfo {year}
  {2022}{\natexlab{b}})}\BibitemShut {NoStop}%
\bibitem [{\citenamefont {Amico}\ \emph {et~al.}(2008)\citenamefont {Amico},
  \citenamefont {Fazio}, \citenamefont {Osterloh},\ and\ \citenamefont
  {Vedral}}]{amicoRMP2008}%
  \BibitemOpen
  \bibfield  {author} {\bibinfo {author} {\bibfnamefont {Luigi}\ \bibnamefont
  {Amico}}, \bibinfo {author} {\bibfnamefont {Rosario}\ \bibnamefont {Fazio}},
  \bibinfo {author} {\bibfnamefont {Andreas}\ \bibnamefont {Osterloh}}, \ and\
  \bibinfo {author} {\bibfnamefont {Vlatko}\ \bibnamefont {Vedral}},\
  }\bibfield  {title} {\enquote {\bibinfo {title} {Entanglement in many-body
  systems},}\ }\href {\doibase 10.1103/RevModPhys.80.517} {\bibfield  {journal}
  {\bibinfo  {journal} {Rev. Mod. Phys.}\ }\textbf {\bibinfo {volume} {80}},\
  \bibinfo {pages} {517--576} (\bibinfo {year} {2008})}\BibitemShut {NoStop}%
\bibitem [{\citenamefont {Zeng}\ \emph {et~al.}(2019)\citenamefont {Zeng},
  \citenamefont {Chen}, \citenamefont {Zhou},\ and\ \citenamefont
  {Wen}}]{zhengBOOK}%
  \BibitemOpen
  \bibfield  {author} {\bibinfo {author} {\bibfnamefont {B.}~\bibnamefont
  {Zeng}}, \bibinfo {author} {\bibfnamefont {X.}~\bibnamefont {Chen}}, \bibinfo
  {author} {\bibfnamefont {D.-L.}\ \bibnamefont {Zhou}}, \ and\ \bibinfo
  {author} {\bibfnamefont {X.-G.}\ \bibnamefont {Wen}},\ }\href {\doibase
  10.1007/978-1-4939-9084-9} {\emph {\bibinfo {title} {{Quantum Information
  Meets Quantum Matter}}}}\ (\bibinfo  {publisher} {Springer-Verlag},\ \bibinfo
  {address} {New York},\ \bibinfo {year} {2019})\BibitemShut {NoStop}%
\bibitem [{\citenamefont {Walsh}\ \emph {et~al.}(2021)\citenamefont {Walsh},
  \citenamefont {Charlebois}, \citenamefont {Sémon}, \citenamefont {Sordi},\
  and\ \citenamefont {Tremblay}}]{CaitlinPNAS2021}%
  \BibitemOpen
  \bibfield  {author} {\bibinfo {author} {\bibfnamefont {Caitlin}\ \bibnamefont
  {Walsh}}, \bibinfo {author} {\bibfnamefont {Maxime}\ \bibnamefont
  {Charlebois}}, \bibinfo {author} {\bibfnamefont {Patrick}\ \bibnamefont
  {Sémon}}, \bibinfo {author} {\bibfnamefont {Giovanni}\ \bibnamefont
  {Sordi}}, \ and\ \bibinfo {author} {\bibfnamefont {André-Marie~S.}\
  \bibnamefont {Tremblay}},\ }\bibfield  {title} {\enquote {\bibinfo {title}
  {Information-theoretic measures of superconductivity in a two-dimensional
  doped mott insulator},}\ }\href {\doibase 10.1073/pnas.2104114118} {\bibfield
   {journal} {\bibinfo  {journal} {Proceedings of the National Academy of
  Sciences}\ }\textbf {\bibinfo {volume} {118}},\ \bibinfo {pages}
  {e2104114118} (\bibinfo {year} {2021})}\BibitemShut {NoStop}%
\bibitem [{\citenamefont {Haule}(2007)}]{hauleCTQMC}%
  \BibitemOpen
  \bibfield  {author} {\bibinfo {author} {\bibfnamefont {Kristjan}\
  \bibnamefont {Haule}},\ }\bibfield  {title} {\enquote {\bibinfo {title}
  {{Quantum Monte Carlo impurity solver for cluster dynamical mean-field theory
  and electronic structure calculations with adjustable cluster base}},}\
  }\href {\doibase 10.1103/PhysRevB.75.155113} {\bibfield  {journal} {\bibinfo
  {journal} {Phys. Rev. B}\ }\textbf {\bibinfo {volume} {75}},\ \bibinfo {eid}
  {155113} (\bibinfo {year} {2007})}\BibitemShut {NoStop}%
\bibitem [{\citenamefont {Shim}\ \emph {et~al.}(2007)\citenamefont {Shim},
  \citenamefont {Haule},\ and\ \citenamefont {Kotliar}}]{shim:nature}%
  \BibitemOpen
  \bibfield  {author} {\bibinfo {author} {\bibfnamefont {J.~H.}\ \bibnamefont
  {Shim}}, \bibinfo {author} {\bibfnamefont {K.}~\bibnamefont {Haule}}, \ and\
  \bibinfo {author} {\bibfnamefont {G.}~\bibnamefont {Kotliar}},\ }\bibfield
  {title} {\enquote {\bibinfo {title} {Fluctuating valence in a correlated
  solid and the anomalous properties of d-plutonium},}\ }\href {\doibase
  http://dx.doi.org/10.1038/nature05647} {\bibfield  {journal} {\bibinfo
  {journal} {Nature}\ }\textbf {\bibinfo {volume} {446}},\ \bibinfo {pages}
  {513--516} (\bibinfo {year} {2007})}\BibitemShut {NoStop}%
\bibitem [{\citenamefont {Kent}\ and\ \citenamefont
  {Kotliar}(2018)}]{Gabi:Science2018}%
  \BibitemOpen
  \bibfield  {author} {\bibinfo {author} {\bibfnamefont {Paul R.~C.}\
  \bibnamefont {Kent}}\ and\ \bibinfo {author} {\bibfnamefont {Gabriel}\
  \bibnamefont {Kotliar}},\ }\bibfield  {title} {\enquote {\bibinfo {title}
  {Toward a predictive theory of correlated materials},}\ }\href {\doibase
  10.1126/science.aat5975} {\bibfield  {journal} {\bibinfo  {journal}
  {Science}\ }\textbf {\bibinfo {volume} {361}},\ \bibinfo {pages} {348--354}
  (\bibinfo {year} {2018})}\BibitemShut {NoStop}%
\bibitem [{\citenamefont {Gull}\ \emph {et~al.}(2008)\citenamefont {Gull},
  \citenamefont {Werner}, \citenamefont {Wang}, \citenamefont {Troyer},\ and\
  \citenamefont {Millis}}]{gullEPL}%
  \BibitemOpen
  \bibfield  {author} {\bibinfo {author} {\bibfnamefont {E.}~\bibnamefont
  {Gull}}, \bibinfo {author} {\bibfnamefont {P.}~\bibnamefont {Werner}},
  \bibinfo {author} {\bibfnamefont {X.}~\bibnamefont {Wang}}, \bibinfo {author}
  {\bibfnamefont {M.}~\bibnamefont {Troyer}}, \ and\ \bibinfo {author}
  {\bibfnamefont {A.~J.}\ \bibnamefont {Millis}},\ }\bibfield  {title}
  {\enquote {\bibinfo {title} {{Local order and the gapped phase of the Hubbard
  model: A plaquette dynamical mean-field investigation}},}\ }\href {\doibase
  10.1209/0295-5075/84/37009} {\bibfield  {journal} {\bibinfo  {journal}
  {Europhys. Lett.}\ }\textbf {\bibinfo {volume} {84}},\ \bibinfo {pages}
  {37009} (\bibinfo {year} {2008})}\BibitemShut {NoStop}%
\bibitem [{\citenamefont {Ferrero}\ \emph
  {et~al.}(2009{\natexlab{a}})\citenamefont {Ferrero}, \citenamefont
  {Cornaglia}, \citenamefont {Leo}, \citenamefont {Parcollet}, \citenamefont
  {Kotliar},\ and\ \citenamefont {Georges}}]{michelEPL}%
  \BibitemOpen
  \bibfield  {author} {\bibinfo {author} {\bibfnamefont {M.}~\bibnamefont
  {Ferrero}}, \bibinfo {author} {\bibfnamefont {P.~S.}\ \bibnamefont
  {Cornaglia}}, \bibinfo {author} {\bibfnamefont {L.~De}\ \bibnamefont {Leo}},
  \bibinfo {author} {\bibfnamefont {O.}~\bibnamefont {Parcollet}}, \bibinfo
  {author} {\bibfnamefont {G.}~\bibnamefont {Kotliar}}, \ and\ \bibinfo
  {author} {\bibfnamefont {A.}~\bibnamefont {Georges}},\ }\bibfield  {title}
  {\enquote {\bibinfo {title} {Valence bond dynamical mean-field theory of
  doped mott insulators with nodal/antinodal differentiation},}\ }\href
  {\doibase 10.1209/0295-5075/85/57009} {\bibfield  {journal} {\bibinfo
  {journal} {Europhys. Lett.}\ }\textbf {\bibinfo {volume} {85}},\ \bibinfo
  {pages} {57009} (\bibinfo {year} {2009}{\natexlab{a}})}\BibitemShut {NoStop}%
\bibitem [{\citenamefont {Ferrero}\ \emph
  {et~al.}(2009{\natexlab{b}})\citenamefont {Ferrero}, \citenamefont
  {Cornaglia}, \citenamefont {De~Leo}, \citenamefont {Parcollet}, \citenamefont
  {Kotliar},\ and\ \citenamefont {Georges}}]{michelPRB}%
  \BibitemOpen
  \bibfield  {author} {\bibinfo {author} {\bibfnamefont {Michel}\ \bibnamefont
  {Ferrero}}, \bibinfo {author} {\bibfnamefont {Pablo~S.}\ \bibnamefont
  {Cornaglia}}, \bibinfo {author} {\bibfnamefont {Lorenzo}\ \bibnamefont
  {De~Leo}}, \bibinfo {author} {\bibfnamefont {Olivier}\ \bibnamefont
  {Parcollet}}, \bibinfo {author} {\bibfnamefont {Gabriel}\ \bibnamefont
  {Kotliar}}, \ and\ \bibinfo {author} {\bibfnamefont {Antoine}\ \bibnamefont
  {Georges}},\ }\bibfield  {title} {\enquote {\bibinfo {title} {Pseudogap
  opening and formation of fermi arcs as an orbital-selective mott transition
  in momentum space},}\ }\href {\doibase 10.1103/PhysRevB.80.064501} {\bibfield
   {journal} {\bibinfo  {journal} {Phys. Rev. B}\ }\textbf {\bibinfo {volume}
  {80}},\ \bibinfo {pages} {064501} (\bibinfo {year}
  {2009}{\natexlab{b}})}\BibitemShut {NoStop}%
\bibitem [{\citenamefont {Sordi}\ \emph {et~al.}(2010)\citenamefont {Sordi},
  \citenamefont {Haule},\ and\ \citenamefont {Tremblay}}]{sht}%
  \BibitemOpen
  \bibfield  {author} {\bibinfo {author} {\bibfnamefont {G.}~\bibnamefont
  {Sordi}}, \bibinfo {author} {\bibfnamefont {K.}~\bibnamefont {Haule}}, \ and\
  \bibinfo {author} {\bibfnamefont {A.-M.~S.}\ \bibnamefont {Tremblay}},\
  }\bibfield  {title} {\enquote {\bibinfo {title} {{Finite Doping Signatures of
  the Mott Transition in the Two-Dimensional Hubbard Model}},}\ }\href
  {\doibase 10.1103/PhysRevLett.104.226402} {\bibfield  {journal} {\bibinfo
  {journal} {Phys. Rev. Lett.}\ }\textbf {\bibinfo {volume} {104}},\ \bibinfo
  {pages} {226402} (\bibinfo {year} {2010})}\BibitemShut {NoStop}%
\bibitem [{\citenamefont {Sordi}\ \emph {et~al.}(2011)\citenamefont {Sordi},
  \citenamefont {Haule},\ and\ \citenamefont {Tremblay}}]{sht2}%
  \BibitemOpen
  \bibfield  {author} {\bibinfo {author} {\bibfnamefont {G.}~\bibnamefont
  {Sordi}}, \bibinfo {author} {\bibfnamefont {K.}~\bibnamefont {Haule}}, \ and\
  \bibinfo {author} {\bibfnamefont {A.-M.~S.}\ \bibnamefont {Tremblay}},\
  }\bibfield  {title} {\enquote {\bibinfo {title} {{Mott physics and
  first-order transition between two metals in the normal-state phase diagram
  of the two-dimensional Hubbard model}},}\ }\href {\doibase
  10.1103/PhysRevB.84.075161} {\bibfield  {journal} {\bibinfo  {journal} {Phys.
  Rev. B}\ }\textbf {\bibinfo {volume} {84}},\ \bibinfo {pages} {075161}
  (\bibinfo {year} {2011})}\BibitemShut {NoStop}%
\bibitem [{\citenamefont {Fratino}\ \emph {et~al.}(2022)\citenamefont
  {Fratino}, \citenamefont {Bag}, \citenamefont {Camjayi}, \citenamefont
  {Civelli},\ and\ \citenamefont {Rozenberg}}]{LorenzoPRB2022}%
  \BibitemOpen
  \bibfield  {author} {\bibinfo {author} {\bibfnamefont {L.}~\bibnamefont
  {Fratino}}, \bibinfo {author} {\bibfnamefont {S.}~\bibnamefont {Bag}},
  \bibinfo {author} {\bibfnamefont {A.}~\bibnamefont {Camjayi}}, \bibinfo
  {author} {\bibfnamefont {M.}~\bibnamefont {Civelli}}, \ and\ \bibinfo
  {author} {\bibfnamefont {M.}~\bibnamefont {Rozenberg}},\ }\bibfield  {title}
  {\enquote {\bibinfo {title} {Doping-driven resistive collapse of the mott
  insulator in a minimal model for ${\mathrm{vo}}_{2}$},}\ }\href {\doibase
  10.1103/PhysRevB.105.125140} {\bibfield  {journal} {\bibinfo  {journal}
  {Phys. Rev. B}\ }\textbf {\bibinfo {volume} {105}},\ \bibinfo {pages}
  {125140} (\bibinfo {year} {2022})}\BibitemShut {NoStop}%
\bibitem [{\citenamefont {Haule}\ and\ \citenamefont
  {Kotliar}(2007{\natexlab{b}})}]{hauleAVOIDED}%
  \BibitemOpen
  \bibfield  {author} {\bibinfo {author} {\bibfnamefont {Kristjan}\
  \bibnamefont {Haule}}\ and\ \bibinfo {author} {\bibfnamefont {Gabriel}\
  \bibnamefont {Kotliar}},\ }\bibfield  {title} {\enquote {\bibinfo {title}
  {Avoided criticality in near-optimally doped high-temperature
  superconductors},}\ }\href {\doibase 10.1103/PhysRevB.76.092503} {\bibfield
  {journal} {\bibinfo  {journal} {Phys. Rev. B}\ }\textbf {\bibinfo {volume}
  {76}},\ \bibinfo {pages} {092503} (\bibinfo {year}
  {2007}{\natexlab{b}})}\BibitemShut {NoStop}%
\bibitem [{\citenamefont {Gull}\ \emph {et~al.}(2011)\citenamefont {Gull},
  \citenamefont {Millis}, \citenamefont {Lichtenstein}, \citenamefont
  {Rubtsov}, \citenamefont {Troyer},\ and\ \citenamefont {Werner}}]{millisRMP}%
  \BibitemOpen
  \bibfield  {author} {\bibinfo {author} {\bibfnamefont {Emanuel}\ \bibnamefont
  {Gull}}, \bibinfo {author} {\bibfnamefont {Andrew~J.}\ \bibnamefont
  {Millis}}, \bibinfo {author} {\bibfnamefont {Alexander~I.}\ \bibnamefont
  {Lichtenstein}}, \bibinfo {author} {\bibfnamefont {Alexey~N.}\ \bibnamefont
  {Rubtsov}}, \bibinfo {author} {\bibfnamefont {Matthias}\ \bibnamefont
  {Troyer}}, \ and\ \bibinfo {author} {\bibfnamefont {Philipp}\ \bibnamefont
  {Werner}},\ }\bibfield  {title} {\enquote {\bibinfo {title} {{Continuous-time
  Monte~Carlo methods for quantum impurity models}},}\ }\href {\doibase
  10.1103/RevModPhys.83.349} {\bibfield  {journal} {\bibinfo  {journal} {Rev.
  Mod. Phys.}\ }\textbf {\bibinfo {volume} {83}},\ \bibinfo {pages} {349--404}
  (\bibinfo {year} {2011})}\BibitemShut {NoStop}%
\bibitem [{\citenamefont {Werner}\ \emph {et~al.}(2006)\citenamefont {Werner},
  \citenamefont {Comanac}, \citenamefont {de~Medici}, \citenamefont {Troyer},\
  and\ \citenamefont {Millis}}]{Werner:2006}%
  \BibitemOpen
  \bibfield  {author} {\bibinfo {author} {\bibfnamefont {Philipp}\ \bibnamefont
  {Werner}}, \bibinfo {author} {\bibfnamefont {Armin}\ \bibnamefont {Comanac}},
  \bibinfo {author} {\bibfnamefont {Luca}\ \bibnamefont {de~Medici}}, \bibinfo
  {author} {\bibfnamefont {Matthias}\ \bibnamefont {Troyer}}, \ and\ \bibinfo
  {author} {\bibfnamefont {Andrew~J.}\ \bibnamefont {Millis}},\ }\bibfield
  {title} {\enquote {\bibinfo {title} {Continuous-time solver for quantum
  impurity models},}\ }\href {\doibase 10.1103/PhysRevLett.97.076405}
  {\bibfield  {journal} {\bibinfo  {journal} {Phys. Rev. Lett.}\ }\textbf
  {\bibinfo {volume} {97}},\ \bibinfo {pages} {076405} (\bibinfo {year}
  {2006})}\BibitemShut {NoStop}%
\bibitem [{\citenamefont {S\'emon}\ \emph
  {et~al.}(2014{\natexlab{a}})\citenamefont {S\'emon}, \citenamefont {Yee},
  \citenamefont {Haule},\ and\ \citenamefont {Tremblay}}]{patrickSkipList}%
  \BibitemOpen
  \bibfield  {author} {\bibinfo {author} {\bibfnamefont {P.}~\bibnamefont
  {S\'emon}}, \bibinfo {author} {\bibfnamefont {Chuck-Hou}\ \bibnamefont
  {Yee}}, \bibinfo {author} {\bibfnamefont {Kristjan}\ \bibnamefont {Haule}}, \
  and\ \bibinfo {author} {\bibfnamefont {A.-M.~S.}\ \bibnamefont {Tremblay}},\
  }\bibfield  {title} {\enquote {\bibinfo {title} {{Lazy skip-lists: An
  algorithm for fast hybridization-expansion quantum Monte Carlo}},}\ }\href
  {\doibase 10.1103/PhysRevB.90.075149} {\bibfield  {journal} {\bibinfo
  {journal} {Phys. Rev. B}\ }\textbf {\bibinfo {volume} {90}},\ \bibinfo
  {pages} {075149} (\bibinfo {year} {2014}{\natexlab{a}})}\BibitemShut
  {NoStop}%
\bibitem [{\citenamefont {H\'ebert}\ \emph {et~al.}(2015)\citenamefont
  {H\'ebert}, \citenamefont {S\'emon},\ and\ \citenamefont
  {Tremblay}}]{Hebert:2015}%
  \BibitemOpen
  \bibfield  {author} {\bibinfo {author} {\bibfnamefont {Charles-David}\
  \bibnamefont {H\'ebert}}, \bibinfo {author} {\bibfnamefont {Patrick}\
  \bibnamefont {S\'emon}}, \ and\ \bibinfo {author} {\bibfnamefont {A.-M.~S.}\
  \bibnamefont {Tremblay}},\ }\bibfield  {title} {\enquote {\bibinfo {title}
  {Superconducting dome in doped quasi-two-dimensional organic mott insulators:
  A paradigm for strongly correlated superconductivity},}\ }\href {\doibase
  10.1103/PhysRevB.92.195112} {\bibfield  {journal} {\bibinfo  {journal} {Phys.
  Rev. B}\ }\textbf {\bibinfo {volume} {92}},\ \bibinfo {pages} {195112}
  (\bibinfo {year} {2015})}\BibitemShut {NoStop}%
\bibitem [{\citenamefont {Melnick}\ \emph {et~al.}(2021)\citenamefont
  {Melnick}, \citenamefont {Sémon}, \citenamefont {Yu}, \citenamefont
  {D'Imperio}, \citenamefont {Tremblay},\ and\ \citenamefont
  {Kotliar}}]{patrick21}%
  \BibitemOpen
  \bibfield  {author} {\bibinfo {author} {\bibfnamefont {Corey}\ \bibnamefont
  {Melnick}}, \bibinfo {author} {\bibfnamefont {Patrick}\ \bibnamefont
  {Sémon}}, \bibinfo {author} {\bibfnamefont {Kwangmin}\ \bibnamefont {Yu}},
  \bibinfo {author} {\bibfnamefont {Nicholas}\ \bibnamefont {D'Imperio}},
  \bibinfo {author} {\bibfnamefont {André-Marie}\ \bibnamefont {Tremblay}}, \
  and\ \bibinfo {author} {\bibfnamefont {Gabriel}\ \bibnamefont {Kotliar}},\
  }\bibfield  {title} {\enquote {\bibinfo {title} {Accelerated impurity solver
  for dmft and its diagrammatic extensions},}\ }\href {\doibase
  https://doi.org/10.1016/j.cpc.2021.108075} {\bibfield  {journal} {\bibinfo
  {journal} {Computer Physics Communications}\ }\textbf {\bibinfo {volume}
  {267}},\ \bibinfo {pages} {108075} (\bibinfo {year} {2021})}\BibitemShut
  {NoStop}%
\bibitem [{\citenamefont {S\'emon}\ \emph
  {et~al.}(2014{\natexlab{b}})\citenamefont {S\'emon}, \citenamefont {Sordi},\
  and\ \citenamefont {Tremblay}}]{patrickERG}%
  \BibitemOpen
  \bibfield  {author} {\bibinfo {author} {\bibfnamefont {P.}~\bibnamefont
  {S\'emon}}, \bibinfo {author} {\bibfnamefont {G.}~\bibnamefont {Sordi}}, \
  and\ \bibinfo {author} {\bibfnamefont {A.-M.~S.}\ \bibnamefont {Tremblay}},\
  }\bibfield  {title} {\enquote {\bibinfo {title} {Ergodicity of the
  hybridization-expansion monte carlo algorithm for broken-symmetry states},}\
  }\href {\doibase 10.1103/PhysRevB.89.165113} {\bibfield  {journal} {\bibinfo
  {journal} {Phys. Rev. B}\ }\textbf {\bibinfo {volume} {89}},\ \bibinfo
  {pages} {165113} (\bibinfo {year} {2014}{\natexlab{b}})}\BibitemShut
  {NoStop}%
\bibitem [{Note1()}]{Note1}%
  \BibitemOpen
  \bibinfo {note} {This has to be contrasted with Ref.~\protect \rev@citealpnum
  {Hebert:2015}, where, for the anisotropic Hubbard model within $2 \times 2$
  CDMFT, the $C_{2v}$ symmetry with mirrors along the plaquette diagonals was
  used as a representation for the one particle basis. This choice dictates
  that the superconducting order parameter transforms in space as the $A_2$
  representation of the $C_{2 v}$ symmetry group with mirrors along the
  plaquette diagonals.}\BibitemShut {Stop}%
\bibitem [{\citenamefont {Xu}\ \emph {et~al.}(2005)\citenamefont {Xu},
  \citenamefont {Kumar}, \citenamefont {Buldyrev}, \citenamefont {Chen},
  \citenamefont {Poole}, \citenamefont {Sciortino},\ and\ \citenamefont
  {Stanley}}]{water1}%
  \BibitemOpen
  \bibfield  {author} {\bibinfo {author} {\bibfnamefont {Limei}\ \bibnamefont
  {Xu}}, \bibinfo {author} {\bibfnamefont {Pradeep}\ \bibnamefont {Kumar}},
  \bibinfo {author} {\bibfnamefont {S.~V.}\ \bibnamefont {Buldyrev}}, \bibinfo
  {author} {\bibfnamefont {S.-H.}\ \bibnamefont {Chen}}, \bibinfo {author}
  {\bibfnamefont {P.~H.}\ \bibnamefont {Poole}}, \bibinfo {author}
  {\bibfnamefont {F.}~\bibnamefont {Sciortino}}, \ and\ \bibinfo {author}
  {\bibfnamefont {H.~E.}\ \bibnamefont {Stanley}},\ }\bibfield  {title}
  {\enquote {\bibinfo {title} {{Relation between the Widom line and the dynamic
  crossover in systems with a liquid liquid phase transition}},}\ }\href
  {\doibase 10.1073/pnas.0507870102} {\bibfield  {journal} {\bibinfo  {journal}
  {Proc. Natl. Acad. Sci. USA}\ }\textbf {\bibinfo {volume} {102}},\ \bibinfo
  {pages} {16558--16562} (\bibinfo {year} {2005})}\BibitemShut {NoStop}%
\bibitem [{\citenamefont {McMillan}\ and\ \citenamefont
  {Stanley}(2010)}]{supercritical}%
  \BibitemOpen
  \bibfield  {author} {\bibinfo {author} {\bibfnamefont {Paul~F.}\ \bibnamefont
  {McMillan}}\ and\ \bibinfo {author} {\bibfnamefont {H.~Eugene}\ \bibnamefont
  {Stanley}},\ }\bibfield  {title} {\enquote {\bibinfo {title} {Fluid phases:
  Going supercritical},}\ }\href {\doibase doi:10.1038/nphys1711} {\bibfield
  {journal} {\bibinfo  {journal} {Nat Phys}\ }\textbf {\bibinfo {volume} {6}},\
  \bibinfo {pages} {479--480} (\bibinfo {year} {2010})}\BibitemShut {NoStop}%
\bibitem [{\citenamefont {Sordi}\ \emph
  {et~al.}(2012{\natexlab{b}})\citenamefont {Sordi}, \citenamefont {S\'emon},
  \citenamefont {Haule},\ and\ \citenamefont {Tremblay}}]{ssht}%
  \BibitemOpen
  \bibfield  {author} {\bibinfo {author} {\bibfnamefont {G.}~\bibnamefont
  {Sordi}}, \bibinfo {author} {\bibfnamefont {P.}~\bibnamefont {S\'emon}},
  \bibinfo {author} {\bibfnamefont {K.}~\bibnamefont {Haule}}, \ and\ \bibinfo
  {author} {\bibfnamefont {A.-M.~S.}\ \bibnamefont {Tremblay}},\ }\bibfield
  {title} {\enquote {\bibinfo {title} {{Pseudogap temperature as a Widom line
  in doped Mott insulators}},}\ }\href {\doibase doi:10.1038/srep00547}
  {\bibfield  {journal} {\bibinfo  {journal} {Sci. Rep.}\ }\textbf {\bibinfo
  {volume} {2}},\ \bibinfo {pages} {547} (\bibinfo {year}
  {2012}{\natexlab{b}})}\BibitemShut {NoStop}%
\bibitem [{\citenamefont {Walsh}\ \emph
  {et~al.}(2019{\natexlab{a}})\citenamefont {Walsh}, \citenamefont {S\'emon},
  \citenamefont {Poulin}, \citenamefont {Sordi},\ and\ \citenamefont
  {Tremblay}}]{CaitlinSb}%
  \BibitemOpen
  \bibfield  {author} {\bibinfo {author} {\bibfnamefont {C.}~\bibnamefont
  {Walsh}}, \bibinfo {author} {\bibfnamefont {P.}~\bibnamefont {S\'emon}},
  \bibinfo {author} {\bibfnamefont {D.}~\bibnamefont {Poulin}}, \bibinfo
  {author} {\bibfnamefont {G.}~\bibnamefont {Sordi}}, \ and\ \bibinfo {author}
  {\bibfnamefont {A.-M.~S.}\ \bibnamefont {Tremblay}},\ }\bibfield  {title}
  {\enquote {\bibinfo {title} {{Thermodynamic and information-theoretic
  description of the Mott transition in the two-dimensional Hubbard model}},}\
  }\href {\doibase 10.1103/PhysRevB.99.075122} {\bibfield  {journal} {\bibinfo
  {journal} {Phys. Rev. B}\ }\textbf {\bibinfo {volume} {99}},\ \bibinfo
  {pages} {075122} (\bibinfo {year} {2019}{\natexlab{a}})}\BibitemShut
  {NoStop}%
\bibitem [{\citenamefont {Sordi}\ \emph {et~al.}(2013)\citenamefont {Sordi},
  \citenamefont {S\'emon}, \citenamefont {Haule},\ and\ \citenamefont
  {Tremblay}}]{sshtRHO}%
  \BibitemOpen
  \bibfield  {author} {\bibinfo {author} {\bibfnamefont {G.}~\bibnamefont
  {Sordi}}, \bibinfo {author} {\bibfnamefont {P.}~\bibnamefont {S\'emon}},
  \bibinfo {author} {\bibfnamefont {K.}~\bibnamefont {Haule}}, \ and\ \bibinfo
  {author} {\bibfnamefont {A.-M.~S.}\ \bibnamefont {Tremblay}},\ }\bibfield
  {title} {\enquote {\bibinfo {title} {{$c$-axis resistivity, pseudogap,
  superconductivity, and Widom line in doped Mott insulators }},}\ }\href
  {\doibase 10.1103/PhysRevB.87.041101} {\bibfield  {journal} {\bibinfo
  {journal} {Phys. Rev. B}\ }\textbf {\bibinfo {volume} {87}},\ \bibinfo
  {pages} {041101} (\bibinfo {year} {2013})}\BibitemShut {NoStop}%
\bibitem [{\citenamefont {Sordi}\ \emph {et~al.}(2019)\citenamefont {Sordi},
  \citenamefont {Walsh}, \citenamefont {S\'emon},\ and\ \citenamefont
  {Tremblay}}]{Giovanni:PRBcv}%
  \BibitemOpen
  \bibfield  {author} {\bibinfo {author} {\bibfnamefont {G.}~\bibnamefont
  {Sordi}}, \bibinfo {author} {\bibfnamefont {C.}~\bibnamefont {Walsh}},
  \bibinfo {author} {\bibfnamefont {P.}~\bibnamefont {S\'emon}}, \ and\
  \bibinfo {author} {\bibfnamefont {A.-M.~S.}\ \bibnamefont {Tremblay}},\
  }\bibfield  {title} {\enquote {\bibinfo {title} {Specific heat maximum as a
  signature of mott physics in the two-dimensional hubbard model},}\ }\href
  {\doibase 10.1103/PhysRevB.100.121105} {\bibfield  {journal} {\bibinfo
  {journal} {Phys. Rev. B}\ }\textbf {\bibinfo {volume} {100}},\ \bibinfo
  {pages} {121105} (\bibinfo {year} {2019})}\BibitemShut {NoStop}%
\bibitem [{\citenamefont {Walsh}\ \emph
  {et~al.}(2019{\natexlab{b}})\citenamefont {Walsh}, \citenamefont {S\'emon},
  \citenamefont {Sordi},\ and\ \citenamefont {Tremblay}}]{CaitlinOpalescence}%
  \BibitemOpen
  \bibfield  {author} {\bibinfo {author} {\bibfnamefont {C.}~\bibnamefont
  {Walsh}}, \bibinfo {author} {\bibfnamefont {P.}~\bibnamefont {S\'emon}},
  \bibinfo {author} {\bibfnamefont {G.}~\bibnamefont {Sordi}}, \ and\ \bibinfo
  {author} {\bibfnamefont {A.-M.~S.}\ \bibnamefont {Tremblay}},\ }\bibfield
  {title} {\enquote {\bibinfo {title} {Critical opalescence across the
  doping-driven mott transition in optical lattices of ultracold atoms},}\
  }\href {\doibase 10.1103/PhysRevB.99.165151} {\bibfield  {journal} {\bibinfo
  {journal} {Phys. Rev. B}\ }\textbf {\bibinfo {volume} {99}},\ \bibinfo
  {pages} {165151} (\bibinfo {year} {2019}{\natexlab{b}})}\BibitemShut
  {NoStop}%
\bibitem [{\citenamefont {Walsh}\ \emph {et~al.}(2020)\citenamefont {Walsh},
  \citenamefont {S\'emon}, \citenamefont {Poulin}, \citenamefont {Sordi},\ and\
  \citenamefont {Tremblay}}]{Caitlin:PRXQ2020}%
  \BibitemOpen
  \bibfield  {author} {\bibinfo {author} {\bibfnamefont {C.}~\bibnamefont
  {Walsh}}, \bibinfo {author} {\bibfnamefont {P.}~\bibnamefont {S\'emon}},
  \bibinfo {author} {\bibfnamefont {D.}~\bibnamefont {Poulin}}, \bibinfo
  {author} {\bibfnamefont {G.}~\bibnamefont {Sordi}}, \ and\ \bibinfo {author}
  {\bibfnamefont {A.-M.~S.}\ \bibnamefont {Tremblay}},\ }\bibfield  {title}
  {\enquote {\bibinfo {title} {Entanglement and classical correlations at the
  doping-driven mott transition in the two-dimensional hubbard model},}\ }\href
  {\doibase 10.1103/PRXQuantum.1.020310} {\bibfield  {journal} {\bibinfo
  {journal} {PRX Quantum}\ }\textbf {\bibinfo {volume} {1}},\ \bibinfo {pages}
  {020310} (\bibinfo {year} {2020})}\BibitemShut {NoStop}%
\bibitem [{\citenamefont {Walsh}\ \emph {et~al.}(2022)\citenamefont {Walsh},
  \citenamefont {Charlebois}, \citenamefont {S\'emon}, \citenamefont {Sordi},\
  and\ \citenamefont {Tremblay}}]{CaitlinSoundVelocity}%
  \BibitemOpen
  \bibfield  {author} {\bibinfo {author} {\bibfnamefont {C.}~\bibnamefont
  {Walsh}}, \bibinfo {author} {\bibfnamefont {M.}~\bibnamefont {Charlebois}},
  \bibinfo {author} {\bibfnamefont {P.}~\bibnamefont {S\'emon}}, \bibinfo
  {author} {\bibfnamefont {G.}~\bibnamefont {Sordi}}, \ and\ \bibinfo {author}
  {\bibfnamefont {A.-M.~S.}\ \bibnamefont {Tremblay}},\ }\bibfield  {title}
  {\enquote {\bibinfo {title} {Prediction of anomalies in the velocity of sound
  for the pseudogap of hole-doped cuprates},}\ }\href {\doibase
  10.1103/PhysRevB.106.235134} {\bibfield  {journal} {\bibinfo  {journal}
  {Phys. Rev. B}\ }\textbf {\bibinfo {volume} {106}},\ \bibinfo {pages}
  {235134} (\bibinfo {year} {2022})}\BibitemShut {NoStop}%
\bibitem [{\citenamefont {Fratino}\ \emph
  {et~al.}(2016{\natexlab{b}})\citenamefont {Fratino}, \citenamefont {S\'emon},
  \citenamefont {Sordi},\ and\ \citenamefont {Tremblay}}]{Lorenzo3band}%
  \BibitemOpen
  \bibfield  {author} {\bibinfo {author} {\bibfnamefont {L.}~\bibnamefont
  {Fratino}}, \bibinfo {author} {\bibfnamefont {P.}~\bibnamefont {S\'emon}},
  \bibinfo {author} {\bibfnamefont {G.}~\bibnamefont {Sordi}}, \ and\ \bibinfo
  {author} {\bibfnamefont {A.-M.~S.}\ \bibnamefont {Tremblay}},\ }\bibfield
  {title} {\enquote {\bibinfo {title} {Pseudogap and superconductivity in
  two-dimensional doped charge-transfer insulators},}\ }\href {\doibase
  10.1103/PhysRevB.93.245147} {\bibfield  {journal} {\bibinfo  {journal} {Phys.
  Rev. B}\ }\textbf {\bibinfo {volume} {93}},\ \bibinfo {pages} {245147}
  (\bibinfo {year} {2016}{\natexlab{b}})}\BibitemShut {NoStop}%
\bibitem [{\citenamefont {Scalapino}\ and\ \citenamefont
  {Trugman}(1996)}]{scalapino1996}%
  \BibitemOpen
  \bibfield  {author} {\bibinfo {author} {\bibfnamefont {D.~J.}\ \bibnamefont
  {Scalapino}}\ and\ \bibinfo {author} {\bibfnamefont {S.~A.}\ \bibnamefont
  {Trugman}},\ }\bibfield  {title} {\enquote {\bibinfo {title} {Local
  antiferromagnetic correlations and dx2-y2 pairing},}\ }\href {\doibase
  10.1080/01418639608240361} {\bibfield  {journal} {\bibinfo  {journal}
  {Philosophical Magazine B}\ }\textbf {\bibinfo {volume} {74}},\ \bibinfo
  {pages} {607--610} (\bibinfo {year} {1996})}\BibitemShut {NoStop}%
\bibitem [{\citenamefont {{Danilov}}\ \emph {et~al.}(2022)\citenamefont
  {{Danilov}}, \citenamefont {{van Loon}}, \citenamefont {{Brener}},
  \citenamefont {{Iskakov}}, \citenamefont {{Katsnelson}},\ and\ \citenamefont
  {{Lichtenstein}}}]{Danilov:2022}%
  \BibitemOpen
  \bibfield  {author} {\bibinfo {author} {\bibfnamefont {Michael}\ \bibnamefont
  {{Danilov}}}, \bibinfo {author} {\bibfnamefont {Erik G.~C.~P.}\ \bibnamefont
  {{van Loon}}}, \bibinfo {author} {\bibfnamefont {Sergey}\ \bibnamefont
  {{Brener}}}, \bibinfo {author} {\bibfnamefont {Sergei}\ \bibnamefont
  {{Iskakov}}}, \bibinfo {author} {\bibfnamefont {Mikhail~I.}\ \bibnamefont
  {{Katsnelson}}}, \ and\ \bibinfo {author} {\bibfnamefont {Alexander~I.}\
  \bibnamefont {{Lichtenstein}}},\ }\bibfield  {title} {\enquote {\bibinfo
  {title} {{Degenerate plaquette physics as key ingredient of high-temperature
  superconductivity in cuprates}},}\ }\href {\doibase
  10.1038/s41535-022-00454-6} {\bibfield  {journal} {\bibinfo  {journal} {npj
  Quantum Materials}\ }\textbf {\bibinfo {volume} {7}},\ \bibinfo {eid} {50}
  (\bibinfo {year} {2022})}\BibitemShut {NoStop}%
\bibitem [{\citenamefont {Bergeron}\ and\ \citenamefont
  {Tremblay}(2016)}]{DominicMEM}%
  \BibitemOpen
  \bibfield  {author} {\bibinfo {author} {\bibfnamefont {Dominic}\ \bibnamefont
  {Bergeron}}\ and\ \bibinfo {author} {\bibfnamefont {A.-M.~S.}\ \bibnamefont
  {Tremblay}},\ }\bibfield  {title} {\enquote {\bibinfo {title} {Algorithms for
  optimized maximum entropy and diagnostic tools for analytic continuation},}\
  }\href {\doibase 10.1103/PhysRevE.94.023303} {\bibfield  {journal} {\bibinfo
  {journal} {Phys. Rev. E}\ }\textbf {\bibinfo {volume} {94}},\ \bibinfo
  {pages} {023303} (\bibinfo {year} {2016})}\BibitemShut {NoStop}%
\bibitem [{\citenamefont {Reymbaut}\ \emph {et~al.}(2015)\citenamefont
  {Reymbaut}, \citenamefont {Bergeron},\ and\ \citenamefont
  {Tremblay}}]{AlexisPRB2015MEM}%
  \BibitemOpen
  \bibfield  {author} {\bibinfo {author} {\bibfnamefont {A.}~\bibnamefont
  {Reymbaut}}, \bibinfo {author} {\bibfnamefont {D.}~\bibnamefont {Bergeron}},
  \ and\ \bibinfo {author} {\bibfnamefont {A.-M.~S.}\ \bibnamefont
  {Tremblay}},\ }\bibfield  {title} {\enquote {\bibinfo {title} {Maximum
  entropy analytic continuation for spectral functions with nonpositive
  spectral weight},}\ }\href {\doibase 10.1103/PhysRevB.92.060509} {\bibfield
  {journal} {\bibinfo  {journal} {Phys. Rev. B}\ }\textbf {\bibinfo {volume}
  {92}},\ \bibinfo {pages} {060509} (\bibinfo {year} {2015})}\BibitemShut
  {NoStop}%
\bibitem [{\citenamefont {Yue}\ and\ \citenamefont
  {Werner}(2023)}]{yue2023maximum}%
  \BibitemOpen
  \bibfield  {author} {\bibinfo {author} {\bibfnamefont {Changming}\
  \bibnamefont {Yue}}\ and\ \bibinfo {author} {\bibfnamefont {Philipp}\
  \bibnamefont {Werner}},\ }\href@noop {} {\enquote {\bibinfo {title} {Maximum
  entropy analytic continuation of anomalous self-energies},}\ } (\bibinfo
  {year} {2023}),\ \Eprint {http://arxiv.org/abs/2303.16888} {arXiv:2303.16888
  [cond-mat.supr-con]} \BibitemShut {NoStop}%
\bibitem [{\citenamefont {Verret}\ \emph {et~al.}(2019)\citenamefont {Verret},
  \citenamefont {Roy}, \citenamefont {Foley}, \citenamefont {Charlebois},
  \citenamefont {S\'en\'echal},\ and\ \citenamefont
  {Tremblay}}]{Verret:PRB2019}%
  \BibitemOpen
  \bibfield  {author} {\bibinfo {author} {\bibfnamefont {S.}~\bibnamefont
  {Verret}}, \bibinfo {author} {\bibfnamefont {J.}~\bibnamefont {Roy}},
  \bibinfo {author} {\bibfnamefont {A.}~\bibnamefont {Foley}}, \bibinfo
  {author} {\bibfnamefont {M.}~\bibnamefont {Charlebois}}, \bibinfo {author}
  {\bibfnamefont {D.}~\bibnamefont {S\'en\'echal}}, \ and\ \bibinfo {author}
  {\bibfnamefont {A.-M.~S.}\ \bibnamefont {Tremblay}},\ }\bibfield  {title}
  {\enquote {\bibinfo {title} {Intrinsic cluster-shaped density waves in
  cellular dynamical mean-field theory},}\ }\href {\doibase
  10.1103/PhysRevB.100.224520} {\bibfield  {journal} {\bibinfo  {journal}
  {Phys. Rev. B}\ }\textbf {\bibinfo {volume} {100}},\ \bibinfo {pages}
  {224520} (\bibinfo {year} {2019})}\BibitemShut {NoStop}%
\bibitem [{\citenamefont {Alloul}\ \emph {et~al.}(1989)\citenamefont {Alloul},
  \citenamefont {Ohno},\ and\ \citenamefont {Mendels}}]{Alloul:1989}%
  \BibitemOpen
  \bibfield  {author} {\bibinfo {author} {\bibfnamefont {H.}~\bibnamefont
  {Alloul}}, \bibinfo {author} {\bibfnamefont {T.}~\bibnamefont {Ohno}}, \ and\
  \bibinfo {author} {\bibfnamefont {P.}~\bibnamefont {Mendels}},\ }\bibfield
  {title} {\enquote {\bibinfo {title} {{$^{89}\mathrm{Y}$ NMR evidence for a
  fermi-liquid behavior in
  ${\mathrm{YBa}}_{2}$${\mathrm{Cu}}_{3}$${\mathrm{O}}_{6+\mathrm{x}}$}},}\
  }\href {\doibase 10.1103/PhysRevLett.63.1700} {\bibfield  {journal} {\bibinfo
   {journal} {Phys. Rev. Lett.}\ }\textbf {\bibinfo {volume} {63}},\ \bibinfo
  {pages} {1700--1703} (\bibinfo {year} {1989})}\BibitemShut {NoStop}%
\bibitem [{\citenamefont {Takigawa}\ \emph {et~al.}(1989)\citenamefont
  {Takigawa}, \citenamefont {Hammel}, \citenamefont {Heffner},\ and\
  \citenamefont {Fisk}}]{Takigawa:1989}%
  \BibitemOpen
  \bibfield  {author} {\bibinfo {author} {\bibfnamefont {M.}~\bibnamefont
  {Takigawa}}, \bibinfo {author} {\bibfnamefont {P.~C.}\ \bibnamefont
  {Hammel}}, \bibinfo {author} {\bibfnamefont {R.~H.}\ \bibnamefont {Heffner}},
  \ and\ \bibinfo {author} {\bibfnamefont {Z.}~\bibnamefont {Fisk}},\
  }\bibfield  {title} {\enquote {\bibinfo {title} {Spin susceptibility in
  superconducting ${\mathrm{yba}}_{2}$${\mathrm{cu}}_{3}$${\mathrm{o}}_{7}$
  from $^{63}\mathrm{Cu}$ knight shift},}\ }\href {\doibase
  10.1103/PhysRevB.39.7371} {\bibfield  {journal} {\bibinfo  {journal} {Phys.
  Rev. B}\ }\textbf {\bibinfo {volume} {39}},\ \bibinfo {pages} {7371--7374}
  (\bibinfo {year} {1989})}\BibitemShut {NoStop}%
\bibitem [{\citenamefont {Ferrero}\ \emph {et~al.}(2007)\citenamefont
  {Ferrero}, \citenamefont {Leo}, \citenamefont {Lecheminant},\ and\
  \citenamefont {Fabrizio}}]{micheleJPCM2007}%
  \BibitemOpen
  \bibfield  {author} {\bibinfo {author} {\bibfnamefont {Michel}\ \bibnamefont
  {Ferrero}}, \bibinfo {author} {\bibfnamefont {Lorenzo~De}\ \bibnamefont
  {Leo}}, \bibinfo {author} {\bibfnamefont {Philippe}\ \bibnamefont
  {Lecheminant}}, \ and\ \bibinfo {author} {\bibfnamefont {Michele}\
  \bibnamefont {Fabrizio}},\ }\bibfield  {title} {\enquote {\bibinfo {title}
  {Strong correlations in a nutshell},}\ }\href {\doibase
  10.1088/0953-8984/19/43/433201} {\bibfield  {journal} {\bibinfo  {journal}
  {Journal of Physics: Condensed Matter}\ }\textbf {\bibinfo {volume} {19}},\
  \bibinfo {pages} {433201} (\bibinfo {year} {2007})}\BibitemShut {NoStop}%
\bibitem [{\citenamefont {Walsh}\ \emph
  {et~al.}(2019{\natexlab{c}})\citenamefont {Walsh}, \citenamefont {S\'emon},
  \citenamefont {Poulin}, \citenamefont {Sordi},\ and\ \citenamefont
  {Tremblay}}]{Caitlin:PRL2019}%
  \BibitemOpen
  \bibfield  {author} {\bibinfo {author} {\bibfnamefont {C.}~\bibnamefont
  {Walsh}}, \bibinfo {author} {\bibfnamefont {P.}~\bibnamefont {S\'emon}},
  \bibinfo {author} {\bibfnamefont {D.}~\bibnamefont {Poulin}}, \bibinfo
  {author} {\bibfnamefont {G.}~\bibnamefont {Sordi}}, \ and\ \bibinfo {author}
  {\bibfnamefont {A.-M.~S.}\ \bibnamefont {Tremblay}},\ }\bibfield  {title}
  {\enquote {\bibinfo {title} {Local entanglement entropy and mutual
  information across the mott transition in the two-dimensional hubbard
  model},}\ }\href {\doibase 10.1103/PhysRevLett.122.067203} {\bibfield
  {journal} {\bibinfo  {journal} {Phys. Rev. Lett.}\ }\textbf {\bibinfo
  {volume} {122}},\ \bibinfo {pages} {067203} (\bibinfo {year}
  {2019}{\natexlab{c}})}\BibitemShut {NoStop}%
\bibitem [{\citenamefont {Udagawa}\ and\ \citenamefont
  {Motome}(2015)}]{Udagawa_Motome:2015}%
  \BibitemOpen
  \bibfield  {author} {\bibinfo {author} {\bibfnamefont {Masafumi}\
  \bibnamefont {Udagawa}}\ and\ \bibinfo {author} {\bibfnamefont {Yukitoshi}\
  \bibnamefont {Motome}},\ }\bibfield  {title} {\enquote {\bibinfo {title}
  {Entanglement spectrum in cluster dynamical mean-field theory},}\ }\href
  {http://stacks.iop.org/1742-5468/2015/i=1/a=P01016} {\bibfield  {journal}
  {\bibinfo  {journal} {Journal of Statistical Mechanics: Theory and
  Experiment}\ }\textbf {\bibinfo {volume} {2015}},\ \bibinfo {pages} {P01016}
  (\bibinfo {year} {2015})}\BibitemShut {NoStop}%
\bibitem [{\citenamefont {Humeniuk}(2019)}]{Humeniuk2019}%
  \BibitemOpen
  \bibfield  {author} {\bibinfo {author} {\bibfnamefont {Stephan}\ \bibnamefont
  {Humeniuk}},\ }\bibfield  {title} {\enquote {\bibinfo {title} {Quantum state
  tomography on a plaquette in the two-dimensional hubbard model},}\ }\href
  {\doibase 10.1103/PhysRevB.100.115121} {\bibfield  {journal} {\bibinfo
  {journal} {Phys. Rev. B}\ }\textbf {\bibinfo {volume} {100}},\ \bibinfo
  {pages} {115121} (\bibinfo {year} {2019})}\BibitemShut {NoStop}%
\end{thebibliography}
\end{document}